\documentclass[12pt]{article}
\pdfoutput=1

\usepackage{putex}
\usepackage{graphicx}
\usepackage{caption}
\usepackage{amsmath}
\usepackage{array}
\usepackage{subcaption}
\usepackage{epstopdf}
\usepackage{enumerate}
\usepackage{cite}
\usepackage{youngtab}
\usepackage{tensor}
\usepackage{slashed}
\usepackage[aligntableaux=center]{ytableau}
\usepackage[utf8]{inputenc}
\usepackage{rotating}
\usepackage{bigfoot}
\usepackage[
      colorlinks=true,
      linkcolor=blue,
      urlcolor=blue,
      filecolor=black,
      citecolor=red,
      ]{hyperref}

\newcommand {\be} {\begin {equation}}
\newcommand {\ee} {\end {equation}}

\newcommand {\bes} {\begin {equation*}}
\newcommand {\ees} {\end {equation*}}

\newcommand{\es}[2] {\begin{equation} \label{#1} \begin{split} #2 \end{split} \end{equation}}

\newcommand{\cC}{{\mathcal C}}
\newcommand{\cD}{{\mathcal D}}

\newcommand{\cJ}{{\mathcal J}}

\newcommand{\cO}{{\mathcal O}}
\newcommand{\cP}{{\mathcal P}}

\newcommand{\cR}{{\mathcal R}}

\newcommand{\cW}{{\mathcal W}}

\newcommand{\beq}{\begin{equation}}
\newcommand{\eeq}{\end{equation}}

\def\ie{\begin{equation}\begin{aligned}}
\def\fe{\end{aligned}\end{equation}}

\numberwithin{equation}{section}

\def\<{\langle}
\def\>{\rangle}

\begin{document}

\institution{Exile}{Department of Particle Physics and Astrophysics, Weizmann Institute of Science, Rehovot, Israel}

\title{
Weizmann Lectures on the Numerical Conformal Bootstrap
}

\authors{Shai M.~Chester,\worksat{\Exile} }

\abstract{
These lectures were given at the Weizmann Institute in the spring of 2019. They are intended to familiarize students with the nuts and bolts of the numerical bootstrap as efficiently as possible. After a brief review of the basics of conformal field theory in $d>2$ spacetime dimensions, we discuss how to compute conformal blocks, formulate the crossing equations as a semi-definite programming problem, solve this problem using \texttt{SDPB} on a personal computer, and interpret the results. We include worked examples for all steps, including bounds for 3d CFTs with $\mathbb{Z}_2$ or $O(N)$ global symmetries. Each lecture includes a problem set, which culminate in a precise computation of the 3d Ising model critical exponents using the mixed correlator $\mathbb{Z}_2$ bootstrap. A \texttt{Mathematica} file is included that transforms crossing equations into the proper input form for \texttt{SDPB}.
}
\date{}

\maketitle

\tableofcontents

\section{Conformal Symmetry I}
\label{symmetry}
In the first lecture we will explain why conformal field theories are important, and then discuss the classical representation theory of operators as well as constraints on quantum correlation functions in a conformal field theory.\footnote{These lecture notes are heavily based on the previous lecture notes \cite{Rychkov:2016iqz,Simmons-Duffin:2016gjk}, for other reviews of the bootstrap approach to CFT see \cite{Poland:2018epd,Qualls:2015qjb}. Previous numerical bootstrap studies include \cite{Go:2019lke,Caracciolo:2009bx,Poland:2010wg,Rattazzi:2010gj,Rattazzi:2010yc,Vichi:2011ux,Poland:2011ey,Rychkov:2011et,ElShowk:2012ht,Liendo:2012hy,ElShowk:2012hu,Beem:2013qxa,Kos:2013tga,Alday:2013opa,Gaiotto:2013nva,Berkooz:2014yda,El-Showk:2014dwa,Nakayama:2014lva,Nakayama:2014yia,Alday:2014qfa,Chester:2014fya,Kos:2014bka,Caracciolo:2014cxa,Nakayama:2014sba,Beem:2014zpa,Chester:2014gqa,Simmons-Duffin:2015qma,Bobev:2015jxa,Kos:2015mba,Chester:2015qca,Beem:2015aoa,Iliesiu:2015qra,Poland:2015mta,Lemos:2015awa,Lin:2015wcg,Chester:2015lej,Chester:2016wrc,Kos:2016ysd,Lin:2016gcl,Nakayama:2016cim,Nakayama:2016jhq,Iha:2016ppj,Behan:2016dtz,Nakayama:2016knq,Echeverri:2016ztu,Li:2016wdp,Bae:2016jpi,Lemos:2016xke,Beem:2016wfs,Li:2017ddj,Cornagliotto:2017dup,Dymarsky:2017xzb,Chang:2017xmr,Dymarsky:2017yzx,Chang:2017cdx,Nakayama:2017vdd,Li:2017kck,Cornagliotto:2017snu,Agmon:2017xes,Rong:2017cow,Baggio:2017mas,Stergiou:2018gjj,Hasegawa:2018yqg,Liendo:2018ukf,Rong:2018okz,Atanasov:2018kqw,Behan:2018hfx,Kousvos:2018rhl,Cappelli:2018vir,Gowdigere:2018lxz,Li:2018lyb,Karateev:2019pvw,Chester:2017vdh,Stergiou:2019dcv,Agmon:2019imm,Binder:2020ckj,Chester:2019ifh,Chester:2020iyt}.}

\subsection{Why conformal field theory?}
\label{why}

Conformal field theories (CFTs) are quantum field theories (QFTs) that are invariant under the conformal group, which consists of changes of coordinates that preserve angles.\footnote{We will later define CFTs to be invariant only under the subgroup of the conformal group that is generated by the conformal algebra, which excludes conformal transformations that are not continuously connected to the identity, such as inversions. This distinction will not be important in these lectures.} These include translations and Lorentz transformations, that define the Poincare group of standard relativistic QFT, as well as rescalings. Most scale invariant QFTs are also conformally invariant,\footnote{For unitary theories in 2d and 4d, scale invariance implies conformal invariance under weak assumptions, see the reviews \cite{Polchinski:1987dy,Nakayama:2013is}.} so even though the conformal group is strictly bigger, it will guide our intuition to think of CFTs as scale invariant QFTs. As we will now review, conformal symmetry shows up all over physics: from the definition of QFTs, to phase transitions in condensed matter, to holography and quantum gravity. 

QFTs are defined as renormalization group (RG) flows from a microscopic theory at short distance scales (the UV), to a theory at long distance scales (the IR). Let us write the action $S$ for a QFT in $d$ dimensions as
\es{S}{
S=\int d^dx\sum_i g_i\cO_i\,,
}
where we have couplings $g_i$ for operators $\cO_i$. The action has zero mass dimension by definition, so for operators $\cO_i$ with mass dimension $\Delta_i$ we can define the dimensionless couplings $\bar g_i\equiv \mu^{\Delta_i-d}g_i$, where $\mu$ is the mass/energy scale, i.e inverse length scale, at which we study the theory. Under RG flow $\mu$ changes from small length scales (the UV) at $\mu\to\infty$ to long length scales (the IR) at $\mu\to0$, and all parameters of the QFT such as $\Delta_i$ change as well. When $\mu=0$, the theory becomes scale invariant by definition, and thus (as we assumed above) conformally invariant, and the $\Delta_i$ get frozen at particular values. In this limit $\bar g_i$ with $\Delta_i>d$ become very small, so we call the corresponding operator $\cO_i$ irrelevant, while operators with $\Delta_i=d$ and $\Delta_i<d$ are called relevant and marginal, respectively. Conversely, in the UV at $\mu\to\infty$ relevant operators can be ignored and irrelevant operators matter. For instance, consider $\phi^4$ theory in $d=3$ with action
\es{phi4}{
S=\int d^3x\left(\frac12(\partial\phi)^2+\frac12m^2\phi^2+\frac{1}{4!}g\phi^4\right)\,,
}
which has a $\mathbb{Z}_2$ symmetry $\phi\to-\phi$. In the UV, $\Delta_{\phi^2}=1$ and $\Delta_{\phi^4}=2$ are both relevant, so we can ignore them and the theory is free. In the IR the theory depends on the dimensionless ratio $g^2/m^2$. If $m^2\to-\infty$ then the potential $V(\phi)=\frac12m^2\phi^2+\frac{1}{4!}g\phi^4$ has a minimum at $\phi_0=\sqrt{\frac{-6m^2}{g}}$, so the $\mathbb{Z}_2$ symmetry $\phi\to-\phi$ is spontaneously broken, while if $m^2\to\infty$ then the minimum $\phi_0=0$ preserves the $\mathbb{Z}_2$ symmetry. When $\mu=0$, both of these phases have infinite mass, and so correspond to trivial empty CFTs with no local operators. For an intermediate value of $g^2/m^2$, however, we have a phase transition between the symmetry breaking and the symmetry preserving phases, which is described by a nontrivial interacting CFT. The scaling dimensions in the IR CFT are nontrivial numbers that can be very different from their UV values. In particular, we will see that only $\phi^2$ and $\phi$ remain relevant operators in the IR CFT.

We can generalize $\phi^4$ theory to a theory of $N$ real scalars $\phi_i$ with $i=1,\dots N$ by simply setting $\phi\to\phi_i$ in \eqref{phi4}, in which case the theory has an $O(N)$ global symmetry, and $O(1)\cong \mathbb{Z}_2$ recovers the original case. For $N=2$, the IR fixed point of this theory describes the quantum phase transition of thin-film superconductors \cite{PhysRevB.44.6883,Smakov:2005zz}, where Lorentz symmetry is emergent at the phase transition.


CFTs describe phase transitions not just in relativistic quantum field theories, but also in classical statistical systems of large numbers of interacting particles, where thermal fluctuations suppress quantum fluctuations \cite{Polyakov:1970xd}. These systems are described by lattice models in the UV. There are two basic length scales in lattice models: the distance $\ell$ between each lattice site, and the correlation length $\xi$ beyond which interactions between each site become exponentially small. When a lattice model undergoes a phase transition, $\xi$ becomes infinitely bigger than $\ell$, and we can approximate the model by a scale invariant continuum field theory in Euclidean signature, which is often a CFT. The physical content of the phase transition, such as power law scaling of correlation functions of spins, can all be expressed in terms of CFT data such as scaling dimensions.

For instance, the Ising model is a lattice of classical spins $s_i\in\{\pm1\}$ with nearest-neighbor interactions, and is often used to describe magnets. The partition function is \cite{Cardy:1996xt}
\es{Zising}{
Z_\text{Ising}=\sum_{\{s_i\}}e^{-\frac{J}{T}\sum_{\langle ij\rangle}s_is_j}\,,
}
where $T$ is the temperature, $J$ is the coupling between the spins, $i,j$ label lattice points, and $\langle ij\rangle$ denotes nearest neighbors. The correlation length is then defined in terms of the 2-point function of spins as
\es{xi}{
\langle s_is_j\rangle\sim\exp(-|i-j|/\xi)\quad\text{as}\quad |i-j|\to\infty\,.
}
For a certain critical value of $J_c/T_c$ the Ising model also flows to an interacting CFT, and the spins become fields $\sigma(x)$. The correlation length then diverges with temperature\footnote{This temperature should not be confused with the temperature of a $d+1$-dimensional QFT on $\mathbb{R}^{d-1}\times S^\beta$, which can be obtained by Wick rotating a Lorentzian QFT to Euclidean space and then compactifying the imaginary time direction on a circle whose radius is the inverse temperature $\beta$. This describes a QFT at finite temperature, unlike the statistical systems we consider here that are completely classical, and are only described by a field theory near their phase transitions.}   as
\es{nu}{
\xi\sim |(T-T_c)/T_c|^{-\nu}\,,\qquad  \text{as}\qquad T\to T_c\,,
}
while the spin 2-point function then obeys a power law relation
\es{eta}{
\langle \sigma(x_1)\sigma(x_2)\rangle\sim\frac{1}{|x_1-x_2|^{d-2+\eta}}\,,
}
where $d$ is the spacetime dimension. The quantities $\nu$ and $\eta$ are called critical exponents, which are nontrivial numbers that can be related to the scaling dimensions of relevant operators in the associated Euclidean CFT.

%

Another example of a classical statistical theory is the water-vapor phase transition. A volume of water is a collection of randomly moving atoms that on average is translationally and rotationally invariant. A volume of water can be in the liquid, solid, or gas phase depending on its pressure $P$ and temperature $T$. For a certain ratio $P/T$ the phase transition becomes continuous and is described by a Euclidean CFT.


In fact, the water/vapor CFT, the Ising model CFT, and the $\phi^4$ model CFT are precisely the same CFT! This can be explained by the general fact that the number of parameters that need to be tuned to reach an IR CFT corresponds to the number of relevant scalar operators that respect the symmetries of the UV theory.\footnote{This is assuming that the UV theory respects rotational invariance. For UV theories that break rotational invariance, one would also need to tune relevant operators with spin.} For the Ising model and the $\phi^4$ model, whose UV theory respects the $\mathbb{Z}_2$ symmetry, we had to tune one parameter ($m^2$ for $\phi^4$ or $J/T$ for Ising), so there can only be one relevant scalar operator in their IR CFTs that respect $\mathbb{Z}_2$. For the water/vapor case, the UV theory does not respect $\mathbb{Z}_2$, and we had to tune two parameters, the pressure and the temperature, so there must be two relevant scalar operators in the IR CFT. If there is a unique 3d CFT with only two relevant scalar operators (which we will show numerically is probably the case), or with only one relevant scalar operator that respects $\mathbb{Z}_2$, then all these three theories must have the same IR CFT, and indeed they do. This phenomena of IR equivalence between UV theories is known as critical universality in statistical and condensed matter physics, or duality in high energy physics. Critical universality is common because even though UV theories can depend on a huge number of parameters, such as the millions of atoms in water, IR CFTs are found by tuning the couplings of usually just a few relevant parameters. These dualities can even occur between classical statistical fixed points and relativistic quantum fixed points. For instance, the superfluid transition in ${}^4\text{He}$, which is driven by thermal fluctuations, is described by a Euclidean fixed point with $O(2)$ symmetry \cite{PhysRevB.68.174518} that is equivalent to the Lorentzian $O(2)$ fixed point described above for the thin-film superconductor phase transition, which is driven by quantum fluctuations at zero temperature.

One last reason to study CFTs is their relevance to quantum gravity. The only known non-perturbative formulation of quantum gravity is the AdS/CFT correspondence \cite{Maldacena:1997re}, which relates quantum gravity\footnote{Quantum gravity in this context means string theory, M-theory, or higher spin gravity.} in a negatively curved spacetime called Anti-de Sitter (AdS) to certain CFTs that live on the boundary of the spacetime. Quantum gravity is very difficult to study, but CFT is in principle completely well defined, so this correspondence is our best method of learning about quantum gravity in a controlled setting. We can even learn about quantum gravity in regular flat space by taking the radius of AdS to infinity, which corresponds to a certain limit of the CFT \cite{Heemskerk:2009pn,Penedones:2010ue}.

Conformal field theories have more symmetry than standard QFTs, so you might assume that they are easier to study. The problem is that most interacting QFTs become strongly coupled in the IR where the CFT exists, so we cannot study CFT by doing perturbation theory in one of the coupling, as we would for say quantum electrodynamics, where the dimensionless fine structure constant $\alpha\sim1/137$ makes a good expansion coefficient. We can see this in the $\phi^4$ model in 3d, where the coupling $g$ that appears in the Lagrangian scales positively with length, and so becomes big in the long distance limit where the CFT is defined. Several alternative methods are used to study such strongly coupled CFTs:
\begin{enumerate}
\item Perturbation theory in a parameter other than the coupling. CFTs typically become non-interacting when the number $N$ of flavors of matter or colors of a gauge group become large, or when the CFT is defined in a certain dimension $d$. We can then perform a perturbative expansion in large $N$ or $d-\epsilon$, and then extrapolate to the value of $N$ or $\epsilon$ that corresponds to the physical theory. For instance, in $d=4$ the $\phi^4$ coupling becomes marginal and the theory is free, so we can perform an expansion for small $\epsilon=4-d$ and then extrapolate to $\epsilon=1$ \cite{PhysRevLett.28.240}. Of course, such extrapolations are not really justified, but in practice they give answers that are reasonably close to experiment in some cases.
\item Monte Carlo simulations for a lattice model that describes the theory in the UV, which is a non-perturbative formulation by definition. For instance, recall that the Ising model is defined as a lattice model in the UV, so we can simply truncate that lattice to finite volume and put it on a computer. The difficulty is that the computer cost can be enormous, and passing to the continuum and extrapolating to infinite volume can introduce large errors.
\item The numerical conformal bootstrap.\footnote{The modern numerical conformal bootstrap was initiated in \cite{Rattazzi:2008pe}, which was inspired by the earlier works \cite{Ferrara:1973yt,Polyakov:1974gs}.} This method does not use any UV formulation of the theory, like a Lagrangian or a lattice model, but directly studies the constraints of conformal symmetry on the abstract IR CFT. We will find that conformal symmetry imposes infinite consistency constraints on correlations functions in a CFT. We can truncate to a finite number of constraints and optimize them numerically to place bounds on physical quantities in the CFT. These bounds are rigorous and can only improve as we add more constraints, unlike the perturbative methods that required unjustified extrapolations and do not necessarily improve at higher order. The bootstrap also uses the full power of conformal symmetry, which can make it more constraining than lattice studies that do not use conformal symmetry. 
\end{enumerate}

In Table \ref{bootWin} we compare the values $\Delta_{\sigma}$ and $\Delta_{\epsilon}$ of the lowest dimension $\mathbb{Z}_2$ odd and even operators in the Ising CFT as computed using the $\epsilon$ expansion, Monte Carlo simulations, and the numerical conformal bootstrap.\footnote{These scaling dimensions are related to the critical exponents $\eta$ and $\nu$ in \eqref{nu} and \eqref{eta} as $\Delta_\sigma=\frac12+\frac\eta2$ and $\Delta_\epsilon=3-1/\nu$.} While all these methods roughly agree, the bootstrap values are orders of magnitude more accurate. In the course of these lectures, you will in fact rederive these groundbreaking results yourself! 

\begin{table}
\begin{center}
\begin{tabular}{c|c|c|c}
 ref & Method & $\Delta_{\sigma}$ & $\Delta_{\epsilon}$   \\
 \hline 
\cite{Hasenbusch:2011yya}&  Monte Carlo & $0.51813(5)$   & $1.41275(25)$   \\
 \hline
 \cite{Pelissetto:2000ek}& $\epsilon$-expansion & $0.5182(25)$   & $1.4139(62)$    \\
 \hline
\cite{Kos:2016ysd} &Bootstrap&  $0.5181489(10)$   & $1.412625(10)$   \\
\end{tabular}
\caption{The most precise estimate of scaling dimensions for the critical Ising model via different methods.}
\label{bootWin}
\end{center}
\end{table}

\subsection{Conformal transformations}
\label{transforms}
Let us begin by defining the conformal group. The conformal group consists of transformations $ x\to x'$ that leave the metric $g_{\mu\nu}( x)$ unchanged up to an overall position dependent rescaling:
\es{groupDef}{
g'_{\mu\nu}( x')= \Omega^2( x)g_{\mu\nu}( x)\,,
}
where the spacetime indices $\mu,\nu=1,\dots,d$, and we will work in Euclidean space in these lectures. Consider an infinitesimal change of coordinates $x'^\mu=x^\mu+\epsilon^\mu(x)$, then the resulting infinitesimal change of the metric $g_{\mu\nu}$ around the flat Euclidean metric $\delta_{\mu\nu}=\diag(1,\dots,1)$ can be expressed as
\es{metSmall}{
g'_{\mu\nu}&=\frac{\partial x^\alpha}{\partial x'^\mu}\frac{\partial x^\beta}{\partial x'^\nu}g_{\alpha\beta}\\
&=(\delta_\mu^\alpha-\partial_\mu\epsilon^\alpha)(\delta_\nu^\beta-\partial_\nu\epsilon^\beta)g_{\alpha\beta}\\
&=g_{\mu\nu}-(\partial_\mu\epsilon_\nu+\partial_\nu\epsilon_\mu)+O(\epsilon^2)\,.\\
}
Comparing to \eqref{groupDef}, we see that an infinitesimal conformal transformation must obey the so-called conformal Killing equation
\es{kill}{
(\partial_\mu\epsilon_\nu+\partial_\nu\epsilon_\mu)=\frac{2}{d}\partial_\rho\epsilon^\rho\,,
}
where $\Omega^2(x)$ in \eqref{groupDef} was fixed to $1-\frac2d\partial_\rho\epsilon^\rho$ by taking the trace on both sides of \eqref{metSmall}. In $d>2$ dimensions the conformal Killing equation has four kinds of solutions:\footnote{In $d=1$ any smooth function is a conformal transformation (since there are no angles in 1d), while in $d=2$ there is an infinite set of solutions that comprise the larger Virasoro group, of which the global conformal group described here is a subgroup. For more details see \cite{yellow}.}
\es{infSol}{
\epsilon^\mu(x)&=a^\mu\qquad\quad\qquad\qquad\;\;\; \;\text{(translations)}\\
\epsilon^\mu(x)&=m^\mu{}_\nu x^\nu\qquad \qquad\qquad\;\,\text{(rotations)}\\
\epsilon^\mu(x)&=\alpha x^\mu\qquad\qquad\qquad\quad\;\; \text{(dilations)}\\
\epsilon^\mu(x)&=2(x_\nu b^\nu)x^\mu-b^\mu x^2\qquad \text{(special conformal transformation)}\,,\\
}
 where $a^\mu,m^\mu_\nu,\alpha,b^\mu$ are constants and $m_{\mu\nu}$ is antisymmetric. The Poincare subgroup of the conformal group has just translations and rotations,\footnote{In Minkowski space, the rotations would be Lorentz transformations.} while scale invariance includes dilations\footnote{Aka ``dilatations'' for those who aren't into the whole brevity thing.} but not necessarily  special conformal transformations (SCTs). The finite versions of these  transformations are identical to the infinitesimal ones for translation, rotations, and dilations, while for the SCT the finite version is
 \es{SCTfinite}{
 x'^\mu=\frac{x^\mu-b^\mu x^2}{1-2b^\nu x_\nu+b^2x^2}\qquad\text{or}\qquad \frac{x'^\mu}{x'^2}=\frac{x^\mu}{x^2}-b^\mu\,.
 }
 The second formulation shows that an SCT consists of an inversion $x^\mu\to x^\mu/x^2$, a translation by $b^\mu$, and then a second inversion.\footnote{Inversion is an element of the conformal group as defined by \eqref{groupDef}, but it is not continuously connected to the identity, and so is not part of the conformal group generated by the conformal algebra. It is the latter definition that we use to define CFTs, so inversions need not be a symmetry of CFT.}
 
 \subsection{Conformal Algebra}
 \label{algebra}
 
For a symmetry transformation that acts on a function of position as $f(x)\to f'(x')$ with infinitesimal parameter $a$, we define the generator $G\equiv a^{-1}(f'(x')-f(x))$ that represents an infinitesimal transformation at the same point $x$. For a finite transformation $a$ we exponentiate these generators as $e^{aG}$. From the definition of the infinitesimal transformations in \eqref{infSol}, we then define the generators of the conformal algebra as
 \es{generators}{
 P_\mu&=\partial_\mu\qquad\quad\qquad\qquad\quad\;\; \text{(translations)}\\
M_{\mu\nu}&=x_\nu\partial_\mu-x_\mu\partial_\nu\qquad \qquad\;\text{(rotations)}\\
D&= x^\mu\partial_\mu\qquad\qquad\qquad\quad\;\; \text{(dilations)}\\
K_\mu&=2(x_\nu \partial^\nu)x_\mu- x^2\partial_\mu\qquad \text{(special conformal transformation)}\,.\\
 }
 These generators have nonzero commutation relations:
 \es{confAlg}{
 [M_{\mu\nu},P_\rho]&=\delta_{\nu\rho}P_\mu-\delta_{\mu\rho}P_\nu\,,\\
 [M_{\mu\nu},K_\rho]&=\delta_{\nu\rho}K_\mu-\delta_{\mu\rho}K_\nu\,,\\
 [M_{\mu\nu},M_{\rho\sigma}]&=\delta_{\nu\rho}M_{\mu\sigma}-\delta_{\mu\rho}M_{\nu\sigma}+\delta_{\nu\sigma}M_{\rho\mu}-\delta_{\mu\sigma}M_{\rho\nu}\,,\\
 [K_\mu,P_\nu]&=2\delta_{\mu\nu}D-2M_{\mu\nu}\,,\\
  [D,P_\mu]&=P_\mu\,,\\
 [D,K_\mu]&=-K_\mu\,.\\
 }
The $M_{\mu\nu}$ are the familiar generators of the rotation group $SO(d)$ in $d$ Euclidean dimensions. In fact, the entire conformal algebra is isomorphic to $SO(d+1,1)$, i.e. the rotation group in $d+2$ dimensional Lorentzian spacetime. This isomorphism can be shown by writing the generators as
 \es{sodplus1}{
 L_{\mu\nu}=M_{\mu\nu}\,,\qquad  L_{-1,0}=D\,,\qquad  L_{0,\mu}=\frac{P_\mu+K_\mu}{2}\,,\qquad  L_{-1,\mu}=\frac{P_\mu-K_\mu}{2}\,, 
 }
 so that $L_{ab}$ for $a,b=\{-1,0,\dots,d\}$ satisfy the same commutation relations as $M_{\mu\nu}$ above except that there are now $d+2$ indices and the Euclidean metric $\delta_{\mu\nu}$ is replaced by the Lorentzian metric $\diag(-1,1,\dots 1)$. The conformal group $SO(d+1,1)$ must be non-compact since it contains non-compact generators like translations and SCTs, unlike the compact rotation group $SO(d)$.
 
 \subsection{Representations of conformal group}
 \label{reps}
 Let us now see how an operator $\cO(x)$ transforms under conformal symmetry. We will start by reviewing transformations under the Poincare generators $P_\mu$ and $M_{\mu\nu}$, that apply to an operator in any relativistic quantum field theory. We will then add scale invariance by considering $D$. Finally, we consider the full conformal group by adding $K_\mu$.
 
  We begin with the translation generator $P_\mu$, which we can exponentiate to $e^{x^\mu P_\mu}$ to translate any operator as $e^{x^\mu P_\mu}\cO(0)e^{-x^\mu P_\mu}=\cO(x)$. The infinitesimal version of this action is
  \es{xP}{
[P_{\mu},\cO(x)]&=\partial_\mu\cO(x)\,.\\
 }
  For rotation generators $M_{\mu\nu}$, we first consider their action on $\cO(0)$:
 \es{M0}{
 [M_{\mu\nu},\cO^a(0)]&=(S_{\mu\nu})_b{}^a\cO^b(0)\,,\\
 }
 where $a,b$ are indices for some irreducible representation (irrep) of $SO(d)$, which acts on $\cO$ with the matrix $S_{\mu\nu}$ that satisfies the same algebra \eqref{confAlg} as $M_{\mu\nu}$. We will mostly consider traceless symmetric representations of $SO(d)$, whose rank $\ell$ we will refer to as the spin. For instance, in $d=3$, the rotation group is $SO(3)$ and all bosonic irreps are trivially traceless symmetric and labeled by integers $\ell$ with dimension $2\ell+1$.\footnote{Fermionic irreps also exist with half-integer $\ell$.} To find the action of $M_{\mu\nu}$ on $\cO(x)=e^{x^\mu P_\mu}\cO(0)e^{-x^\mu P_\mu}$ starting from \eqref{M0}, we use the commutation relations \eqref{confAlg} and the Hausdorff formula
\es{haus}{
e^{-A}Be^{A}=B+[B,A]+\frac{1}{2!}[[B,A],A]+\frac{1}{3!}[[[B,A],A],A]\dots\,,
}
to get
\es{xM}{
[M_{\mu\nu},\cO^a(x)]&=((x_\nu\partial_\mu-x_\mu\partial_\nu)\delta_b{}^a+(S_{\mu\nu})_b{}^a)\cO^b(x)\,.\\
 }
 
Next, we move on to scale invariant theories by considering the action of the dilation $D$. Since $D$ commutes with $M_{\mu\nu}$, an operator $\cO(0)$ in a rotation irrep defined by \eqref{M0} can be simultaneously diagonalized with $D$ so that
 \es{D0}{
  [D,\cO(0)]&=\Delta\cO(0)\,,\\
 }
 where $\Delta$ is called the scaling dimension. The scaling dimension and spin are the two basic quantum numbers that describe any operator in a CFT. We can find the action on $\cO(x)$ just as we did above for the Poincare group, with the result 
 \es{xD}{
[D,\cO(x)]&=(x^\mu\partial_\mu+\Delta)\cO(x)\,.\\
}

Finally we consider operators transforming under the full conformal group including the generator $K_\mu$ of SCTs. From \eqref{confAlg}, we can compute
\es{raise}{
[D,[P_\mu,\cO(0)]]=(\Delta+1)[P_\mu,\cO(0)]\,,\qquad [D,[K_\mu,\cO(0)]]=(\Delta-1)[K_\mu,\cO(0)]\,,
}
so that $P_\mu$ and $K_\mu$ act as raising and lowering operators, respectively, for the scaling dimension. As we will see in the next lecture, unitary CFTs have a lower bound on $\Delta$, so there must exist an operator $\cO(0)_\text{primary}$, called a primary, that satisfies
\es{primary}{
[K_\mu,\cO(0)_\text{primary}]=0\,.
}
 These primaries generate the irreducible representations of the conformal group. For any primary operator $\cO(0)$ with dimension $\Delta$, there are an infinite tower of so-called descendent operators with dimensions $\Delta+n$ for $n\in\mathbb{N}$, that are obtained by acting $n$ times on $\cO(0)$ with $P_\mu$. For instance, $\cO(x)=e^{P_\mu x^\mu}\cO(0)$ is itself an infinite linear combination of descendent operators, and we can compute its transformation under $K_\mu$ similar to the computations above to get
 \es{xK}{
[K_{\mu},\cO(x)]&=(2(x_\nu \partial^\nu)x_\mu- x^2\partial_\mu+2\Delta x_\mu-2x^\nu S_{\mu\nu})\cO(x)\,.\\
}
 For now on, when we talk about operators in a CFT we will always assume that they are primary operators, unless otherwise specified. These primaries are completely described by specifying their scaling dimension $\Delta$ and spin $\ell$, as well as their possible representation under some non-spacetime symmetry group.
  
In \eqref{xP}, \eqref{xM}, \eqref{xD}, and \eqref{xK} we showed how an operator $\cO(x)$ transforms under an infinitesimal conformal transformation with conformal generator $G$. We can exponentiate these commutation relations to find the action of a finite transformation $U=e^{G}$. Consider a general finite conformal transformation $x\to x'$ that transforms the metric up to a factor $\Omega^2(x)$ as in \eqref{groupDef}. This transformations acts on  an operator $\cO(x)^a$ in a representation $R$ with index $a$ as
  \es{finiteGen}{
  U\cO^a(x)U^{-1}&=\Omega(x')^\Delta D(R(x'))_b{}^a\cO^b(x')\,,\qquad
  \frac{\partial x'^\mu}{\partial x^\nu}\equiv \Omega(x')R^\mu{}_\nu(x')\,,
  }
where $R^\mu{}_\nu(x')$ is an element of $SO(d)$ and $D(R)_b{}^a$ is a matrix implementing the action of $R$ on $\cO$.  For instance, consider the finite dilation $x'=\lambda x$ for constant $\lambda$ acting on a scalar operator $\phi(x)$. Applying \eqref{finiteGen} we get 
\es{finiteD}{
e^D\phi(x)e^{-D}=\lambda^\Delta \phi(\lambda x)\,,
}
  which reduces to \eqref{xD} if we set $\lambda=1+\epsilon$, rescale $D\to \epsilon D$ and match $O(\epsilon)$ terms. We see that the scaling dimension $\Delta$, defined as the eigenvalue of $D$ in \eqref{D0}, indeed corresponds to the usual definition of how a function $\phi(x)$ transforms under rescaling $x$.

  \subsection{Correlation functions}
  \label{correlatorsSec}
So far we have considered the kinematics of operators in CFT. For dynamics, we should consider correlations functions in some quantization, whose details will be unimportant in this lecture. Correlation functions of $n$ operators $\cO(x)^{a_i}_{i}$ for $i=1,\dots, n$ with index $a_i$ in irrep $R_i$ of $SO(d)$ must be invariant  at separated points under finite conformal transformations \eqref{finiteGen}, which define the conformal Ward identity:
\es{ward}{
\langle \cO^{a_1}_{1}(x_1)\dots\cO^{a_n}_{n}(x_n)\rangle=\Omega(x'_1)^{\Delta_1}D(R_1(x'))_{b_1}{}^{a_1}\dots\Omega(x'_n)^{\Delta_n}D(R_n(x'))_{b_n}{}^{a_n}\langle \cO^{b_1}_{1}(x'_1)\dots\cO^{b_n}_{n}(x'_n)\rangle\,.
}
This immediately implies that all 1-point functions must vanish, except for the unit operator which has $\Delta=0$.\footnote{One can also consider CFTs on non-trivial manifolds that are conformally flat, such as $S^d$, which preserve a subset of the full conformal group. 1-point functions can then depend on parameters that characterize the manifold, such as the radius of the $S^d$.} For two point functions, rotations and translations imply that $\langle\phi_i(x_1)\phi_j(x_2)\rangle=f(|x_1-x_2|)$, while scale invariance furthermore fixes $f\propto|x_1-x_2|^{-\Delta_i-\Delta_j}$. We can plug this ansatz into the Ward identity \eqref{ward} for arbitrary conformal transformation, which importantly also includes SCTs, which finally fixes the 2-point function to be
\es{2point}{
\langle\phi_i(x_1)\phi_j(x_2)\rangle=\frac{C\delta_{ij}}{x_{12}^{2\Delta_i}}\,,
}
where $x_{12}\equiv|x_1-x_2|$, and $C$ defines the normalization of the operators $\phi_i$. For three point functions of three different scalar operators, a similar exercise shows that 
\es{3point}{
\langle\phi_i(x_1)\phi_j(x_2)\phi_j(x_3)\rangle=\frac{\lambda_{\phi_i\phi_j\phi_k}}{x_{12}^{\Delta_i+\Delta_j-\Delta_k}x_{23}^{\Delta_j+\Delta_k-\Delta_i}x_{31}^{\Delta_k+\Delta_i-\Delta_j}}\,,
}
where the constant $\lambda_{\phi_i\phi_j\phi_k}$ is called the OPE coefficient, whose name will be justified later. As will be shown in the problem set, we can generalize these calculations to operators with spin. For 2-point functions of operators $\cO^{\mu_1\dots\mu_\ell}$ in the rank-$\ell$ traceless symmetric irrep of $SO(d)$, the operators must have the same dimension and spin, in which case we get
\es{2point2}{
\langle\cO^{\mu_1\dots\mu_\ell}(x_1)\cO_{\nu_1\dots\nu_\ell}(x_2)\rangle&=C^{(\ell)}\left(\frac{I^{(\mu_1}{}_{\nu_1}(x_{12})\dotsb I^{\mu_\ell)}{}_{\nu_\ell}(x_{12})}{x_{12}^{2\Delta}}-\text{traces}\right)\,,\\
&I^\mu{}_\nu\equiv\delta^\mu{}_\nu-2\frac{x^\mu x_\nu}{x^2}\,,
}
where we can symmetrize in either the $\mu$'s or the $\nu$'s, and the traces apply to the $\mu$'s and $\nu$'s separately. For a 3-point function of two scalar operators $\phi_i$ and $\phi_j$ and a spin $\ell$ operator, which is in fact the only $SO(d)$ irrep for which the 3-point function with two other scalars does not vanish, we get
\es{3point2}{
\langle\phi_i(x_1)\phi_j(x_2)\cO_{\Delta,\ell}^{\mu_1\dots\mu_\ell}(x_3)\rangle&=\frac{\lambda_{\phi_i\phi_j\cO_{\Delta,\ell}}(Z^{\mu_1}\dotsb Z^{\mu_\ell}-\text{traces})}{x_{12}^{\Delta_i+\Delta_j-\Delta+\ell}x_{23}^{\Delta_j+\Delta-\Delta_i-\ell}x_{31}^{\Delta+\Delta_i-\Delta_j-\ell}}\,,\\
Z^\mu&\equiv\frac{x_{13}^\mu}{x_{13}^2}-\frac{x_{23}^\mu}{x_{23}^2}\,.
}
The antisymmetry of each $Z^\mu$ under permuting $1\leftrightarrow2$ implies that
\es{OPEsign}{
\lambda_{\phi_i\phi_j\cO_{\Delta,\ell}}=(-1)^\ell \lambda_{\phi_j\phi_i\cO_{\Delta,\ell}}\,,
}
so that the spin $\ell$ must be even when $i=j$. While $C$ and $C^{(\ell)}$ in the two point functions can be set to one by normalizing the operators, the OPE coefficients cannot be defined away and are real physicals parameters of the CFT. 

For $n$-point functions in $d$ dimensions, we can construct the following number of invariants:
\es{invariants}{
n\geq d+1:&\qquad nd-\frac12(d+1)(d+2)\,,\\
n< d+1:&\qquad \frac12n(n-3)\,.\\
}
When $n\geq d+1$, the number of invariants is simply the number of points $nd$ minus the dimension of the conformal group $SO(d+1,1)$. If $n< d+1$, however, then we must add to this the dimension of the nontrivial stability group $SO(d+2-n)$.\footnote{To see this, let us use conformal transformations to fix two of the $n$ points to zero and infinity, respectively, then the remaining $n-2$ points define a hyperplane, and the stability group is the rotation group $SO(d+2-n)$ perpendicular to the hyperplane} For $n=4$, we find that for any $d\geq2$ there are two invariants, called conformal cross ratios, which we can take to be
\es{uv}{
u\equiv\frac{x_{12}^2x_{34}^2}{x_{13}^2x_{24}^2}\,,\qquad v\equiv\frac{x_{23}^2x_{14}^2}{x_{13}^2x_{24}^2}\,.
}
These are manifestly invariant under dilations, translations, rotations, and inversions, and thus SCTs as well. Conformal symmetry then fixes a 4-point function of identical scalar operators to be
 \es{4point}{
\langle \phi(x_1)\phi(x_2)\phi(x_3)\phi(x_4)\rangle=\frac{g(u,v)}{x_{12}^{2\Delta_\phi}x_{34}^{2\Delta_\phi}}\,,
}
where $\Delta_\phi$ is the dimension of the operator and $g(u,v)$ is so far an arbitrary function. The ordering of operators inside a correlation function does not matter in Euclidean space, so the LHS of \eqref{4point} is invariant under simultaneous permutations of $x_i$, but the RHS is not manifestly invariant. To make the RHS invariant we require the so-called crossing equations
\es{crossing2}{
g(u,v)=g(u/v,1/v)\,,\qquad v^{\Delta_\phi}g(u,v)=u^{\Delta_\phi}g(v,u)\,,
}
where the first constraint comes from swapping $1\leftrightarrow2$ and the second from swapping $1\leftrightarrow3$. We will find that one of these constraints can be chosen to be trivially satisfied, while the other imposes infinite non-perturbative constraints on scaling dimensions and OPE coefficients, which are the basis of the conformal bootstrap.

\pagebreak

\subsection{Problem Set 1}
\label{hw1}
\begin{enumerate}
\item  The embedding space formalism \cite{Dirac:1936fq,Mack:1969rr,Boulware:1970ty,Ferrara:1973eg,Weinberg:2010fx,Costa:2011mg} considers the Euclidean conformal group $SO(d+1,1)$ as a $d+2$ Lorentz group with coordinates $X^a$ for $a=1,\dots,d+2$, where $X^{d+2}$ is the time direction. The non-linear constraints of conformal invariance in $d$ dimensions then become the linear constraints of Lorentz invariance in $d+2$ dimensions. We relate $X^a$ to the usual $x^\mu$ coordinates by considering a section of the null lightcone $X^2=0$:
\es{Xtox}{
x^\mu=X^\mu\,,\qquad X^+\equiv X^{d+2}+X^{d+1}=1\,,\qquad X^-\equiv X^{d+2}-X^{d+1}=x^2\,.
}
For operators in the spin $\ell$ traceless symmetric irrep $\cO^{\mu_1\dots\mu_\ell}(x)$ of $SO(d)$, it is useful to contract with $\ell$ null polarization vectors $y_{\mu_i}$ with $y^2=0$ to define $\cO(x,y)\equiv \cO^{\mu_1\dots\mu_\ell}(x) y_{\mu_1}\dots y_{\mu_\ell}$. We can recover $\cO^{\mu_1\dots\mu_\ell}(x) $ from $\cO(x,y)$ by applying the differential operators
\es{OfromO}{
\cO^{\mu_1\dots\mu_\ell}(x) =&\frac{1}{\ell!(d/2-1)_\ell}D_{\mu_1}\cdots D_{\mu_\ell}\cO(x,y)\,,\\
 D_{\mu_i}\equiv& \left(\frac d2-1+y\cdot\frac{\partial}{\partial y}\right)\frac{\partial}{\partial y^{\mu_i}}-\frac12y_{\mu_i}\frac{\partial^2}{\partial y \cdot\partial y}\,.
}
We can similarly define $\cO(X,Y)\equiv \cO^{a_1\dots a_\ell}(X)Y_{\mu_1}\dots Y_{\mu_\ell}$ in the embedding formalism, where now $Y$ are $d+2$-dimensional null vectors $Y^2=0$. We can relate $Y$ to $y$ as
\es{Ytoy}{
y^\mu=Y^\mu\,,\qquad Y^+\equiv Y^{d+2}+Y^{d+1}=0\,,\qquad Y^-\equiv Y^{d+2}-Y^{d+1}=2x\cdot y\,,
}
so that $Y\cdot X=0$.
\begin{enumerate}
\item For scalar operators $\phi(X)$ in embedding space, the conformal ward identity \eqref{ward} is implemented by demanding homogeneity 
\es{homo}{
\phi(\lambda X)=\lambda^{-\Delta}\phi(X)\,,
}
for any real $\lambda$. Use $d+1$ Lorentz invariance and \eqref{homo} to derive 2 and 3-point functions in embedding space, then recover \eqref{2point} and \eqref{3point} using \eqref{Xtox}.
\item  For 3-point functions $\langle\phi(x_1)\phi(x_2)\cO(x_3)\rangle$, where $\phi$ are scalars and $\cO$ transforms in $SO(d)$ irrep $R$, show that $R$ must be in the traceless symmetric irrep of $SO(d)$.
\item For a spin $\ell$ operator $\cO(X,Y)$ in embedding space, we impose the constraints
\es{homo2}{
\cO(\lambda X,\alpha Y)=\lambda^{-\Delta}\alpha^\ell\cO(X,Y)\,,\qquad \cO(X,Y+\beta X)=\cO(X,Y)\,,
}
for any real $\lambda,\alpha,\beta$. The $\lambda$ constraint is the homogeneity condition \eqref{homo}, the $\alpha$ constraint is by definition of the spin $\ell$ polarization vectors, and the $\beta$ constraint is a transversality condition that ensures that $\cO(x,y)$ describes a symmetric traceless irrep of $SO(d)$. Use \eqref{homo2} and $d+1$ Lorentz invariance to derive the 2-point function of two spin $\ell$ operators, and the 3-point function of two scalars and a spin $\ell$ operator, then recover \eqref{2point2} and \eqref{3point2} using \eqref{Xtox} and \eqref{Ytoy}.
\end{enumerate} 

\item Consider the generalization of \eqref{4point} to a 4-point function $\langle\phi_i(x_1)\phi_j(x_2)\phi_k(x_3)\phi_l(x_4)\rangle$ of scalars $\phi$ with different scaling dimensions, so that $ g_{ijkl}(u,v)$ now depends on both the labelling and ordering of the operators. 
\begin{enumerate}
\item Compute the $x_i$-dependent prefactor of $g_{ijkl}(u,v)$ as fixed by conformal symmetry, up to convention dependent factors of $u$ and $v$. This generalizes the factor $x_{12}^{-2\Delta_\phi}x_{34}^{-2\Delta_\phi}$ in the identical scalar case.
\item Derive the crossing relations between the $g_{ijkl}(u,v)$ for different permutations of $ijkl$, which generalizes \eqref{crossing2}. 
\item Now fix $i=j$, and show in this case that one of the crossing relations becomes a constraint on $g_{iikl}$.
\end{enumerate}
\end{enumerate}
 
 \pagebreak
 
 \section{Conformal Symmetry II}
 \label{symmetry2}
 In the previous lecture we did not need to specify what quantization we chose for the states we used to define correlators. We will now show how a good choice of quantization will help us derive nontrivial constraints on scaling dimensions and OPE coefficients.
 
 \subsection{Radial quantization}
 \label{rad}

 For any quantum theory in $d$ dimensions, states are defined as functions of a one-dimensional subspace, which we call the ``time'' direction, and for each point in this subspace we define a different Hilbert space over the orthogonal $d-1$ directions, which we call a slice. Expectation values of states in different slices are defined by connecting the different Hilbert spaces using an evolution operator $U$ that changes the ``time'' of a state. Correlation functions of operators sandwiched between arbitrary states are then defined by ``time'' ordering in the usual sense, which ensures that these functions are always finite. 
 
 It is convenient to choose the ``time'' direction to respect the spacetime symmetries of the theory, so that $U=e^G$ for some generator $G$ of the spacetime group. The Hilbert spaces are then related by symmetry, and states are labelled by eigenvalues of $G$ as well as generators of symmetries that preserve each slice. For instance, in standard relativistic QFT the ``time'' direction is the usual time direction, the slices live in the space directions, and states at different times are related by the time evolution operator $e^{iP_0}$, where $P_0$ is the Hamiltonian.\footnote{It is conventional to put an $i$ here so that $P_0$ is Hermitian.} States are then labelled by their energy-momenta $P^\mu|k\rangle=k^\mu|k\rangle$, and correlation functions are defined to be time ordered.
 
 For a scale invariant QFT, we will find it more convenient to define the ``time'' direction as the radial direction $r$ around the origin $x=0$, so that slices are $(d-1)$-dimensional spheres $S^{d-1}$ around $x=0$ with a given radius $r$. Radial evolution is then generated by the dilation operator as $e^{D}$, and correlation function are radially ordered. States $|\cO\rangle$ are labelled by their eigenvalues under $D$ as well as the $SO(d)$ rotations that preserve each $S^{d-1}$:
 \es{stateLabel}{
 D|\cO\rangle=\Delta|\cO\rangle\,,\qquad M_{\mu\nu}|\cO\rangle=S_{\mu\nu}|\cO\rangle\,,
 }
 where $\Delta$ is the scaling dimension and $S_{\mu\nu}$ implements the transformation under rotation for a given representation $R$ of $SO(d)$. For instance, the vacuum $|0\rangle$ is defined to be an $SO(d)$ singlet that vanishes under dilations
 \es{vacuum}{
 D|0\rangle=0\,.
 }
 
 A remarkable property of scale invariant theories is that local operators are in one-to-one correspondence with states.\footnote{Non-local operators like Wilson lines do not have this correspondance.} For a general QFT, operators are defined at points in spacetime while states are defined along an entire spatial slice. For scale invariant theories, though, an entire spatial slice can be transformed using dilations to a single point, the origin, where it can be identified with an operator at that point.
 
 In more detail, a local operator $\cO(0)$ at the origin with scaling dimension $\Delta$ as defined in \eqref{D0} generates a state according to the correspondance
 \es{stateOp}{
\cO(0)\leftrightarrow \cO(0)|0\rangle=|\cO\rangle\,,
 }
 which satisfies the definition \eqref{stateLabel} of $|\cO\rangle$ as
 \es{stateOp2}{
D\cO(0)|0\rangle=[D,\cO(0)]|0\rangle+\cO D|0\rangle=\Delta\cO(0)|0\rangle\,,
 }
 where in the second equality we used \eqref{D0} as well as the fact that $D$ kills the vacuum. Conversely, given a state $|\cO\rangle$, we can construct an operator $\cO$ by defining its correlators with all other local operators $\cO_i$ as 
 \es{makeOp}{
 \langle0| \cO_i(x_1)\cO_i(x_2)\dots \cO(0)|0\rangle= \langle0| \cO_i(x_1)\cO_i(x_2)\dots |\cO\rangle\,.
 }
 This is called the state-operator correspondance.

\subsection{Unitarity and reflection positivity}
\label{UandP}

An important aspect of a quantization scheme is how we define conjugation of operators. For instance, in a unitary Lorentzian QFT in the standard quantization described before, all spacetime generators $G$ are Hermitian operators so that they generate unitary transformation $e^{iG}$ that preserve the positivity of correlation functions. For a scalar Hermitian operator $\phi(0,0)$ at the spacetime origin, we can transform to an arbitrary spacetime point
\es{unitary}{
\phi(t,\vec x)=e^{iP_0t-i\vec x\cdot \vec P}\phi(0,0)e^{-iP_0t+i\vec x\cdot \vec P}
}
using the Hermitian energy-momentum generators $(P_0,\vec P)$. The LHS is now also Hermitian, since conjugating the RHS both changes the sign of $i$ and reverses the order of the operators. 

If we now Wick rotate to Euclidean space by defining $t_E\equiv -it$ and $\phi_E(t_E,\vec x)\equiv\phi(-it_E,\vec x)$, we find
\es{unitaryE}{
\phi_E(t_E,\vec x)=e^{P_0t_E-i\vec x\cdot \vec P}\phi(0,0)e^{-P_0t_E+i\vec x\cdot \vec P}\,.
}
The Euclidean scalar operator only remains invariant under conjugation if we define
\es{unitaryE2}{
\phi_E(t_E,\vec x)^\dagger=\phi_E(-t_E,\vec x)\,.
}
The generalization to operators with spin $\ell$ is
\es{unitaryEspin}{
\cO_E^{\mu_1\dots\mu_\ell}(t_E,\vec x)^\dagger=\Theta^{\mu_1}{}_{\nu_1}\dotsb\Theta^{\mu_\ell}{}_{\nu_\ell}\cO_E^{\nu_1\dots\nu_\ell}(-t_E,\vec x)\,,
}
where $\Theta^\mu{}_\nu=\delta^\mu{}_\nu-2\delta^\mu{}_0\delta^0{}_\nu$ is a reflection in the Euclidean time direction. We thus see that unitarity in Lorentzian space corresponds under Wick rotation to a reality condition that includes flipping the sign of the $t_E$ direction in Euclidean space, which we refer to as reflection positivity or unitarity interchangeabley. Conversely, the Osterwalder-Schrader theorem states that a reflection positive Euclidean QFT can be analytically continued to a unitary Lorentzian QFT under mild assumptions \cite{yellow2}. 

One important implication of unitarity for a CFT is that the OPE coefficients $\lambda_{\phi_i\phi_j\cO_{\Delta,\ell}}$ of two real scalars $\phi_i$ and $\phi_j$ and a real spin $\ell$ operator $\cO_{\Delta,\ell}$ is real. This is easiest to see in Lorentzian signature where real operators are just Hermitian. We can then consider the three operators at spacelike separated $x_{ij}^2>0$ so that they commute, in which case 
\es{realOPE}{
\langle0|\phi_i(x_1)\phi_j(x_2)\cO_{\Delta,\ell}^{\mu_1\dots\mu_\ell}(x_3)|0\rangle^\dagger=\langle0|\phi_i(x_1)\phi_j(x_2)\cO_{\Delta,\ell}^{\mu_1\dots\mu_\ell}(x_3)|0\rangle\,,
}
which implies that $\lambda_{\phi_i\phi_j\cO_{\Delta,\ell}}$ in \eqref{3point2} must be real.

We will also find it useful to consider unitarity in radial quantization. We can relate radial quantization to the conventional Euclidean time quantization considered above by Weyl rescaling $\mathbb{R}^d$ to $\mathbb{R}\times S^{d-1}$ as
\es{rescale}{
ds^2_{\mathbb{R}^d}&=dr^2+r^2ds^2_{S^{d-1}}\\
&=r^2\left(\frac{dr^2}{r^2}+ds^2_{S^{d-1}}\right)\\
&=e^{2\tau}(d\tau^2+ds^2_{S^{d-1}})=e^{2\tau}ds^2_{\mathbb{R}\times S^{d-1}}\,,
}
where $r=e^\tau$. Dilations $r\to\lambda r$ on $\mathbb{R}^d$ are thus identified with time shifts $\tau\to\tau+\log\lambda$ on $\mathbb{R}\times S^{d-1}$. Under a general Weyl rescaling $\Omega$, correlators in a CFT on a space with metric $g$ transform as\footnote{This remains true even in even dimensions where Weyl anomalies can occur.}
\es{weyl}{
\langle\cO_{i_1}(x_1)\dots\cO_{i_n}(x_n)\rangle_g=\frac{1}{\Omega(x_1)^{\Delta_1}}\dotsb \frac{1}{\Omega(x_n)^{\Delta_n}}\langle\cO_{i_1}(x_1)\dots\cO_{i_n}(x_n)\rangle_{\Omega^2 g}\,.
}
We then define cylinder operators 
\es{cylinder}{
\cO(\tau,\vec n)\equiv e^{\Delta\tau}\cO(x=e^\tau\vec n)\,,
}
where $\vec n$ is a unit vector on $S^{d-1}$, so that correlation functions of cylinder operators are equivalent to those of flat space operators without the extra factors in \eqref{weyl}. In sum, radial quantization of operators on $\mathbb{R}^d$ becomes Euclidean time quantization of cylinder operators on $\mathbb{R}\times S^{d-1}$. The standard Euclidean conjugation for scalar cylinder operators $\phi(\tau,\vec n)^\dagger=\phi(-\tau,\vec n)$ then corresponds according to \eqref{cylinder} to conjugation of scalar flat space operators in radial quantization as
\es{radConj}{
\phi(x)^\dagger=x^{-2\Delta}\phi\left(\frac{x^\mu}{x^2}\right)\,,
}
where the RHS is just the action of an inversion on an operator as defined for a general conformal transformation in \eqref{finiteGen}, and an analogous formula holds for operators with spin. This implies that conjugation in radial quantization acts as an inversion. Recall that SCTs were defined in \eqref{SCTfinite} as a translation sandwiched between two inversions, which implies that the infinitesimal generators are conjugate by inversion. In radial quantization, we then find that
\es{PtoK}{
P_\mu^\dagger=K_\mu\,.
}


\subsection{Unitarity bounds}
\label{unitarity}
We will now use the relation $P_\mu^\dagger=K_\mu$ that holds in radial quantization to derive bounds on scaling dimensions in CFTs from unitarity \cite{Mack:1975je,Minwalla:1997ka,Jantzen1977}. Consider an operator $\cO^a$ normalized as
\es{norm}{
\langle\cO_b|\cO^a\rangle=\delta_b^a\,,
}
in a nontrivial irrep $R_\cO$ of $SO(d)$ that acts on $\cO^a$ with matrix $(S_{\mu\nu})_b{}^a$ as in \eqref{M0}. We now compute the positive definite norm $|P|\cO^a\rangle|^2$ as
\es{KPnorm}{
0&\leq (P_\nu|\cO^b\rangle)(P_\mu|\cO^a\rangle)^\dagger\\
&= \langle\cO_a|K_\mu P_\nu|\cO^b\rangle\\
&= \langle\cO_a|[K_\mu P_\nu]|\cO^b\rangle\\
&=2\Delta\delta_{\mu\nu}\delta_a{}^b-2(S_{\mu\nu})_a{}^b\,,
}
where in the first equation we used \eqref{PtoK}, in the second we used the fact that $K_\mu$ kills primaries, and in the third the conformal algebra \eqref{confAlg}, the normalization \eqref{norm}, and the action of $M_{\mu\nu}$ and $D$ on $\cO^a$ as in \eqref{M0} and \eqref{D0}. This result implies that 
\es{maxEig}{
\Delta\geq\text{max-eigenvalue}((S_{\mu\nu})_a{}^b)\,.
}
To find the eigenvalues of $(S_{\mu\nu})_a{}^b$ we write it as
\es{Swrite}{
(S_{\mu\nu})_a{}^b=\frac12(L^{\alpha\beta})_{\mu\nu}(S_{\alpha\beta})_a{}^b\,,\qquad (L^{\alpha\beta})_{\mu\nu}\equiv\delta_\mu^\alpha\delta_\nu^\beta-\delta_\nu^\alpha\delta_\mu^\beta\,,
} 
where $(L^{\alpha\beta})_{\mu\nu}$ is the generator of rotations in the vector irrep $V_1$ of $SO(d)$. We can then use the standard trick to write $L^AS_A$, where $A=[\alpha,\beta]$ is an adjoint index of $SO(d)$, as
\es{QMtrick}{
L^AS_A=\frac12\left((L+S)^2-L^2-S^2\right)\,,
}
where $-L^2$, $-S^2$, and $-(S+L)^2$ are quadratic Casimirs \footnote{These have the opposite sign of the usual Casimirs because our generators differ from the standard one by a factor of $i$.} of the $V_1$, $R$ and tensor product $V_1\otimes R$ irreps of $SO(d)$, respectively. We now restrict $R=V_\ell$ to rank $\ell$ traceless symmetric irreps $V_\ell$ of $SO(d)$, which have Casimir 
\es{cas}{
\text{Cas}(V_\ell)=\ell(\ell+d-2)\,.
}
We are interested in the maximal eigenvalue of $L^AS_A$, so we want the irrep in $V_1\otimes V_\ell$ with minimal eigenvalue, which is $V_{\ell-1}$. Putting everything together we get
\es{Ubound}{
\Delta&\geq\frac12(-\text{Cas}(V_{\ell-1})+\text{Cas}(V_1)+\text{Cas}(V_\ell))\\
&=\ell+d-2\qquad\qquad\text{for $\ell>0$}\,.
}
Note that this bound does not hold for $\ell=0$, because in that case the tensor product $V_1\otimes V_0=V_1$ so we must replace $\text{Cas}(V_{\ell-1})$ in \eqref{Ubound} by $\text{Cas}(V_{1})$, which gives the bound $\Delta\geq0$ that is weaker for $d>2$. 
To get a stronger bound for $d>2$ we demand the positivity of the next descendent norm $|P^2|\cO\rangle|^2$, which for scalar operators gives
\es{Ubound0}{
\Delta&\geq\frac{d-2}{2}\qquad\qquad\text{for $\ell=0$}\,,\\
\Delta&=0\qquad\qquad\qquad\text{for identity operator}\,,
}
as will be shown in the problem set. No further constraints can be derived by demanding positivity of descendents at higher levels.

\subsection{Conserved currents}
\label{cons}

When the unitarity bounds are saturated, this implies that the norm of the relevant descendent vanishes, so we call it a null state. For instance, an $\ell=0$ state $|\cO\rangle$ with $\Delta=\frac{d-2}{2}$ saturates \eqref{Ubound0}, which implies that 
\es{nullScal}{
P^2|\cO\rangle=0\,.
}
In operator language, this shows that $\cO$ satisfies the Klein-Gordon equation $\partial^2\cO=0$, and so must be part of a free theory. Note that the Klein-Gordon equations appears without having to think of it as an equation of motion of a Lagrangian.\footnote{A similar argument applies to fermionic operators that satisfy their unitarity bound and therefore satisfy the Dirac equation for a free theory of fermions.}

For a spin $\ell$ operator $\cO^{\mu_1\dots\mu_\ell}$ with $\Delta=\ell+d-2$ that saturates \eqref{Ubound}, we have the null state
\es{nullSpin}{
P_{\mu_1}|\cO^{\mu_1\dots\mu_\ell}\rangle=0\,.
}
In operator language, this shows that $\cO^{\mu_1\dots\mu_\ell}$ is conserved $\partial_{\mu_1}\cO^{\mu_1\dots\mu_\ell}=0$. We see that a conserved current with spin $\ell>0$ is defined in a non-Lagrangian language by saturation of the unitarity bound. For instance, $J^\mu\equiv \cO^{\mu}$ with $\Delta_J=d-1$ is a conserved current for a global symmetry, $T^{\mu\nu}\equiv\cO^{\mu\nu}$ with $\Delta_T=d$ is a the conserved stress tensor, while higher spin conserved currents in $d>2$ can only exist in a free theory.

If a CFT has a stress tensor operator, then we call it a local CFT. In that case we can define the infinitesimal generators of the conformal group as
\es{genFromT}{
Q=-\int_\Sigma dS_\mu\epsilon^Q_\nu(x)T^{\mu\nu}(x)\,,
}
where $\epsilon\equiv \epsilon_\nu\partial^\nu$ are the solutions of the conformal Killing equation \eqref{infSol} for each generator $Q_\epsilon$ \eqref{generators} that satisfy the conformal algebra \eqref{confAlg}, and the integral is the same for any $\Sigma$ that surrounds the origin. For instance, for dilations $\epsilon_\nu^D=x_\nu$, so we can define $ D=-\int_\Sigma dS_\mu x_\nu T^{\mu\nu}(x)$. As shown in the problem set, we can then apply this formula to $\langle \phi(x_1)\phi(x_2)T_{\mu\nu}(x_3)\rangle$ for scalars $\phi$ to derive \cite{Osborn:1993cr}
\es{cTtoOPE}{
\lambda_{\phi\phi T}=-\frac{d\Delta_\phi}{(d-1)S_d}\,,\qquad S_d\equiv\frac{2\pi^{d/2}}{\Gamma(d/2)}\,,
}
where $S_d$ is the volume of the $d$-dimensional sphere. The canonical normalization of $T^{\mu\nu}$ in \eqref{genFromT} implies that the 2-point function coefficient in \eqref{2point2} is 
\es{cTDef}{
C^{(2)}_\text{stress}=\frac{c_T}{S_d^2}\,,
}
where $c_T$ is a theory dependent constant called the central charge. If we want all our operators to have unit normalized 2-point functions, which is the standard convention in the bootstrap literature, then we define the unit normalized $\cO_{d,2}\equiv T\frac{S_d}{\sqrt{c_T}}$, so that \eqref{cTtoOPE} becomes
\es{cTtoOPE2}{
\lambda_{\phi\phi \cO_{d,2}}=-\frac{d\Delta_\phi}{(d-1)\sqrt{c_T}}\,.
}

Note that the abstract properties that define a CFT, such as classification of operators as primaries and descendents, unitarity bounds, or the constraints on correlation functions, only required generators that satisfy the conformal algebra \eqref{confAlg}, and did not require that those generators be defined as an integral of the stress tensor. We can thus consistently define non-local CFTs that do not contain a stress tensor. An important example are generalized free field theories (GFFTs), whose $n$-point functions can be computed in terms of 2-point function using Wick contractions like free theories, but we allow the scaling dimension of the lowest non-identity scalar operator $\phi$ to be $\Delta_\phi>\frac{d-2}{2}$ instead of $\Delta=\frac{d-2}{2}$ as in a free theory. Composite operators in this theory are defined as 
\es{composite}{
\cO_{\Delta,\ell}\equiv\phi\partial^{2n}\partial_{\mu_1}\dots\partial_{\mu_\ell}\phi\qquad \Delta=2\Delta_\phi+2n+\ell\,,
}
which excludes the stress tensor for $\Delta_\phi>\frac{d-2}{2}$.

\subsection{Operator product expansion}
\label{OPEsec}

The final application of radial quantization in a CFT that we will discuss is a method of writing $n$-point correlators in terms of lower point correlators, which again will apply equally well to local and non-local CFTs. Consider two operators $\cO^a_i(x_1)$ and $\cO^b_j(x_2)$, where $a$ and $b$ are indices for possibly different irreps of $SO(d)$, inserted inside a sphere $S$, which generate a state $\cO^a_i(x_1)\cO^b_j(x_2)|0\rangle=|\Psi^{ab}\rangle$ on the surface of $S$. By the state-operator correspondence we know that all states correspond to operators, which in a CFT must be primaries or descendents. We can thus write $|\Psi^{ab}\rangle$ as some infinite sum of primaries and descendents, which yields
\es{OPE}{
\cO^a_i(x_1)\cO^b_j(x_2)=\sum_{k}C^{ab}_{ijk}{}_c(x_{12},\partial_2)\cO^c_k(x_2)\,,
}
where the derivatives in $C^{ab}_{ijk}{}_c(x_{12},\partial_2)$ create all the descendents when acting on the primary $\cO^c_k(x_2)$, and this equation is valid in any correlation function provided that the other operators are outside $S$. This operator equation is called the Operator Product Expansion (OPE). Note that a similar expansion exists in any QFT, but for a general QFT the expansion only holds asymptotically in the small $x_{12}$ limit. The novelty of CFT is that the OPE is a convergent expansion that holds for finite $x_{12}$, which is because the state-operator correspondence allowed use to write the operators in $S$ as states on the boundary of $S$, and it is a general theorem that any state in a Hilbert space can be written as a convergent sum of other states in the Hilbert space.

Let us apply the OPE to a three point function of two scalars $\phi_i$ and $\phi_j$ and a spin $\ell$ operators to get
\es{3OPE}{
\langle\phi_i(x_1)\phi_j(x_2)\cO_{\Delta,\ell}^{\mu_1\dots\mu_\ell}(x_3)\rangle=C^{\nu_1\dots\nu_\ell}_{ij}(x_{12},\partial_2)\langle\cO_{\Delta,\ell}{}_{\nu_1\dots\nu_\ell}(x_2)\cO_{\Delta,\ell}^{\mu_1\dots\mu_\ell}(x_3)\rangle\,,
}
where we assume $|x_{23}|\geq|x_{12}|$ so that the OPE is valid, and we chose an orthonormal basis of operators as in \eqref{2point2} with unit normalizations $C^{(\ell)}=1$, so that only a single operator appears on the RHS. Using the equation \eqref{3point2} for this 3-point function as fixed by conformal invariance, we find
\es{OPEto3}{
\frac{\lambda_{\phi_i\phi_j\cO_{\Delta,\ell}}(Z^{\mu_1}\dotsb Z^{\mu_\ell}-\text{traces})}{x_{12}^{\Delta_i+\Delta_j-\Delta+\ell}x_{23}^{\Delta_j+\Delta-\Delta_i-\ell}x_{31}^{\Delta+\Delta_i-\Delta_j-\ell}}=C^{\nu_1\dots\nu_\ell}_{ij}(x_{12},\partial_2)
\left(\frac{I^{(\mu_1}{}_{\nu_1}(x_{12})\dotsb I^{\mu_\ell)}{}_{\nu_\ell}(x_{12})}{x_{23}^{2\Delta}}-\text{traces}\right)\,,
}
so we see that $C^{\mu_1\dots\mu_\ell}_{ij}(x_{12},\partial_2)$ is proportional to the OPE coefficient $\lambda_{\phi_i\phi_j\cO_{\Delta,\ell}}$, which justifies its name. The functional dependence of $C^{\mu_1\dots\mu_\ell}_{ij}(x_{12},\partial_2)$ can then be completely fixed by conformal invariance in terms of the scaling dimensions by expanding each side. 

The OPE can be used to recursively define any $n$-point function in terms of one point functions, scaling dimensions of operators that appear in each OPE, and $n-2$ OPE coefficients. Recall that $\langle\cO\rangle=0$ unless $\cO$ is the unit operator, so we see that all $n$-point functions, and thus all local physical data in a CFT,\footnote{This excludes non-local physical quantities like partition functions and Wilson loops.} are completely fixed in terms of the scaling dimensions and OPE coefficients of the theory, which we call the CFT data. For instance, for a 4-point function of identical real scalar operators $\phi$ we can take the OPE twice in the $12$ and $34$ channels to get
\es{4pointOPE}{
&\langle\phi(x_1)\phi(x_2)\phi(x_3)\phi(x_4)\rangle=\frac{1 }{x_{12}^{2\Delta_\phi}x_{34}^{2\Delta_\phi}}   \sum_{\Delta,\ell} \lambda_{\phi\phi\cO_{\Delta,\ell}}^2g_{\Delta,\ell}(u,v)\,,
}
where convergence of the OPE requires that $x_{12}$ be small enough that we can draw a sphere separating $x_1,x_2$ from $x_3,x_4$, and the sum runs over all primary operators that appear in the OPE $\phi\times\phi$ with dimensions $\Delta$ and spin $\ell$, where as shown before only traceless symmetric irreps of $SO(d)$ with even spin $\ell$ can appear in a 3-point function with two identical scalars. The sum over the descendents of each primary are called conformal blocks $g_{\Delta,\ell}(u,v)$, which are defined from the OPE to be 
\es{blockDef}{
g_{\Delta,\ell}(u,v)&\equiv x_{12}^{2\Delta_\phi}x_{34}^{2\Delta_\phi} \frac{C_{\mu_1\dots\mu_\ell}(x_{12},\partial_2)C_{\nu_1\dots\nu_\ell}(x_{34},\partial_4)I^{\mu_1\dots\mu_\ell\nu_1\dots\nu_\ell}(x_{24})}{\lambda_{\phi\phi\cO_{\Delta,\ell}}^2   x_{24}^{2\Delta}}\,,\\
}
where $I$ is defined in \eqref{2point2}, and we divided by the OPE coefficients because the $C(x,\partial)$ are proportional to them. It is not obvious from the definition here that the conformal blocks are functions of conformal cross ratios $u,v$, but we already know this is required from conformal invariance of the 4-point function \eqref{4point}, where we identify
\es{gtog}{
g(u,v)=\sum_{\Delta,\ell}\lambda_{\phi\phi\cO_{\Delta,\ell}}^2 g_{\Delta,\ell}(u,v)\,.
} 
We could have also taken the OPE in other channels, which leads to the crossing equations given in \eqref{crossing2} for $g(u,v)$. We now see that these crossing equations give a whole function of constraints on the OPE coefficients and scaling dimensions that appear in the OPE $\phi\times\phi$. To study these constraints we need explicit functions for the conformal blocks, which will be the subject of the next lecture.

\pagebreak

\subsection{Problem Set 2}
\label{hw2}
\begin{enumerate}
\item Derive the unitarity bound \eqref{Ubound0} for a spin $\ell=0$ operator $\phi$ by demanding positivity of the norm $|P^2\phi\rangle$.
\item Use \eqref{2point2}, \eqref{3point2}, and \eqref{genFromT} for $Q=D$ to derive the relation \eqref{cTtoOPE} between the canonically normalized stress tensor $T$ and the OPE coefficient $\lambda_{\phi\phi T}$ of two scalars $\phi$ with $T$.
\item Consider a free theory in $d>2$ dimensions with scalar operator $\phi(x)$ with $\Delta_\phi=\frac{d-2}{2}$ and normalized as
\es{2pointfree}{
\langle\phi(x_1)\phi(x_2)\rangle=\frac{1}{|x_{12}|^{d-2}}\,,
}
as well as the canonically normalized stress tensor $T_{\mu\nu}$ with $C^{(\ell)}=c_T/S_d^2$ in \eqref{2point2}. 
\begin{enumerate}
\item Compute $c_T$ by writing $T_{\mu\nu}$ in terms of $\phi$ and computing the 2-point function $\langle TT\rangle$ using Wick contractions.
\item Now redefine $\cO_{d,2}\equiv T\frac{S_d}{\sqrt{c_T}}$ such that $c_T$ appears in $\langle \phi(x_1)\phi(x_2)\cO_{d,2}(x_3)\rangle $ as in \eqref{cTtoOPE2}, then compute this 3-point function using Wick contractions and see that you get the same value of $c_T$ as before.
\end{enumerate}
\end{enumerate}

\pagebreak
 
\section{Conformal blocks}
\label{blockSec}

In this lecture we will discuss how to compute conformal blocks,\footnote{Our discussion follows the original work \cite{Dolan:2003hv}, see \cite{Fortin:2019gck} for a different approach.} that capture the contribution of descendent operators to a 4-point function. We begin by considering the definition of the conformal blocks $g_{\Delta,\ell}(u,v)$ in the first line of \eqref{blockDef}, which was derived in the previous lecture by taking the OPE twice and thereby expressing the blocks in terms of the functions $C^{\mu_1\dots\mu_\ell}(x,\partial)$ that appear in the OPE \eqref{OPE} applied to two scalars operators. As shown in \eqref{OPEto3}, these $C^{\mu_1\dots\mu_\ell}(x,\partial)$ can be computed by comparing to the three point function, from which we find 
\es{Cfunf}{
C^{\mu_1\dots\mu_\ell}(x,\partial)\underset{x\to0}{\sim}\lambda_{\phi\phi\cO_{\Delta,\ell}}\frac{x^{\mu_1}\cdots x^{\mu_\ell}}{x^{2\Delta_\phi-\Delta+\ell}}\,,
}
for $x=x_{12},x_{34}$, where the $x_{12},x_{34}\to0$ limit corresponds to $u\to0$ and $v\to1$ as we see from \eqref{uv}. We then plug into \eqref{blockDef} and contract the tensors to get \footnote{The blocks are normalized here so that the OPE coefficients are the same in the 4-point function \eqref{4pointOPE} as in the 3-point function \eqref{3point2}.} 
\es{blockAss}{
g_{\Delta,\ell}(u,v)=(-2)^{-\ell}u^{\frac{\Delta-\ell}{2}}(1-v)^\ell+\dots\,.
}
where note that to leading order the block does not depend on the spacetime dimension $d$. Computing higher order corrections is very inefficient using this definition of the block. Instead, we can expand the 4-point function in a complete set of states to get
\es{completeSet}{
\sum_{\Delta,\ell}\sum_{m,s} \frac{\langle0|\phi(x_1)\phi(x_2)|{\Delta+m,s}\rangle\langle{\Delta+m,s}|\phi(x_3)\phi(x_4)|0\rangle}{\langle{\Delta+m,s}|{\Delta+m,s}\rangle}\,,
} 
where $\Delta,\ell$ denote the primaries that can appear in these 3-point functions, and $m,s$ denotes the descendent states $P^{\mu_1}\cdots P^{\mu_m}|\Delta,\ell\rangle$ with allowed spins 
\es{s}{
s=\{\max(0,\ell-m),\dots,\ell+m\}\,.
} 
Comparing to \eqref{4pointOPE}, we see that the conformal blocks are then defined as
\es{blockDef2}{
g_{\Delta,\ell}(u,v)=x_{12}^{2\Delta_\phi}x_{34}^{2\Delta_\phi} \sum_{m,s} \frac{\langle0|\phi(x_1)\phi(x_2)|{\Delta+m,s}\rangle\langle{\Delta+m,s}|\phi(x_3)\phi(x_4)|0\rangle}{\lambda^2_{\phi\phi\cO_{\Delta,\ell}}\langle{\Delta+m,s}|{\Delta+m,s}\rangle}\,,
}
where we divided by the OPE coefficients to cancel those factors that appear implicitly in the numerator. This formulation in terms of three point functions of two identical scalars and a spin $\ell$ operator shows that the blocks satisfy
\es{blockId}{
g_{\Delta,\ell}(u,v)=(-1)^\ell g_{\Delta,\ell}(u/v,1/v)\,,
}
so the first crossing equation $g(u,v)=g(u/v,1/v)$ in \eqref{crossing2}, which came from swapping $1\leftrightarrow2$, is trivially satisfied by each block separately in the 4-point function as long as $\ell$ is even. Equivalently, this 4-point function crossing equation is satisfied since the 3-point function $\langle0|\phi(x_1)\phi(x_2)|{\Delta+m,s}\rangle$ itself is already invariant under swapping the two identical scalars as long as the spin is even. In the rest of this lecture, we will use various methods to compute the blocks from this definition.

\subsection{Conformal Casimir}
\label{conCas}

For many groups, the most efficient way to derive explicit expressions for their irreps in a given basis is using the quadratic Casimir, which was done for the conformal group in \cite{Dolan:2003hv}. Recall that the conformal group is isomorphic to $SO(d+1,1)$ with generators $L_{ab}$ given in \eqref{sodplus1} with $a,b=0,1,\dots d$. The quadratic Casimir is then defined as 
\es{conformalCas}{
C=-\frac12L^{ab}L_{ab}\,,
}
where the minus sign is because we defined our generators to be anti-Hermitian, so they differ from the usual convention by a factor of $i$. By definition the quadratic Casimir has the same eigenvalue on every state in an irrep.  Using the conformal algebra \eqref{confAlg}, we derive
\es{conf}{
\mathcal{C}|\cO\rangle=c_{\Delta,\ell}|\cO\rangle\,,\qquad c_{\Delta,\ell}=\Delta(\Delta-d)+\ell(\ell+d-2)\,.
}
The 4-point function is fixed by conformal invariance in terms of two points, so we consider $\cC$ acting on two operators in the 4-point function. A useful identity is
\es{identity}{
L_{ab}\phi_1(x_1)\phi_2(x_2)|0\rangle=&([L_{ab},\phi_1(x_1)]\phi_2(x_2)+\phi_1(x_1)[L_{ab},\phi_2(x_2)])|0\rangle\\
=&(\mathcal{L}_{ab,1}+\mathcal{L}_{ab,2})\phi_1(x_1)\phi_2(x_2)|0\rangle\,,
}
where the second equality is because $\cC$ has the same eigenvalue acting on the left or right of an operator, and $\mathcal{L}^{ab}_i(x,\partial)$ is defined by the differential presentation \eqref{generators} of the conformal generators acting on the $i$th coordinate. Using this identity twice on the definition \eqref{blockDef2} of the conformal block in terms of a sum over descendents, we find
\es{Cason4}{
\left[-\frac12(\mathcal{L}_1^{ab}+\mathcal{L}_2^{ab})(\mathcal{L}_{ab,1}+\mathcal{L}_{ab,2})-c_{\Delta,\ell}\right]x_{12}^{2\Delta_\phi}x_{34}^{2\Delta_\phi}g_{\Delta,\ell}(u,v)=0\,,
} 
where we used the fact that the quadratic Casimir acts the same on all operators in a conformal multiplet. After moving the differential operator through the $x_i$-dependent prefactors we find that the conformal block satisfies the differential equation
\es{ConfCasFinal}{
&\cD g_{\Delta,\ell}(u,v)=c_{\Delta,\ell}g_{\Delta,\ell}(u,v)\,,\\
&\cD\equiv(1-u-v)\partial_vv\partial_v+u\partial_u(2u\partial_u-d)-(1+u-v)\left(u\partial_u+v\partial_v\right)\left(u\partial_u+v\partial_v\right)\,,
}
where note that $\Delta_\phi$ does not appear, which is a special feature of 4-point functions of identical operators. This differential equation can be solved exactly in any even dimension using the initial condition \eqref{blockAss}. 

\begin{figure}[t!]
\begin{center}
   \includegraphics[width=0.5\textwidth]{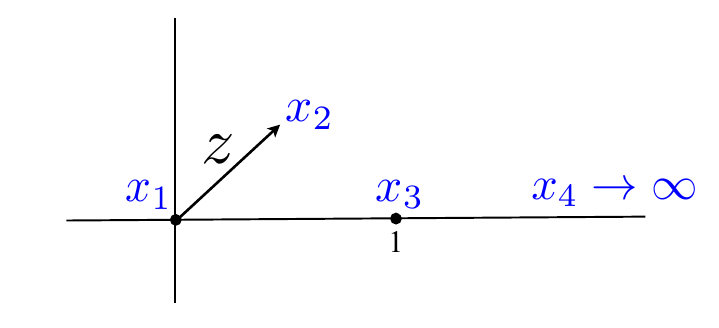}
 \caption{This figure was adapted from \cite{Hogervorst:2013sma}.}
\label{frame0}
\end{center}
\end{figure}  

To solve this differential equation it is useful to introduce the variables $z$ and $\bar z$ that are defined in terms of the four points $x_i$ using the following conformal transformations:
\begin{enumerate}
\item Use an SCT consisting of a translation $x_i\to x_i-x_4$ conjugated by inversions to fix $x_4\to\infty$. 
\item Translate $x_i\to x_i-x_1$ to fix $x_1=(0,0, \vec 0)$, where $\vec0$ is a $(d-2)$-dimensional vector.
\item Rotate $x_i\to R_i{}^j x_j$ with $R$ such that $x_3=(|x_3|,0,\vec 0)$.
\item Dilate $x_i\to x_i/|x_3|$ so that $x_3=(1,0,\vec 0)$.
\item Rotate $x_i\to R'_i{}^j x_j$ with $R'$ that preserves $(1,0,\vec 0)$ but fixes $x_2=(a,b,\vec0)$ such that $a^2+b^2=x_2^2$.
\end{enumerate}
At this point we cannot use any more conformal transformation to fix the points, and we define $z=a+ib$ and $\bar z=a-ib$.\footnote{This derivation was performed in Euclidean signature, for which we see that $z$ and $\bar z$ are complex variables satisfying $\bar z=z^\dagger$. If we analytically continue to Lorentzian signature using $x_i^0\to ix_i^0$, then $z$ and $\bar z$ become independent real variables.} In  Figure \ref{frame0} we show this conformal frame, which in general is defined as a choice of $x_i$ that exhausts the constraints from conformal symmetry, and so can be thought of as gauge fixing the conformal symmetry. We can then relate $z$ and $\bar z$ to $u$ and $v$ using the definition \eqref{uv} to get
\es{ztouv}{
u=z\bar z\,,\qquad v=(1-z)(1-\bar z)\,.
}

We can now solve \eqref{ConfCasFinal} in terms of $z$ and $\bar z$ exactly in any even dimension using the initial condition \eqref{blockAss}. For instance, in $d=2,4$ one finds
\es{2and4}{
2d:\quad g_{\Delta,\ell}(z,\bar z)&=\frac{(-2)^{-\ell}}{1+\delta_{\ell,0}}\left({k_{\Delta+\ell}(z) k_{\Delta-\ell}(\bar z)+k_{\Delta-\ell}(z) k_{\Delta+\ell} (\bar z)}\right)\,,   \\
4d:\quad g_{\Delta,\ell}(z,\bar z)&=(-2)^{-\ell}\frac{z \bar z}{z-\bar z}\left({k_{\Delta+\ell}(z) k_{\Delta-\ell-2}(\bar z)-k_{\Delta-\ell-2}(z) k_{\Delta+\ell} (\bar z)}\right)\,,   \\
k_\beta(x)&\equiv x^{\beta/2}{}_2F_1\left(\frac{\beta}{2},\frac{\beta}{2},\beta,x\right)\,.
}
One can check that these blocks are invariant under the crossing $g_{\Delta,\ell}(u,v)=g_{\Delta,\ell}(u/v,1/v)$ when $\ell$ is even, as discussed before.

In odd dimensions no such exact solution exists, so we must look for a series solution of the conformal Casimir equation \eqref{ConfCasFinal} in terms of a convenient expansion parameter.

\subsection{Radial expansion}
\label{radExp}

\begin{figure}[t!]
\begin{center}
   \includegraphics[width=0.5\textwidth]{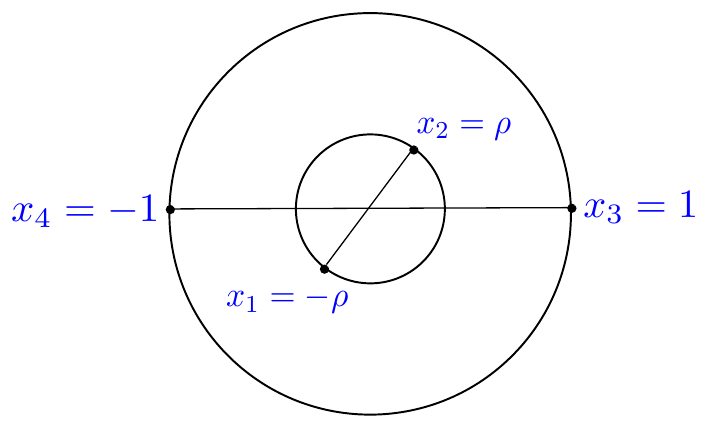}
 \caption{This figure was adapted from \cite{Hogervorst:2013sma}.}
\label{figrho}
\end{center}
\end{figure}  

We will now discuss a series solution to \eqref{ConfCasFinal} following \cite{Hogervorst:2013sma} that takes the form of an expansion in large $\Delta$ and that works for any $d$. This expansion is naturally written in the conformal frame depicted in Figure \ref{figrho}, where the complex coordinate $\rho$ is related to complex $z$ in \eqref{ztouv} as
\es{rhoToz}{
\rho=\frac{z}{(1+\sqrt{1-z})^2}\,\qquad\text{or}\qquad z=\frac{4\rho}{(1+\rho)^2}\,,
}
and we define $r\equiv|\rho|$ and $\eta\equiv \vec n\cdot\vec n'$ where the unit vectors $\vec n,\vec n'$ are in the directions of $x_2,x_3$, respectively. Recall that the OPEs $\phi(x_1)\times\phi(x_2)$ and $\phi(x_3)\times\phi(x_4)$ that define the conformal block converge if $x_{12}$ is chosen so that we can draw a sphere separating $x_1,x_2$ from $x_3,x_4$, which corresponds to $r<1$ in this conformal frame.

Consider radial quantization in this conformal frame on the cylinder, as described in Section \ref{UandP}. The conformal blocks as defined in \eqref{blockDef2} then take the form
\es{blockDef3}{
&g_{\Delta,\ell}(u,v)=2^{\Delta_\phi}\sum_{m,s} \frac{\langle0|\phi(r,\vec n)\phi(r,-\vec n)|{\Delta+m,s}\rangle\langle{\Delta+m,s}|\phi(1,\vec n')\phi(1,-\vec n')|0\rangle}{\lambda^2_{\phi\phi\cO_{\Delta,\ell}}\langle{\Delta+m,s}|{\Delta+m,s}\rangle}\,,\\
&=2^{\Delta_\phi}\sum_{m,s} r^{\Delta+m}\frac{\langle0|\phi(1,\vec n)\phi(1,-\vec n)|{\Delta+m,s}\rangle\langle{\Delta+m,s}|\phi(1,\vec n')\phi(1,-\vec n')|0\rangle}{\lambda^2_{\phi\phi\cO_{\Delta,\ell}}\langle{\Delta+m,s}|{\Delta+m,s}\rangle}\,,\\
}
where in the second line we used the evolution operator $r^D$ on the cylinder to equate the radii of the operators, and note that these expressions are $\Delta_\phi$-independent due to cancellation between the $2^{\Delta_\phi}$ prefactor and the $\Delta_\phi$-dependence of the 3-point functions. By rotational invariance we can write each spin $s$ 3-point function as
\es{rot3}{
\langle0|\phi(1,\vec n)\phi(1,-\vec n)|{\Delta+m,s}\rangle^{\mu_1\dots \mu_s}\propto n^{\mu_1}\cdots n^{\mu_s}-\text{traces}\,.
}
The contraction of two traceless symmetric tensors in the directions $\vec n,\vec n'$ is a Gegenbauer polynomial $C_s^{\frac{d-2}{2}}(\eta)$, which is the generalization of zonal spherical harmonics beyond $d=3$, in which case they are the familiar Legendre polynomials. The conformal block can thus be written as a sum
\es{blockSum}{
g_{\Delta,\ell}(u,v)=\sum_{m=0}^\infty r^{\Delta+m}\sum_s B_{s,m}(\Delta,\ell,d)C_s^{\frac{d-2}{2}}(\eta)\,,
}
where $B_{s,m}(\Delta,\ell,d)$ are coefficients that can be fixed using the conformal Casimir. For instance, we can fix the leading term $B_{\ell,0}(\Delta,\ell,d)$ by comparing with \eqref{blockAss} using \eqref{rhoToz} and \eqref{ztouv}, from which we find
\es{B00}{
B_{\ell,0}(\Delta,\ell,d)=\frac{4^\Delta\ell!}{(-2)^\ell(d/2-1)_\ell}\,.
}
It is inefficient to compute all the $B_{s,m}(\Delta,\ell,d)$ this way, so we will introduce a better method in the next section.

\subsection{Zamolodchikov recursion relations}
\label{Zam}

An efficient way of computing the series expansion in \eqref{blockSum} when $d$ is {\it not} an even number is a using a recursion relation originally developed for Virasoro blocks in 2d by Zamolodchikov \cite{Zamolodchikov:1985ie,dumb}, and then adapted to global conformal blocks in \cite{Kos:2013tga,Penedones:2015aga}. We start by observing that when the scaling dimension $\Delta$ of a conformal primary $\cO_{\Delta,\ell}$ is at or below the unitarity bound, then there exist certain descendents $P^{m}\cO_{\Delta,\ell}$ for $m\in\mathbb{B}_I$ that become primaries for certain values $\Delta_{I,m}$ and $\ell_{I,m}$, which implies that their norm must vanish. From the definition of $g_{\Delta,\ell}$ as a sum of descendents in \eqref{blockDef2}, we see that the block develops a pole at $\Delta_{I,m}$ because the norm of the descendent vanishes. The descendents of $P^{m}\cO_{\Delta_{I,m},\ell_{I,m}}$ also have zero norm and in fact form a sub-representation of the now reducible conformal multiplet, and so the residue of this pole must be proportional to another conformal block $g_{\Delta_{I,m}+m,\ell_{I,m}}$. Since poles in $\Delta$ determine $g_{\Delta,\ell}$ up to an an entire function of $\Delta$, we can then write the recursion formula
\es{recurse}{
g_{\Delta,\ell}(r,\eta)= g_{\infty,\ell}(\Delta,r,\eta)+\sum_{I}\sum_{m\in\mathbb{B}_I} \frac{c_{I,m}}{\Delta-\Delta_{I,m}}g_{\Delta_{I,m}+m,\ell_{I,m}}(r,\eta)\,,
}
where the $\Delta\to\infty$ term $ g_{\infty,\ell}(\Delta,r,\eta)$ is an entire function of $\Delta$. Recall from \eqref{blockSum} and \eqref{B00} that the blocks have an essential singularity of the form $(4r)^\Delta$ as $\Delta\to\infty$, which we can strip off to define the meromorphic function of $\Delta$:
\es{h}{
h_{\Delta,\ell}(r,\eta)&\equiv (4r)^{-\Delta}g_{\Delta,\ell}(r,\eta)\,,\\
h_{\Delta,\ell}(r,\eta)&= h_{\infty,\ell}(r,\eta)+\sum_{I}\sum_{m\in\mathbb{B}_I}\frac{c_{I,m}}{\Delta-\Delta_{I,m}}(4r)^{m}h_{\Delta_{I,m}+m,\ell_{I,m}}(r,\eta)\,,\\
}
where $ h_{\infty,\ell}(r,\eta)$ is no longer a function of $\Delta$ and the recursion formula is now an expansion in $r$. We can compute $ h_{\infty,\ell}(r,\eta)$ by writing the Casimir equations \eqref{ConfCasFinal} in terms of $r$ and $\eta$, replacing $g_{\Delta,\ell}(r,\eta)$ by $h_{\Delta,\ell}(r,\eta)$, and then expanding to leading order in $\Delta$, to get a simple differential equation
\es{hdiff}{
\partial_r h_{\infty,\ell}(r,\eta)=\frac{-4\eta^2 r((1-d)r^2+1)-(r^3+r)(d(r^2+1)-4)}{(r^2-1)((r^2+1)^2-4\eta^2 r^2)}h_{\infty,\ell}(r,\eta)\,.
}
We then solve this using the initial condition \eqref{B00} to get
\es{leadh}{
 h_{\infty,\ell}(r,\eta) =\frac{\ell!C_\ell^{\frac{d-2}{2}}(\eta)}{(-2)^\ell(d/2-1)_\ell(1-r^2)^{\frac{d-2}{2}}\sqrt{(1+r^2)^2-4r^2\eta^2}}\,.
}

Next, we can find the null descendents $P^{m}\cO_{\Delta_{I,m},\ell_{I,m}}$ by looking at sub-leading in $r$ terms in the Casimir equations with the ansatz \eqref{h}. We find three families of null descendents labelled by $I=1,2,3$. The first family of descendents has the maximal value $\ell_{m,1}=\ell+m$ of the possible spins \eqref{s}, which implies that $m\in\mathbb{B}_1$ must be even since the spin must be even. These descendents become null when $\Delta_{m,1}=1-\ell-m$. The second family has the minimal value of spin $\ell_{m,2}=\ell-m$ in \eqref{s}, which implies that $m\in\mathbb{B}_2$ are even and $m\leq\ell$ since the spin is positive and even. These descendents become null when $\Delta_{m,2}=\ell+d-1-m$. Finally, the third family has the same spin $\ell_{m,3}=\ell$ as the primary, which means that we must act on the primary with an even number of $P^\mu$'s, so $m\in\mathbb{B}_3$ must be even. These descendents become null when $\Delta_{m,3}=(d-m)/2$. This information is summarized in Table \ref{tab1}. Note that all these poles are below the unitarity bound except when $\ell=0$, where the pole $\Delta_{3,2}=\frac{d-2}{2}$ appears.

\begin{table}
\begin{center}
\begin{tabular}{c|c|c|c}
 $I$& $m\in\mathbb{B}_I$&  $\Delta_{I,m}$ & $\ell_{I,m}$  \\
 \hline 
1&  $2,4,6,\dots$&  $1-\ell-m$ & $\ell+m$   \\
 \hline
 2& $2,4,\dots \ell$  & $\ell+d-1-m$ & $\ell-m$   \\
  \hline
 3& $2,4,6,\dots$  & $(d-m)/2$ & $\ell$   \\
\end{tabular}
\caption{Positions of the poles $\Delta_{I,m}$ in $g_{\Delta,\ell}$, shifts of the residue conformal blocks $g_{\Delta_{I,m}+m,\ell_{I,m}}$, and range $\mathbb{B}_{I}$ of possible values of $m$ for each $I=1,2,3$.}
\label{tab1}
\end{center}
\end{table}

 The coefficients $c_{I,m}$ of each family of descendents can be computed by taking the residues of \eqref{blockDef3} at $\Delta_{I,m}$ for each descendent $m$, for which we find
 \es{mixedBlockCoeff}{
c_{1,m}&=-\frac{m(-2)^m}{(m!)^2}\left(\frac{1-m}{2}\right)^2_m \,,\\
c_{2,m}&=-\frac{m\ell!  (d/2+\ell)_{-m}(d/2+\ell-1)_{-m} }{(-2)^m(m!)^2(\ell-m)!(d-2+\ell)_{-m}}\left(\frac{1-m}{2}\right)^2_m \,,\\
c_{3,m}&=  
\frac{-m(-1)^{m/2}((d-m)/2-1)_{m}}{2((m/2)!)^2((d-m)/2+\ell-1)_{m}((d-m)/2+\ell)_{m}} \left(\frac{2-(d+m)/2-\ell}{2}\right)^2_{\frac m2}\\
&\qquad\qquad\qquad\qquad\qquad\qquad\qquad\qquad\qquad\qquad\qquad\qquad\times\left(\frac{(d-m)/2+\ell}{2}\right)^2_{\frac m2}  \,.\\
}

Now that we have defined all the ingredients, we can expand $h_{\infty,\ell}(r,\eta)$ to a desired order in $r$ and $\eta$ and restrict the ranges $\mathbb{B}_I$ to that order, so that the recursion formula \eqref{h} will give $h_{\Delta,\ell}(r,\eta)$ as polynomials in $r$ and $\eta$ that converges for $r<1$ and arbitrary $-1\leq\eta\leq1$. Finally, we can multiply by $(4r)^\Delta$ to get the block $g_{\Delta,\ell}(r,\eta)$.\footnote{Note that when $d$ is even, the coefficients \eqref{mixedBlockCoeff} develop poles and the expansion in single poles that we assumed here does not apply. Thankfully, in the even $d$ case we have exact expressions for the blocks \eqref{2and4}, so we can find their large $\Delta$ expansion by simply expanding these formulae.} For example, let us work out $h_{\Delta,0}(r,\eta)$ in $d=3$ to order $r^6$. We find
\es{zeroBlock}{
h_{\Delta,0}(r,\eta)=&1+r^2\frac{\Delta
     \left(\Delta  \left(4 \eta ^2-1\right)-2 \eta ^2+1\right)}{2(\Delta+1)(\Delta-\frac12)}+\frac{r^4}{8 (\Delta +1) (\Delta +3) ( \Delta -\frac12) ( \Delta +\frac12)}\\
     &\times\Big[48 \Delta ^4 \eta ^4-40 \Delta ^4 \eta ^2+7 \Delta ^4+96 \Delta ^3 \eta ^4-76 \Delta ^3 \eta
   ^2+15 \Delta ^3-12 \Delta ^2 \eta ^4\\
   &\qquad+20 \Delta ^2 \eta ^2-7 \Delta ^2-24 \Delta  \eta ^4+24
   \Delta  \eta ^2-3 \Delta\Big]\\
   &+\frac{r^6 \Delta }{16 (\Delta +1) (\Delta +3) (\Delta +5) (
   \Delta -\frac12) ( \Delta +\frac12) ( \Delta +\frac32)}\\
   &\times \Big[   \Delta ^5 \left(320 \eta ^6-432 \eta ^4+156 \eta ^2-9\right)+2 \Delta
   ^4 \left(1200 \eta ^6-1608 \eta ^4+579 \eta ^2-31\right)\\
   &+\Delta ^3 \left(5360 \eta ^6-7044
   \eta ^4+2454 \eta ^2-98\right)+2 \Delta ^2 \left(1620 \eta ^6-1938 \eta ^4+555 \eta
   ^2+14\right)\\
   &+\Delta  \left(-1360 \eta ^6+2328 \eta ^4-1098 \eta ^2+111\right)-30 \left(32
   \eta ^6-48 \eta ^4+18 \eta ^2-1\right) \Big] +O(r^8)\,.
}
To this order we find the poles
\es{poles0}{
\Delta_{1,2}=-1\,,\quad \Delta_{1,4}=-3\,,\quad \Delta_{1,4}=-5\,,\quad \Delta_{3,2}=\frac12\,,\quad \Delta_{3,4}=-\frac12\,,\quad \Delta_{1,4}=-\frac32\,,
}
where note that the $I=2$ family of poles does not appear since $\ell=0$. In the next lecture, we will use these blocks to formulate the numerical bootstrap algorithm.

\pagebreak

\subsection{Problem Set 3}
\label{hw3}

\begin{enumerate}
\item Use the Zamolodchikov recursion relation to compute the stress tensor block $g_{d,2}(r,\eta)$ to $O(r^{10})$ in $d=2.00001$ and $d=3.99999$, and show that your answers match the exact formulae \eqref{2and4} up to a small error. This demonstrates that if you want to avoid double poles in even dimensions, then for numerical purposes it is sufficient to use the Zamolodchikov recursion formula for $d$ close to an even dimension.
\item In $d=3$, compute $g_{\Delta,\ell}(r,\eta)$ to order $O(r^{20})$ for $\ell=0,1,2$ and arbitrary $\Delta$. Show that to this order all the coefficients $B_{s,m}(\Delta,\ell,d)$ in \eqref{blockSum} have the same sign as the lowest term $B_{0,0}(\Delta,\ell,d)$ as long each $\Delta$ obey its unitarity bound. 
\\Hints: First expand $\tilde h_{\ell}(r,\eta)$ to order $O(r^{20})$, so that $h_{\Delta,\ell}(r,\eta)$ will be a polynomial in $r$ and $\eta$ that can be expressed as a $2\times2$ matrix of coefficients of $r$ and $\eta$. Each step in the Zamolodchikov recursion relations, such as multiplying by a power of $r$, can then be implemented as a matrix transformation, which is more efficient than multiplying polynomials. Make sure that at each recursive step you only use a range $\mathbb{B}_I$ and $\tilde h_{\ell}(r,\eta)$ expanded to the order in $r$ necessary to maintain $O(r^{20})$ at this step.
\item Consider a free theory in $d>2$ dimension with scalar operator $\phi(x)$, where $\Delta_\phi=\frac{d-2}{2}$ and we normalize its 2-point function to be \eqref{2pointfree}. Compute the 4-point function $\langle \phi(x_1)\phi(x_2)\phi(x_3)\phi(x_4)\rangle$ using Wick contractions and expand in conformal blocks to order $r^3$ to read off the OPE coefficients squared that you find to this order. Confirm that $\lambda^2_{\phi\phi T}$ is related by \eqref{cTtoOPE} to $c_T$, as computed in the previous problem set.
\end{enumerate}

\pagebreak

\section{Conformal bootstrap basics}
\label{bootBas}

We now have all the ingredients to start bootstrapping a 4-point function of identical scalars. The 4-point function is expanded in conformal blocks as
\es{4pointAgain}{
\langle\phi(x_1)\phi(x_2)\phi(x_3)\phi(x_4)\rangle=\frac{1}{x_{12}^{2\Delta_\phi}x_{34}^{2\Delta_\phi}}\sum_{\Delta,\ell}\lambda^2_{\phi\phi\cO_{\Delta,\ell}}g_{\Delta,\ell}(u,v)\,,
}
which has the following properties:
\begin{itemize}
\item The sum runs runs over primary operators $\cO_{\Delta,\ell}$ with dimension $\Delta$ and spin $\ell$ that appear in the OPE $\phi\times\phi$, and so must have even spin $\ell$ and include the identity operator $\cO_{0,0}$, and for a local CFT also the stress tensor $\cO_{d,2}$.
\item All the $\cO_{\Delta,\ell}$ except $\cO_{0,0}$ obey the unitarity bounds
\es{unitarityAgain}{
&\Delta\geq d-2+\ell\qquad \text{for}\qquad\ell>0\,,\qquad\qquad\Delta\geq \frac{d-2}{2}\qquad \text{for}\qquad\ell=0\,.\\
}
\item We can choose a real orthonormal basis for the operators such that OPE coefficients $\lambda_{\phi\phi\cO_{\Delta,\ell}}$ are required to be real by unitarity, so that $\lambda^2_{\phi\phi\cO_{\Delta,\ell}}\geq0$. The identity OPE coefficient is normalized so that $\lambda_{\phi\phi{\cO_{0,0}}}=1$.
\item The blocks have the following simple $\Delta$ dependence:
\es{blockAgain}{
g_{\Delta,\ell}(r,\eta)= (4r)^\Delta\sum_{m=0}^\infty r^m \sum_n\frac{f_{m,n}(\eta)}{\Delta-\Delta_{m,n}}\,,
}
where the positions of the poles $\Delta_{m,n}$ and the coefficients $f_{m,n}(\eta)$ can be computed recursively as in section \ref{Zam}. 
\item The four point function satisfies the crossing relations
\es{crossingAgain}{
g(u,v)=g(u/v,1/v)\,,\qquad g(u,v)=(v/u)^{\Delta_\phi}g(v,u)\,,
}
where the first equation is trivially satisfied by each block and the second imposes infinite constraints on the infinite $\Delta$ and $\lambda_{\phi\phi\cO_{\Delta,\ell}}$ in \eqref{4pointAgain}. 
\end{itemize}
We will now use these properties to formulate a numerical algorithm to bound scaling dimensions and OPE coefficients of $\cO_{\Delta,\ell}$ from the second constraint in \eqref{4pointAgain}, following the original work \cite{Rattazzi:2008pe}.

\subsection{Bootstrap algorithms}
\label{bootAlg}

We begin by writing the second crossing equation in \eqref{crossingAgain} as applied to the block expansion \eqref{4pointAgain} as
\es{F}{
\sum_{\Delta,\ell}\lambda^2_{\phi\phi\cO_{\Delta,\ell}}F^{\Delta_\phi}_{\Delta,\ell}(u,v)=0\,,\qquad F^{\Delta_\phi}_{\Delta,\ell}(u,v)\equiv v^{\Delta_\phi}g_{\Delta,\ell}(u,v)-u^{\Delta_\phi}g_{\Delta,\ell}(v,u)\,.
}
We can think of $F^{\Delta_\phi}_{\Delta,\ell}(u,v)$ as infinite-dimensional vectors, labeled by $\Delta,\ell$, in the vector-space $u,v$, so that \eqref{F} imposes that an infinite sum of vectors with positive coefficients must equal zero. Consider a functional $\alpha$ on this vector space. We can use $\alpha$ to bound scaling dimensions $\Delta$ by the following algorithm:
\\
\\
{\bf Scaling dimension bound:}
\begin{enumerate}
\item Normalize $\alpha$ such that $\alpha[F^{\Delta_\phi}_{0,0}(u,v)]=1$.
\item Assume that the scaling dimensions $\Delta_\ell$ of all spin $\ell$ operators in $\phi\times\phi$ except $\cO_{0,0}$ obey lower bounds $\Delta_\ell\geq\Delta_\ell^B$. The unitarity bound \eqref{unitarityAgain} provides a minimal choice.
\item Search for $\alpha$ that satisfies $\alpha[F^{\Delta_\phi}_{\Delta,\ell}(u,v)]\geq0$ for all $\cO_{\Delta,\ell}$ except $\cO_{0,0}$.
\item If such $\alpha$ exists, then by positivity of $\lambda^2_{\phi\phi\cO_{\Delta,\ell}}$ we have
\es{alphContra}{
\alpha\left[\sum_{\Delta,\ell}\lambda^2_{\phi\phi\cO_{\Delta,\ell}}F^{\Delta_\phi}_{\Delta,\ell}(u,v)\right]>0\,,
}
which contradicts $\alpha$ acting on \eqref{F}, so our assumptions $\Delta\geq\Delta_\ell^B$ must be false.\footnote{Note that the normalization $\alpha[F^{\Delta_\phi}_{0,0}(u,v)]=1$ is equivalent to $\alpha[F^{\Delta_\phi}_{0,0}(u,v)]>0$ for this algorithm, since we can rescale \eqref{alphContra} by any positive factor. } If we cannot find such an $\alpha$, then we conclude nothing.
\end{enumerate}
For instance, we can set the lower bounds $\Delta_\ell^B$ to their unitarity values for all $\ell$ except a certain $\ell^G$, then by varying $\Delta_{\ell^G}^B$ and $\Delta_\phi$ this algorithm can be used to find an upper bound on $\Delta_{\ell^G}^B$, i.e. the scaling dimension of the lowest dimension operator with spin $\ell^G$, as a function of $\Delta_\phi$.

Without further assumptions we cannot do better than upper bonds on scaling dimensions. If we furthermore assume that only a few operators have dimensions below a certain bound, which must necessarily be above the unitarity bound, then we can get both upper and lower bounds. Consider adding the additional constraint on $\alpha$ to the algorithm above:
\es{gap}{
\text{\bf{Gap assumption:}}\qquad\qquad\alpha[F^{\Delta_\phi}_{\Delta^I,\ell^G}(u,v)]\geq0\qquad\text{for}\quad\Delta_{\ell^G}^I<\Delta^B_{\ell^G}\,,
}
for some spin $\ell^G$ and dimension $\Delta^I_{\ell^G}$. Only specific values of $\Delta^I_{\ell^G}$ correspond to a physical CFT, so as we vary $\Delta^I_{\ell^G}$ for a given $\Delta_\phi$ we expect to find $\alpha$'s that satisfy the conditions of the algorithm, and thus correspond to disallowed points in the space of $(\Delta^I_{\ell^G},\Delta_{\phi})$, for everywhere except some region that contains the physical values of $\Delta^I_{\ell^G}$. For instance, we could assume that there only exists one relevant scalar in $\phi\times\phi$, for which we would set $\ell^G=0$, $\Delta^B_0=d$, and then vary $\Delta^I_0$ in the range $\frac{d-2}{2}\leq \Delta^I_0\leq d$ that is allowed by the unitarity bound and the gap above it.

We can also get upper bounds on a certain $\lambda^2_{\phi\phi\cO_{\Delta^O,\ell^O}}$, and also lower bounds if we assume a gap, by a slight variation of the scaling dimension algorithm \cite{Caracciolo:2009bx}:
\\
\\
{\bf OPE coefficient bound:}
\begin{enumerate}
\item Normalize $\alpha$ such that $\alpha[F^{\Delta_\phi}_{\Delta^O,\ell^O}(u,v)]=s$, where $s=\pm1$ for upper/lower bounds.
\item Assume that the scaling dimensions $\Delta_\ell$ of all operators in $\phi\times\phi$ except $\cO_{0,0}$ and $\cO_{\Delta^O,\ell^O}$ obey lower bounds $\Delta_\ell\geq\Delta_\ell^B$. 
\item Require that $\alpha[F^{\Delta_\phi}_{\Delta,\ell}(u,v)]\geq0$ for all $\cO_{\Delta,\ell}$ except $\cO_{0,0}$ and $\cO_{\Delta^O,\ell^O}$.
\item Maximize $\alpha[ F^{\Delta_\phi}_{0,0}(u,v)]$ to get the upper/lower bounds
\es{OPEbound}{
&\text{Upper}:\qquad\qquad\lambda^2_{\phi\phi\cO_{\Delta^O,\ell^O}}\leq -\alpha[F^{\Delta_\phi}_{0,0}(u,v)]\,,\\
&\text{Lower}:\qquad\qquad\lambda^2_{\phi\phi\cO_{\Delta^O,\ell^O}}\geq \alpha[F^{\Delta_\phi}_{0,0}(u,v)]\,,\\
}
which follows from \eqref{F}, positivity of $\lambda^2_{\phi\phi\cO_{\Delta,\ell}}$, and steps 1 and 3.
\end{enumerate}
If we just set $\Delta_\ell^B$ to the unitarity bounds, then we can get upper bounds but not lower bounds, because the condition $\alpha[F^{\Delta_\phi}_{\Delta^O,\ell^O}(u,v)]=-1$ in step 1 is then inconsistent with $\alpha[F^{\Delta_\phi}_{\Delta,\ell}(u,v)]\geq0$ from step 3, since without further assumptions there exists a continuum of operators $\cO_{\Delta,\ell^O}$ with $\Delta$ arbitrarily close to $\Delta^O$. To avoid this we must set a gap above unitarity for $\Delta^B_\ell$ and insert the operator $\cO_{\Delta^O,\ell^O}$ whose OPE coefficient we seek to bound, just as in \eqref{gap}. Note that this algorithm involved minimizing $\alpha$, whereas the scaling dimension algorithm only involved the existence of $\alpha$

\subsection{Truncating crossing}
\label{approx}

Before we can implement the algorithms in the previous section in practice, we must account for four infinite properties of the crossed blocks $F^{\Delta_\phi}_{\Delta,\ell}(u,v)$:
\begin{enumerate}
\item Spins $\ell$, which can be any even positive number.
\item Scaling dimensions $\Delta$, which can be any real number greater than the unitarity bounds.
\item The infinite series expansion in $r$ that we used to compute the conformal blocks.
\item The infinite vector space of real $u,v$ (or $r,\eta$) that we act on with the functional $\alpha$.
\end{enumerate}

To make the spins and the expansion in $r$ finite we can simply impose a cutoff $\ell_\text{max}$ and $r_\text{max}$ for each quantity. To make the $u,v$ vector space finite, we can define it as the coefficients of the Taylor expansion of $F^{\Delta_\phi}_{\Delta,\ell}(z,\bar z)$ in terms of the variables $z,\bar z$ defined in \eqref{ztouv}, after converting from the $r,\eta$ coordinates in which we computed the blocks, up to a certain order $\Lambda$ around the crossing symmetric point $z=\bar z=\frac12$, which corresponds to $\eta_c=1,r_c=3-2\sqrt{2}$. The crossing equation then takes the form
\es{F2}{
\sum_\Delta\sum_{\ell\leq\ell_\text{max}}\lambda^2_{\phi\phi\cO_{\Delta,\ell}}\partial_z^m\partial_{\bar z}^nF^{\Delta_\phi,r_\text{max}}_{\Delta,\ell}(z,\bar z)\vert_{z=\bar z=\frac12}=0\qquad\text{for}\qquad m+n\leq\Lambda\,,
}
where the functionals $\alpha_{m,n}$ now act on the elements of the finite vector space $V_\Lambda$ of these derivatives. Since $F^{\Delta_\phi}_{\Delta,\ell}(z,\bar z)$ is symmetric in $z\leftrightarrow\bar z$ and odd under the crossing $(z,\bar z)\to(1-z,1-\bar z)$, we can restrict to $m+n$ odd and $m\leq n$, so $V_\Lambda$ has dimension 
\es{dimV}{
|V_\Lambda|=\frac12\left\lfloor\frac{\Lambda+1}{2}\right\rfloor\left(\left\lfloor\frac{\Lambda+1}{2}\right\rfloor+1\right)\,.
}

The last infinity that we must consider is the continuum of dimensions $\Delta$. If we can approximate $\partial^m_z\partial^n_{\bar z}F^{\Delta_\phi,r_\text{max}}_{\Delta,\ell}(r_c,1)$ as a polynomial in $\Delta$ for each $\ell$, then the truncated crossing equation \eqref{F2} takes the form of positivity constraints on a finite number of polynomials, which can be solved using semidefinite programming. We will postpone the discussion of semidefinite programming until the next lecture, and instead discuss how to efficiently write $\partial^m_z\partial^n_{\bar z}F^{\Delta_\phi}_{\Delta,\ell}(r_c,1)$ as a polynomial following \cite{Poland:2011ey}.

\subsection{Polynomial formulation of crossing} 
\label{polyCross}

Recall from \eqref{blockAgain} that each block $g_{\Delta,\ell}(r,\eta)$ takes the form $(4r)^\Delta$ times a finite sum of poles in $\Delta$. If we truncate the $r$ expansion at $r_\text{max}$, then from the $r$ expansion of the blocks in \eqref{h} and the list of poles in Table \ref{tab1}, we see that the following list of poles will appear in $g^{r_\text{max}}_{\Delta,\ell}(r,\eta)$ expanded to $r_\text{max}$:
\es{poles}{
\mathbb{P}^\ell_{r_\text{max}}=\begin{cases}\Delta_{1,b}=1-\ell-b\qquad\text{for}\qquad b=2,4,6,\dots r_\text{max}\,,\\
\Delta_{2,b}=d/2-b\qquad\text{for}\qquad b=1,2,3,\dots r_\text{max}/2\,,\\
\Delta_{3,b}=\ell+d-1-b\qquad\text{for}\qquad b=2,4,6,\dots \min\{r_\text{max},\ell\}\,.\\
\end{cases}
}

For each $g^{r_\text{max}}_{\Delta,\ell}(r,\eta)$, we can then define the polynomial in $\Delta$:
\es{gTop}{
p_{\Delta,\ell}(r,\eta)=(4r)^{-\Delta}g^{r_\text{max}}_{\Delta,\ell}(r,\eta)\prod_{I\in  \mathbb{P}^\ell_{r_\text{max}}}(\Delta-\Delta_I)\,.
}
We would like the degree of these polynomials to be as small as possible, which improves the efficiency of semidefinite programming. To decrease the degree we can approximate poles in $\mathbb{P}^\ell_{r_\text{max}}/\mathbb{P}^\ell_{\kappa}$ that appear in $g^{r_\text{max}}_{\Delta,\ell}(r,\eta)$ in terms of poles in $\mathbb{P}^\ell_{\kappa}$ for some cutoff $\kappa$ as
\es{replacePoles}{
\frac{1}{\Delta-\Delta_I}\approx\sum_{J\in \mathbb{P}^\ell_{\kappa}}\frac{\mathfrak{c}_{I,J}}{\Delta-\Delta_J}\qquad\text{for}\qquad I\in\mathbb{P}^\ell_{r_\text{max}}/\mathbb{P}^\ell_{\kappa}\,,
}
where coefficients $\mathfrak{c}_{I,J}$ are chosen to make the approximation as accurate as possible in the range $\Delta\geq\Delta_\text{unitarity}$ were these polynomials will be defined. One way of doing this \cite{Kos:2013tga} is to demand that \eqref{replacePoles} and its first $\lfloor\frac{| \mathbb{P}^\ell_{\kappa}|-1}{2}\rfloor$ derivatives hold exactly at $\Delta=\Delta_\text{unitarity}$ (we do not include poles which occur at unitarity, such as for $\ell=0$), and the first $\lceil\frac{| \mathbb{P}^\ell_{\kappa}|-1}{2}\rceil$ derivatives hold at $\Delta\to\infty$, which gives $| \mathbb{P}^\ell_{\kappa}|$ independent linear equations for the $| \mathbb{P}^\ell_{\kappa}|$ coefficients. We can then define the reduced polynomials
\es{gTop2}{
\bar p_{\Delta,\ell}(r,\eta)=(4r)^{-\Delta}g^{r_\text{max},\kappa}_{\Delta,\ell}(r,\eta)\prod_{I\in  \mathbb{P}^\ell_{\kappa}}(\Delta-\Delta_I)\,,
}
where $g^{r_\text{max},\kappa}_{\Delta,\ell}(r,\eta)$ denotes that we have replaced the poles with parameter $\kappa$, and for simplicity we have not written $r_\text{max}$ and $\kappa$ on the LHS. We can now approximate $\partial^m_z\partial^n_{\bar z}F^{\Delta_\phi}_{\Delta,\ell}(r_c,1)$ as a polynomial in $\Delta$ by writing it as
\es{Ftop}{
\partial^m_z\partial^n_{\bar z}F^{\Delta_\phi}_{\Delta,\ell}(r_c,1)&\approx \chi^\ell_\kappa(\Delta)P^{m,n}_{\ell}(\Delta,\Delta_\phi)\,,\qquad  \chi^\ell_\kappa(\Delta)\equiv\frac{(4r_c)^\Delta}{\prod_{I\in  \mathbb{P}^\ell_{\kappa}}(\Delta-\Delta_I)}\,,\\
P_\ell^{m,n}(\Delta,\Delta_\phi)&\equiv 2\left(r^{-\Delta}\partial^m_z\partial^n_{\bar z}\left[v^{\Delta_\phi}r^\Delta\bar p_{\Delta,\ell}(r,\eta) \right]\right)\big\vert_{r=r_c,\eta=1}\,,
}
where $P^{m,n}_{\ell}(\Delta,\Delta_\phi)$ are polynomials in $\Delta$. The crossing equations can then be expressed as polynomials in $\Delta$ by absorbing the  positive factor $ \chi^\ell_\kappa(\Delta)$ into the definition of each $\lambda^2_{\phi\phi\cO_{\Delta,\ell}}$. This does not affect the scaling dimension bound algorithm, but for the OPE coefficient bound algorithm one should be careful to multiply the normalization in step 1 by $ \chi^\ell_\kappa(\Delta)$, so that one computes bounds on the standard OPE coefficients.

Let us consider this pole-replacement for $\ell=0$, $\Delta_\phi=\frac12$, and $r_\text{max}=6$. The $\Delta$ poles $\mathbb{P}^0_{6}$ that appear in $g_{\Delta,0}(r,\eta)$ were given in \eqref{poles0}, while $h_{\Delta,0}(r,\eta)\equiv(4r)^{-\Delta}g_{\Delta,0}(r,\eta)$ was given in \eqref{zeroBlock}. We can use these blocks to compute $\partial^m_z\partial^n_{\bar z}F^{\Delta_\phi}_{\Delta,\ell}(r_c,1)$ for the lowest nonzero case $m=0,n=1$:
\es{Fexample}{
&\partial_{\bar z}F^{1/2}_{\Delta,0}(r_c,1)=2(\partial_{\bar z}[v^{\frac12}g_{\Delta,0}(r[z,\bar z],\eta[z,\bar z])])\big\vert_{r=r_c,\eta=1}\\
&=(12-8\sqrt{2})^\Delta\Big[ \frac{1}{16} \left(703769 \sqrt{2}-995256\right) \Delta -\frac{4 \left(114033
   \sqrt{2}-161267\right)}{3 (\Delta +1)}-\frac{144 \left(36824 \sqrt{2}-52077\right)}{35
   (\Delta +3)}\\
&   -\frac{200 \left(33461 \sqrt{2}-47321\right)}{77 (\Delta +5)}+\frac{1739995287
   \sqrt{2}-2460724142}{59136 (2 \Delta -1)}+\frac{32887363 \sqrt{2}-46509754}{640 (2 \Delta
   +1)}\\
   &+\frac{375 \left(204129 \sqrt{2}-288682\right)}{1792 (2 \Delta +3)}+\frac{1}{32}
   \left(5879233 \sqrt{2}-8314522\right)  \Big]\,.
}
We can then factor out the positive quantity $\chi_6^0(\Delta)$ to get the degree 7 polynomial
\es{posEx}{
&P_0^{0,1}(\Delta,1/2)\big\vert_{\text{degree 7}}=\frac{1} {32 \sqrt{2}-48}
\Big[90 \left(2 \sqrt{2}-3\right)
+3 \left(7908112
   \sqrt{2}-11183763\right) \Delta\\
& \qquad  +
   \left(18531005 \sqrt{2}-26206713\right) \Delta ^2
   +
   \left(552938554-390986551 \sqrt{2}\right) \Delta
   ^3\\
   &\qquad
   +
   \left(827449216-585095018
   \sqrt{2}\right) \Delta ^4
   +
   \left(421348947-297938782 \sqrt{2}\right) \Delta ^5\\
&  \qquad +
   \left(85139111-60202471 \sqrt{2}\right)
   \Delta ^6
   +
\left(5800844-4101819 \sqrt{2}\right) \Delta ^7
 \Big]\,.
}
We can reduce the degree of this polynomial by approximating the poles with $\kappa=4$, following the algorithm described below \eqref{replacePoles} where we evaluate the derivatives of \eqref{replacePoles} at $\Delta=\frac12$ and $\Delta\to\infty$. We find that the poles $-5,-\frac32\in\mathbb{P}_6^0/\mathbb{P}_4^0$ can be approximated in terms of the poles $-3,-1,-\frac12\in\mathbb{P}_4^0$ (where we have excluded the pole at unitarity) as
\es{approxPoleEx}{
&\frac{1}{\Delta+5}\approx -\frac{81}{121 (\Delta +1)}+\frac{882}{605 (\Delta +3)}+\frac{128}{605 \left(\Delta
   +\frac{1}{2}\right)}\,,\\
   & \frac{1}{\Delta+\frac32}\approx\frac{27}{32 (\Delta +1)}+\frac{49}{160 (\Delta +3)}-\frac{3}{20 \left(\Delta
   +\frac{1}{2}\right)}\,.
}
After replacing these poles in the conformal block \eqref{zeroBlock} and factoring out the new positive quantity $\chi_4^0(\Delta)$, we get a degree 5 polynomial
\es{posEx2}{
P_0^{0,1}(\Delta,1/2)\big\vert_{\text{degree 5}}=&
\frac{9 \left(14148093995 \sqrt{2}-20004791886\right)}{43614208}
-\frac{\left(568915219077 \sqrt{2}-804566499986\right) \Delta
   }{10903552}\\
&   +\frac{\left(464762925623 \sqrt{2}-657319483718\right) \Delta
   ^2}{10903552}
   +\frac{1}{64} \left(33738733 \sqrt{2}-47713762\right)
   \Delta ^3\\
&   +\frac{5}{32} \left(2301877
   \sqrt{2}-3255314\right) \Delta ^4
   +\frac{1}{16} \left(703769 \sqrt{2}-995256\right) \Delta ^5
\,.
}
At this low level of $r_\text{max}$ and $\kappa$, these approximations are not very accurate, but at higher 
$r_\text{max}$ and $\kappa$ they become extremely precise, as you will show in the problem set.

\subsection{Rigorous bounds from truncated crossing}
\label{rigor}

In the first lecture, we claimed that one advantage of the conformal bootstrap bounds is that they are rigorous and can be improved monotonically, unlike say results from the $\epsilon$-expansion extrapolated to $\epsilon=1$, which are asymptotic and so are not guaranteed to improve as we compute higher orders. In the previous sections we discussed various truncations that are required to put the infinite crossing equations in a form amenable to numerical analysis. In this section we will show why these truncations are justified, which ones matter the most, and how to improve the numerical bootstrap bounds in a controlled way.

The most important truncation was approximating the infinite vector space $u,v$ that $\alpha$ acts on by $\Lambda$ derivatives in $z(u,v)$ and $\bar z(u,v)$ evaluated at the crossing symmetric point $\eta(u,v)_c=1$ and $r(u,v)_c=3-2\sqrt{2}$. When we increase $\Lambda$, we are increasing the coefficients in $\alpha$ as \eqref{dimV}, but the number of constraints (which only depend on the number of operators  with $\Delta$ and $\ell$) in the bootstrap algorithms remain the same, so it becomes easier to find an $\alpha$ that satisfies these constraints. For the scaling dimension algorithm, that means it becomes easier to find disallowed points, while for the OPE coefficient algorithm this means that it becomes easier to maximize $\alpha[F_{0,0}^{\Delta_\phi}(u,v)]$. In both case, this always leads to improved bounds. This is the sense in which the bootstrap bounds can be improved monotonically.

We could have also truncated the $u,v$ vector space by choosing derivatives around another point, or perhaps by sampling many random values of $u,v$. The advantage of derivatives around the crossing symmetric point is that $r_c\approx.17$ is very small, which justifies the truncation on $r$ for the blocks. For the truncation on spin, recall that for $r\ll1$ the conformal blocks scale as $g_{\Delta,\ell}(r,\eta)\sim r^\Delta$, and the unitarity bound for $\ell>0$ requires that $\Delta\geq \ell+d-2$, so conformal blocks at the crossing symmetric point become smaller with large $\ell$. This does not yet justify the spin cutoff though, because recall that $F^{\Delta_\phi}_{\Delta,\ell}(u,v)$ consists of blocks multiplied by OPE coefficients squared, so we must show that the tail $\sum_{\Delta>\Delta^*}\lambda^2_{\phi\phi\cO_{\Delta,\ell}}g_{\Delta,\ell}(r,\eta)$ of the block expansion becomes small for $r_c$ following \cite{Pappadopulo:2012jk}. 

To derive such a property we first consider the small $r$ expansion \eqref{blockSum} of the conformal blocks, where the $\eta$ dependence is captured by Gegenbauer polynomials $C_s^{\frac{d-2}{2}}(\eta)$, and we showed in the problem set that the coefficients $B_{s,m}(\Delta,\ell,d)$ have the same sign as the lowest term $B_{0,0}(\Delta,\ell,d)$ when $\Delta$ obeys the unitarity bounds. When combined with the fact that $C_s^{\frac{d-2}{2}}(\eta)$ takes its maximal value at $\eta=1$, we find that
\es{realR}{
|g_{\Delta,\ell}(r,\eta)|\leq |g_{\Delta,\ell}(r,1)|\,,
}
so its sufficient to study convergence of the block expansion on the real line $\eta=1$:
\es{4pointReal}{
\langle\phi(r)\phi(-r)\phi(1)\phi(-1)\rangle=\frac{g(r,1)}{2^{4\Delta_\phi}}\,,\qquad g(r,1)\equiv\sum_{\Delta,\ell} \lambda^2_{\phi\phi\cO_{\Delta,\ell}}g_{\Delta,\ell}(r,1)\,,
}
where the $2^{-4\Delta_\phi}$ comes from the prefactor in \eqref{4point}. We will find it convenient to rewrite $g(r,1)$ as a pure expansion in $r$:
\es{gr2}{
g(r,1)=\sum_{\delta}\mathcal{B}_{\delta}r^\delta\,,
}
where the coefficients $\mathcal{B}_{\delta}$ are defined by collecting all the powers of $r$ in block expansion \eqref{4pointReal}, and so are also positive when the unitarity bounds are satisfied. The tails of these two expansions for $g(r,1)$ are then related as 
\es{realBlockExp}{
\sum_{\Delta>\Delta^*,\ell} \lambda^2_{\phi\phi\cO_{\Delta,\ell}}g_{\Delta,\ell}(r,1) < \sum_{\delta>\Delta^*}\mathcal{B}_{\delta}r^\delta\,,
} 
since the RHS contains all the terms on the LHS as well as additional contributions of descendents with dimension $\delta>\Delta^*$ that come from primaries of dimension $\Delta\leq \Delta^*$. To bound the RHS we consider the $r\to1$ limit of the 4-point function:
\es{rto1}{
\lim_{r\to1}:\qquad \langle\phi(r)\phi(-r)\phi(1)\phi(-1)\rangle\approx \frac{1}{(1-r)^{4\Delta_\phi}}\quad\Rightarrow\quad g(r,1)\approx\frac{2^{4\Delta_\phi}}{(1-r)^{4\Delta_\phi}}\,,
}
which we computed by simply taking the 2-point functions of each pair of operators $\langle \phi(r)\phi(1)\rangle$ and $\langle\phi(-r)\phi(-1)\rangle$ that become close as $r\to1$. We can match the $g(r,1)$ asymptotic if we assume power law growth for the coefficients
\es{Bass}{
\lim_{\delta\to\infty}\mathcal{B}_\delta=\frac{2^{4\Delta_\phi}\delta^{4\Delta_\phi-1}}{\Gamma(4\Delta_\phi)}\,,
}
and then approximate the sum $g(r,1)=\sum_\delta \mathcal{B}_\delta r^\delta$ as an integral:
\es{sumtoInt}{
\lim_{r\to1}:\qquad g(r,1)\approx&\int_{0}^\infty d\delta \mathcal{B}_\delta r^\delta=\frac{2^{4\Delta_\phi}}{(-\log r)^{4\Delta_\phi}}\approx \frac{2^{4\Delta_\phi}}{(1-r)^{4\Delta_\phi}}\,.
}
The sum over just the coefficients $\mathcal{B}_\delta$ can then be approximated as 
\es{IE}{
\lim_{E\to\infty}:\qquad I(E)\equiv\sum_{\delta\leq E}\mathcal{B}_\delta\approx\int_0^Ed\delta \frac{(2\delta)^{4\Delta_\phi}}{\Gamma(4\Delta_\phi)}=\frac{(2E)^{4\Delta_\phi}}{\Gamma(4\Delta_\phi+1)}\,.
}
The assumption of polynomial growth for $\mathcal{B}_\delta$ and replacing sums by integrals may seem cavalier, but in fact the same result follows rigorously from the Hardy-Littlewood tauberian theorem \cite{yellow3}, the asymptotic \eqref{rto1}, and the fact that $\mathcal{B}_\delta>0$ due to the unitarity bounds. We can now use \eqref{IE} to estimate the tail of the block expansion for $g(r,1)$:
\es{RychkovBound}{
\sum_{\delta>\Delta^*}\mathcal{B}_\delta r^\delta&=\int_{\Delta^*}^\infty dE\, \left[\mathcal{B}_E\sum_{\delta\leq E}\delta(E-\delta)\right]r^E\\
&=-\log r\int_{\Delta^*}^\infty dE\,[I(E)-I(\Delta^*)]r^E\\
&\leq -\log r\int_{\Delta^*}^\infty dE\, I(E)r^E\\
&\approx-\log r\int_{\Delta^*}^\infty dE\,\frac{(2E)^{4\Delta_\phi}}{\Gamma(4\Delta_\phi+1)}r^E\\
&\lesssim\frac{(2\Delta^*)^{4\Delta_\phi}}{\Gamma(4\Delta_\phi+1)}r^{\Delta^*}\,,
}
where in the second line we integrated by parts, in the third line we dropped the negative term $-I(\Delta^*)$, in the fourth line we used \eqref{IE}, and in the last line we used the asymptotic behavior of the incomplete Gamma function. By \eqref{realR} and \eqref{realBlockExp} this asymptotic bound then applies to the tail of the full block expansion $g(r,\eta)$.\footnote{Stronger convergence bounds can be derived by considering blocks in specific dimension, such as $d=3$ \cite{Rychkov:2015lca}.} At the crossing symmetric point where $r_c\approx .17$ is small, we are thus justified in truncating the block expansion in spin, since large spins must have large dimensions by the unitarity bound. 

In fact, the truncations of $r$ and $\ell$ are so accurate at $r_c$, that given a sufficiently high value of $r_\text{max}$ and $\ell_\text{max}$, the numerical bootstrap bounds become insensitive to these values. This is very important, as unlike the truncation of the $u,v$ vector space, the bootstrap bounds do not change monotonically with $r_\text{max}$ and $\ell_\text{max}$ when these values are too small, since in that case the crossing equations is essentially a random function. The rigorous statement is that bootstrap bounds improve monotonically with $\Lambda$ in the limit $r_\text{max},\ell_\text{max}\to\infty$. In practice, for a given $\Lambda$ we experiment with high enough values of $r_\text{max}$ and $\ell_\text{max}$ such that the bounds stop changing. For $r_\text{max}$, the same value, e.g. $r_\text{max}=40$, can be used for a wide variety of $\Lambda$, while for $\ell_\text{max}$ we must increase this value every time we increase $\Lambda$ to get numerically stable results. In the next lecture, we will demonstrate how to run the numerical bootstrap in practice.

\pagebreak

\subsection{Problem Set 4}
\label{hw4}
\begin{enumerate}
\item Consider the $r_\text{max}=20$ conformal blocks that you computed for the previous problem set. Replace poles with $\kappa=10$, and plot $\bar p_{\Delta,\ell}(r_c,1)$ versus $ p_{\Delta,\ell}(r_c,1)$ for $\ell=0,1,2$ and $\ell+1\leq\Delta\leq\ell+4$ to see how accurate the replacement is at the crossing symmetric point.
\item Make a table of derivatives 
\es{Q}{
Q^{m,n}_\ell(\Delta)\equiv\left(r^{-\Delta}\partial^m_z\partial^n_{\bar z}\left[r^\Delta\bar p_{\Delta,\ell}(r,\eta) \right]\right)\big\vert_{r=r_c,\eta=1}
}
 for $m+n\leq3$ and $\ell=0,1,2$, where note that unlike $P_\ell^{m,n}$ in \eqref{Ftop} we have not included the $2v^{\Delta_\phi}$ factor, so that $Q^{m,n}_\ell(\Delta)$ is independent of $\Delta_\phi$. 
\item Install \texttt{Docker} from the website \texttt{https://www.docker.com/get-started}, which includes instructions for Windows, Mac, or Linux. You can run the extremely fast scalar block creation code \texttt{scalar\_blocks} using Docker as detailed at \\ \texttt{https://gitlab.com/bootstrapcollaboration/scalar\_blocks/-/blob/master/Readme.md}, which outputs $\frac{(-2)^\ell(d/2-1)_\ell}{(d-2)_\ell}Q^{m,n}_\ell(x=\Delta-\ell+2-d)$, where the prefactor is the difference in the normalization of the blocks between these notes and that code. Note that the parameters \texttt{--order}, \texttt{--poles}, and \texttt{max-derivs} refer to $r_\text{max}$, $\kappa$, and $\frac{\Lambda+1}{2}$, respectively. Check that you get the same answers you computed yourself in step 2.
\end{enumerate}

\pagebreak

\section{Semidefinite programming}
\label{semi}

The bootstrap algorithms discussed in previous lectures consisted of three ingredients:
\begin{enumerate}
\item A list of polynomial in $\Delta$ constraints
\es{polyCons}{
\alpha_{m,n}P_\ell^{m,n}(\Delta,\Delta_\phi)\geq0\qquad \text{for}\qquad \Delta\geq \Delta^\ell_B\,,
}
for each even spin $\ell\leq\ell_\text{max}$, where $P_\ell^{m,n}(\Delta,\Delta_\phi)$ is defined in \eqref{Ftop}, $\alpha_{m,n}$ is the functional, $m,n$ label derivatives in $z$ and $\bar z$ evaluated at the crossing symmetric point, and the bounds $\Delta^\ell_B$ and the external operator scaling dimension $\Delta_\phi$ are chosen.
\item A normalization constraint of $\alpha$ acting on the contribution to crossing of a certain operator $\cO_{\Delta_\text{norm},\ell_{norm}}$:
\es{normCon}{
\alpha_{m,n}\partial_z^m\partial_{\bar z}^n F^{\Delta_\phi}_{\Delta_\text{norm},\ell_\text{norm}}(r_c,1)=1\,.
}
\item A quantity $O^{m,n}$ that $\alpha$ maximizes, where for the OPE coefficient maximization algorithm $O^{m,n}$ is the identity operator contribution $\partial_z^m\partial_{\bar z}^n F^{\Delta_\phi}_{0,0}(r_c,1)$, while for the scaling dimension algorithm $O^{m,n}$ is just a vector of zeros.
\end{enumerate}

In this lecture we will show following \cite{Simmons-Duffin:2015qma} how these ingredients can be formulated as polynomial matrix programs (PMPs), that can in turn be written as semi-definite programs (SDPs). We then describe how to use the SDPB program to solve these SDPs numerically, and show some scaling dimension and OPE coefficient bounds for the Ising CFT.

\subsection{Polynomial Matrix Programs}
\label{PMPsec}

We begin by writing our algorithms as PMPs. Consider a set of $m_j\times m_j$ symmetric polynomial matrices
\es{symMat}{
M_j^w(x)=\begin{pmatrix}
P_{j,11}^w(x)&\dots&P_{j,1m_j}^w(x)\\
\vdots&\ddots&\vdots\\
P_{j,m_j1}^w(x) & \dots &P_{j,m_jm_j}^w(x)
\end{pmatrix}
}
labeled by $0\leq w\leq \cW$ and $1\leq j\leq \cJ$, where each element $P_{j,rs}^w(x)$ is a polynomial in $x$. For $b\in\mathbb{R}^\cW$, the goal of a PMP is to
\es{PMP}{
&\text{maximize}\quad b^0+\sum_{w=1}^\cW b^w y_w\quad\text{over}\quad y\in\mathbb{R}^\cW\,,\\
&\text{such that}\quad M_j^0(x)+\sum_{w=1}^\cW y_w M_j^w(x)\succeq0\quad\text{for all $x\geq0$ and $1\leq j\leq \cJ$}\,,\\
} 
where $\succeq0$ denotes positive semidefinite. Comparing to our OPE coefficient and scaling dimension algorithms in section \ref{bootAlg}, as written in terms of the truncated crossing equations in section \ref{approx} and summarized above, we can make the following identifications:
\begin{itemize}
\item $y_w$ corresponds to the functionals $\alpha_{m,n}$ subject to a normalization condition, so that $\mathcal{W}$ is one less than the number of derivatives $|V_\Lambda|$ \eqref{dimV} that define the vector space $\alpha$ acts on. The constant in $y$ terms $b^0$ and $M_j^0(x)$ in the PMP correspond to the constant in $\alpha$ terms in our algorithm that we get after solving for one of the $\alpha_{m,n}$ using the normalization condition.
\item Let $m_j=1$, so the single polynomial element $P_{j,11}^w(x)$ for each $M_j^w(x)$ corresponds to the polynomials $P_\ell^{m,n}(\Delta,\Delta_\phi)$ for $\Delta=x+\Delta_\ell^B$ that we defined in \eqref{Ftop} by stripping a positive factor $\chi_\kappa^\ell(\Delta)$ from each $\partial^m_z\partial^n_{\bar z}F^{\Delta_\phi}_{\Delta,\ell}(r_c,1)$. The condition $x\geq0$ in \eqref{PMP} translates to $\Delta_\ell^B$ being lower bounds on $\Delta$ for each spin $\ell$, and $\cJ$ is the number of even spins $\ell_\text{max}/2+1$ that we kept in our truncation. 
\item For the OPE coefficient algorithm, the unit operator contribution $\partial^m_z\partial^n_{\bar z}F^{\Delta_\phi}_{0,0}(r_c,1)$ that we maximize in step 4 corresponds to $b^w$ that we maximize in the PMP. For the scaling dimension algorithm, we do not maximize anything so we just set $b^w=\vec 0$. 
\end{itemize}

\subsection{Semi-definite Programs}
\label{SDP}

We will now reformulate the PMP as an SDP, whose constraints act on matrices instead of polynomials. Any positive semidefinite matrix of polynomials in $x$ with $x\geq0$ can be written as a single positive semidefinite matrix that is constant in $x$, by using some basis of symmetric polynomials in $x$. We use this fact to write the PMP as a ``dual'' SDP
\es{dual}{
\text{Dual}:\qquad&\text{maximize}\quad b^0+\sum_{w=1}^\cW b^w y_w\quad\text{over}\quad y\in\mathbb{R}^\cW\quad\text{and}\quad Y\,,\\
&\text{such that}\quad  P^0_{j,rs}(x_k)+\sum_{w=1}^\cW y_w P^w_{j,rs}(x_k)=\text{Tr}(A_{jrs}(x_k)Y)\quad\text{and}\quad Y\succeq0\,,
}
where in the second line we used a matrix of symmetric polynomials $A_{jrs}(x)$ evaluated at some points $x_k$ to relate $Y$ to a linear combination of $M_j^w(x_k)$ and $y_w$,\footnote{The details are given in \cite{Simmons-Duffin:2015qma}, but will not be needed in these lectures.} and $Y$ and $A$ are symmetric matrices with rank
\es{rankY}{
\text{rank}(Y)=\text{rank}(A)=\sum_{j=1}^\cJ m_j\left[\lfloor \deg(M_j^w(x)/2)\rfloor+\lfloor \deg((M_j^w(x)-1)/2)\rfloor+2\right]\,.
}
The functionals $y_w$ and the objective function $b^w$ are the same as those of the PMP \eqref{PMP}. Instead of a functional $y_w$ that acts on $w$ in $P^w_{j,rs}(x_k)$, we can also consider a ``primal'' functional $\tilde y_{jrsk}$ (i.e. the dual of the dual) that acts on the other labels, which is constrained by a ``primal'' SDP\footnote{The terminology ``dual'' and ``primal'' is unfortunately standard in SDP literature, even though the dual SDP is in fact more directly related to the PMPs that we use to write our bootstrap algorithms.}
\es{primal}{
\text{Primal}:\qquad&\text{minimize}\quad b^0+\sum_{j,r,s,k} P^0_{j,rs}(x_k) \tilde y_{jrsk}\quad\text{over}\quad \tilde y\in\mathbb{R}^\cP\quad\text{and}\quad\tilde Y= \sum_{j,r,s,k}A_{jrs}(x_k)\tilde y_{jrsk}\,,\\
&\text{such that}\quad  \sum_{j,r,s,k} P^w_{j,rs}(x_k)\tilde y_{jrsk}=-b^w\quad\text{and}\quad\tilde Y\succeq0\,,\\
} 
which is dual to the ``dual'' SDP.

The program \texttt{SDPB} solves the primal and dual SDPs \eqref{primal} and \eqref{dual} by starting with a guess for the initial point $(y_0,Y_0,\tilde y_0,\tilde Y_0)$, and then decreasing the deviations from the constraints 
\es{constraints}{
\delta^w_{\text{primal},1}&\equiv|\sum_{j,r,s,k} P^w_{j,rs}(x_k)\tilde y_{jrsk}+b^w|\,,\\
\delta_{\text{primal},2}&\equiv|\tilde Y- \sum_{j,r,s,k}A_{jrs}(x_k)\tilde y_{jrsk}|\,,\\
\delta^{jrsk}_{\text{dual}}&\equiv|P^0_{j,rs}(x_k)+\sum_{w=1}^\cW y_w P^w_{j,rs}(x_k)-\text{Tr}(A_{jrs}(x_k)Y)|\,,\\
}
and extremizing the objective functions 
\es{objs}{
\texttt{dualObjective}&\equiv b^0+\sum_{w=1}^\cW b^w y_w\\
\texttt{primalObjective}&\equiv b^0+\sum_{j,r,s,k} P^0_{j,rs}(x_k) \tilde y_{jrsk}\,,
}
by incrementally changing $(y_0,Y_0,\tilde y_0,\tilde Y_0)$ subject to the conditions $Y\succeq0$ and $\tilde Y\succeq0$. We say a point is ``primal feasible'' or ``dual feasible'' when the $\delta$ in each SDP become smaller than  certain threshold values:
\es{feasible}{
&\text{primal feasible:}\;\;  \texttt{primalError}\;\equiv\;\text{max}\{\delta^w_{\text{primal},1},\delta_{\text{primal},2}\}\;<\;\texttt{primalErrorThreshold}\\
&\text{dual feasible:}\qquad\;\; \texttt{dualError}\;\equiv\qquad\quad\;\text{max}\{\delta^{jrsk}_{\text{dual}}\}\qquad<\;\texttt{dualErrorThreshold}\,,\\
} 
where \texttt{primalErrorThreshold} and \texttt{dualErrorThreshold} are specified by the user. Once a point $(y,Y,\tilde y,\tilde Y)$ becomes both primal and dual feasible, the difference between the objective functions becomes nonnegative:
\es{diffObj}{
\sum_{j,r,s,k} P^0_{j,rs}(x_k) \tilde y_{jrsk}-\sum_{w=1}^\cW b^w y_w&\approx\sum_{j,r,s,k} P^0_{j,rs}(x_k) \tilde y_{jrsk}+\sum_{w=1}^\cW \sum_{j,r,s,k} P^w_{j,rs}(x_k)\tilde y_{jrsk} y_w\\
&\approx\sum_{j,r,s,k} \tilde y_{jrsk}  \text{Tr}(A_{jrs}(x_k)Y)\\
&\approx\text{Tr}(\tilde YY)\geq0\,,
}
where $\approx$ denotes equality up to the threshholds in \eqref{feasible}, the first equality comes from feasibility of the first primal condition in \eqref{primal}, the second comes from feasibility of the dual condition in \eqref{dual}, the third comes from feasibility of the second primal condition, and $\text{Tr}(\tilde YY)$ is nonnegative because $Y$ and $\tilde Y$ are positive semidefinite. We can then define the nonnegative normalized difference between the objective functions
\es{dualityGap}{
\texttt{dualityGap}\equiv\frac{|\texttt{primalObjective}-\texttt{dualObjective}|}{\max\{1,|\texttt{primalObjective}+\texttt{dualObjective}|\}}<\texttt{dualityGapThreshold}\,,
}
which will gradually decrease until a certain user-specified threshold \texttt{dualityGapTreshold} is reached, at which point the objective functions are approximately equal and \eqref{diffObj} then implies that $\text{Tr}(\tilde Y Y)\approx0$, and thus $\tilde Y Y\approx0$, so we call $(y,Y,\tilde y,\tilde Y)$ an optimal solution to the SDP.

\subsection{Using \texttt{SDPB}}
\label{sdpb}

We now describe how to generate SDPs using the wrapper Mathematica script \texttt{Bootstrap.m},\footnote{Adapted from the Mathematica script at \texttt{https://github.com/davidsd/sdpb/tree/master/mathematica}.} and then solve them with \texttt{SDPB}. The first step is to make a table of derivative of the crossing equations evaluated at the crossing symmetric point. These are written in terms of the truncated crossing function $\partial^m_z\partial^n_{\bar z}F^{\Delta_\phi}_{\Delta,\ell}(r_c,1)$,  which we rewrote in \eqref{Ftop} in terms of polynomials $P^{m,n}_{\ell}(x)$ in $x$ and positive factors $ \chi^\ell_\kappa(x)$, where $x\equiv\Delta-\Delta_B^\ell$ for the scaling dimension lower bounds $\Delta_B^\ell$ for each $\ell$. As described in section \ref{approx}, the accuracy of $\partial^m_z\partial^n_{\bar z}F^{\Delta_\phi}_{\Delta,\ell}(r_c,1)$ is given by the number of derivatives parameter $\Lambda$, the number of spins parameter $\ell_\text{max}$, the expansion in $r$ order $r_\text{max}$ for the conformal blocks, and the number of poles parameter $\kappa$ for each block with spin $\ell$. In practice, $\Lambda$ is the only parameter that improves the convergence of the bootstrap, and the rest of the parameters just need to be set to high enough values so that the bootstrap program works for the given $\Lambda$. 

We then use these $P^{m,n}_{\ell}(x)$ and $ \chi^\ell_\kappa(x)$ to construct the three inputs to any SDP algorithm: a list of polynomial matrix constraints, a normalization condition on $\alpha$, and the objective function that SDPB maximizes. Each polynomial matrix constraint is entered into \texttt{Bootstrap.m} as
\es{posMwithP}{
&\texttt{PositiveMatrixWithPrefactor[DampedRational[{\it c,\{p$_1$,\dots,p$_k$\},b},x],}\\
&\qquad\qquad\qquad\qquad\qquad\qquad\qquad\qquad\texttt{{\it $\langle$matrix of polynomials$\rangle$}]}\,,
}
and the positive factors we consider are encoded in \texttt{DampedRational} as
\es{damp}{
\texttt{DampedRational[{\it c,\{p$_1$,\dots,p$_k$\},b},x]}\to c\frac{b^x}{\prod_{i=1}^k(x-p_i)}\,,
}
where $p_i$ can be the poles in $\Delta$ and $b^x$ can be the factor $r_c^\Delta$. For instance, for the single correlator bootstrap algorithms, we have a table of \texttt{PositiveMatrixWithPrefactor} objects for every even $\ell\leq\ell_\text{max}$, each of which consists of a table $P^{m,n}_{\ell}(x)$ for $m+n$ odd and $m+n\leq\Lambda$, and the associated \texttt{DampedRational} is simply $\chi_\kappa^\ell(x)$. Note that for these algorithms the matrix of polynomials is a trivial $1\times1$ matrix, but we will consider nontrivial matrix constraints in a later lecture.

Next, we define an objective function ${\it obj}$ that the functional $\alpha$ will maximize and a normalization ${\it norm}$ that $\alpha$ will normalize to $1$, as discussed in the previous lecture. Like the list of polynomial matrix constraints, both of these quantities are table of derivatives labeled by $m$ and $n$. Unlike the polynomial constraints, though, ${\it obj}$ and ${\it norm}$ are not polynomials in $x$, since they are derivatives of specific operators with fixed dimensions and spins, and so we do not need to remove the positive factor $\chi_\kappa^\ell(\Delta)$ or truncate the poles in $\Delta$ that appear. For instance, for the scalar bound algorithm, ${\it obj}$ is just a vector of zeros of length $|V_\Lambda|$ and {\it norm} is the unit operator contribution $\partial^m_z\partial^n_{\bar z}F^{\Delta_\phi}_{0,0}(r_c,1)$. These quantities along with the matrix of polynomials can be processed into an \texttt{SDPB} input file using the function
\es{writeBoot}{
\texttt{WriteBootstrapSDP[file,SDP[{\it obj},{\it norm},{\it positive matrices with prefactors}]]}\,,
}
which outputs an \texttt{.xml} file named \texttt{file.xml}.\footnote{In the lastest version of \texttt{SDPB} the \texttt{.xml} file must furthermore be passed to a code called \texttt{pvm2sdp} that converts it into a file that can be directly submitted to \texttt{SDPB}.} We can now submit this file to \texttt{SDPB}, along with the following list of \texttt{SDPB} parameters:
\es{SDPBparam}{
&\texttt{dualityGapThreshold\;\quad\qquad\,=\quad1e-5}\\
&\texttt{primalErrorThreshold\;\quad\quad\;\;\,=\quad1e-10}\\
&\texttt{dualErrorThreshold\;\quad\qquad\;\;\,=\quad1e-5}\\
&\texttt{findPrimalFeasible\;\quad\qquad\;\;\,=\quad 1}\\
&\texttt{findDualFeasible\;\quad\qquad\quad\;\;\,=\quad 1}\\
&\texttt{detectPrimalFeasibleJump\quad=\quad 1}\\
&\texttt{detectDualFeasibleJump\qquad=\quad 1}\\
&\texttt{precision(actual)\qquad\qquad\;\,=\quad 576(576)}\\
&\texttt{procsPerNode\qquad\qquad\qquad\;\;\;\,=\quad 4}\,,\\
}
where we wrote some sample values for each, and did not include other parameters discussed in \cite{Simmons-Duffin:2015qma} that can be safely set to their default values for our purposes. The first three parameters are the thresholds described in the previous section that tell \texttt{SDPB} when it has found a dual feasible, primal feasible, or optimal solution. By default, \texttt{SDPB} will successfully end when it finds an optimal solution, unless we turn on the next four parameters (by setting the values to \texttt{1} for true and \texttt{0} for false), that tell \texttt{SDPB} to terminate when it has found, respectively, a dual feasible solution, a primal feasible solution, or a jump that indicates that the step size has increased to one without finding a primal feasible point, or a dual feasible point. The precision that \texttt{SDPB} uses is measured in binary digits, and needs to be set to a fairly high value.\footnote{The most common bug in running \texttt{SDPB} is that the precision has not been set high enough.} Lastly, \texttt{procsPerNode} sets how many cores per node your machine should use when running \texttt{SDPB}.

For the scaling dimension algorithm, we are not maximizing any quantity and are only checking if a solution to the SDP exists. If the \texttt{dualObjective} and \texttt{primalObjective} both decrease toward zero, then we will find a dual feasible point, which means that the functional $\alpha$ exists so the values of $\Delta_B^\ell$ and $\Delta_\phi$ that we are testing are then disallowed. If \texttt{SDPB} cannot find a dual feasible point, then we cannot conclude anything about $\Delta_B^\ell$ and $\Delta_\phi$. In practice, if \texttt{SDPB} cannot find a solution, then it will keep increasing the step size until a primal feasible jump occurs and the program terminates. So to rule out points as fast as possible, we should set the parameters \texttt{findPrimalFeasible}, \texttt{findDualFeasible}, \texttt{detectPrimalFeasibleJump}, and \texttt{detectDualFeasibleJump} all to true. 

For the OPE coefficient algorithm, we are maximizing $\alpha[F_{0,0}^{\Delta_\phi}(u,v)]$ so as to bound an OPE coefficient. We are thus looking for an optimal solution to the SDP, whose precision is determined by \texttt{dualityGapThreshold}, which we should set to a value comparable to how precise we expect to be able to bound the OPE coefficient with the value of $\Lambda$ that we are using. From \eqref{OPEbound}, we see that the OPE coefficient bound then corresponds to the final value of \texttt{primalObjective}$\approx$\texttt{dualObjective}, with an extra minus sign for the upper bound case. The parameters \texttt{findPrimalFeasible}, \texttt{findDualFeasible}, \texttt{detectPrimalFeasibleJump}, and \texttt{detectDualFeasibleJump} should all be set to \texttt{0}, i.e. false, to make sure the solver does not stop before we have found an optimal point. We should also use more precision, e.g. $\texttt{precision}=776$.

\subsection{Single correlator bootstrap bounds}
\label{sampleBounds}

\begin{figure}[t!]
\begin{center}
   \includegraphics[width=0.85\textwidth]{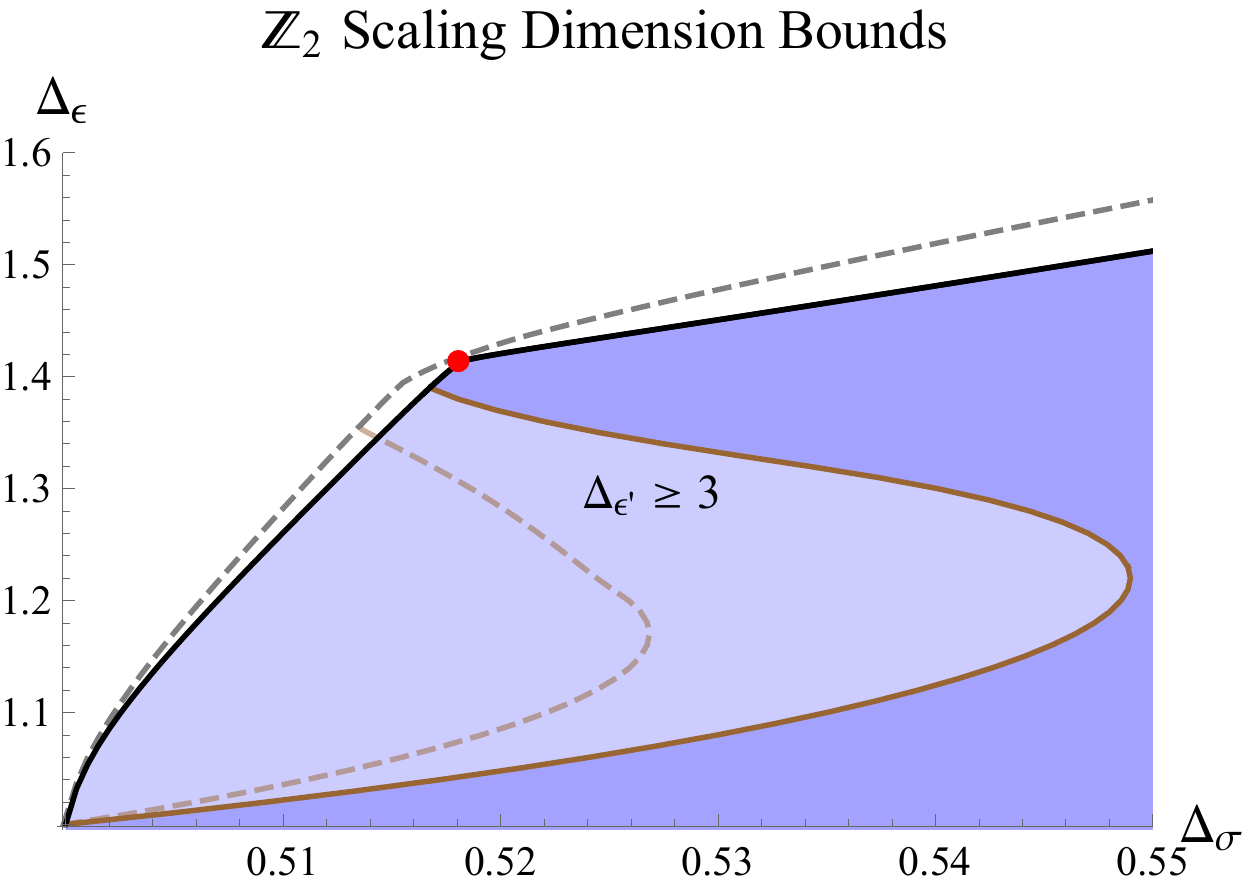}
 \caption{The solid black curve shows numerical bootstrap upper bounds with $\Lambda=19$ on the scaling dimension $\Delta_{\epsilon}$ of the lowest dimension $\mathbb{Z}_2$ even operator as a function of the scaling dimension $\Delta_{\sigma}$ of the lowest dimension $\mathbb{Z}_2$ odd operator. The solid brown curve shows the smaller peninsular allowed region that we get by additionally assuming that the scaling dimension $\Delta_{\epsilon'}$ of the next lowest dimension $\mathbb{Z}_2$ even operator is irrelevant. The blue denotes the allowed regions with and without the gap assumption for $\Lambda=19$. The dashed opaque curves show the analogous results with only $\Lambda=11$. The red dot shows the Monte Carlo value in Table \ref{bootWin} for the Ising CFT.
 }
\label{isingScal}
\end{center}
\end{figure}  

We will now discuss the results of running the scaling dimension and OPE coefficient bootstrap algorithms in 3d for a single correlator, following \cite{ElShowk:2012ht}. In Figure \ref{isingScal} we show upper bounds computed from the scaling dimension algorithm with parameters (in dashed gray)
\es{lam6}{
\Lambda=11\,,\quad \ell_\text{max}=15\,,\quad r_\text{max}=30\,,\quad \kappa=10\,,
}
which can be easily run on a home laptop, as well as the more constraining parameters (in solid black)
\es{lam10}{
\Lambda=19\,,\quad \ell_\text{max}=20\,,\quad r_\text{max}=30\,,\quad \kappa=18\,.
}
Without making any further assumptions, the single correlator bootstrap automatically applies to theories with a $\mathbb{Z}_2$ symmetry, so that that the external operator $\sigma$ is $\mathbb{Z}_2$ odd, while the lowest dimension scalar internal operator $\epsilon$ is $\mathbb{Z}_2$ even.\footnote{The internal operators must be $\mathbb{Z}_2$ even for any 4-point function of operators with the same $\mathbb{Z}_2$ parity. If the external operator was $\mathbb{Z}_2$ even, then we would have to use the same value for external operator scaling dimension as for the lowest dimension scalar internal operator, so we automatically exclude this case by not correlating these values.} The upper bounds start at the free theory point $(\Delta_\sigma,\Delta_\epsilon)=(\frac12,1)$,\footnote{Recall that the blocks have a pole at $\Delta=\frac12$ and $\ell=0$, so in practice we look at points very close to the free theory value.} which must be allowed by our bounds, and then show a distinct kink at the point near the Monte Carlo values in Table \ref{bootWin} for the Ising CFT, shown by a red dot. Note that the $\Lambda=19$ bounds are in general more constraining than the $\Lambda=11$ bounds, but they in fact are very close at the point that we putatively identify with the Ising CFT.\footnote{As far as I am aware, this is the first time this observation has been printed.}

\begin{figure}[t!]
\begin{center}
   \includegraphics[width=0.85\textwidth]{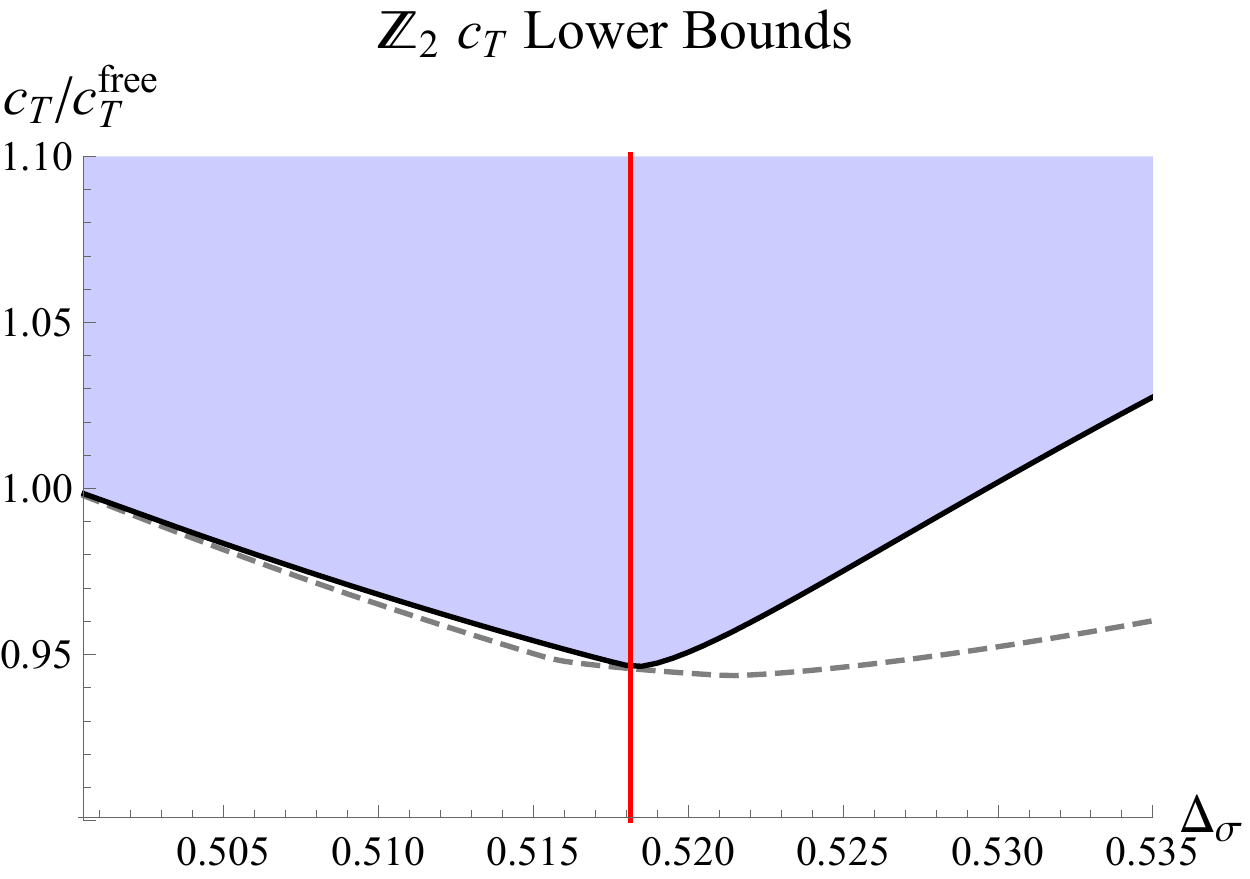}
 \caption{The solid black curve shows numerical bootstrap lower bounds with $\Lambda=19$ on the ratio $c_T/c_T^\text{free}$ of the stress tensor 2-point function and its free value as a function of the scaling dimension $\Delta_{\sigma}$ of the lowest dimension $\mathbb{Z}_2$ odd operator, where the blue denotes the allowed regions. The dashed opaque curve shows the analogous results with only $\Lambda=11$. The red line shows the Monte Carlo value in Table \ref{bootWin} for $\Delta_\sigma$ for the Ising CFT (no similar value is known for $c_T$).}
\label{isingcT}
\end{center}
\end{figure}  

We can make a further assumption that there is only a single relevant $\mathbb{Z}_2$ even operator, which we know to be true for the Ising CFT, by imposing a gap of $\Delta_{\epsilon'}\geq3$ above $\Delta_\epsilon$. This carves out a peninsular region bounded by the brown curve on the left and the original black curve above, whose tip is close to the expected Ising CFT point. This adds further evidence to the conjecture that the kink corresponds to the Ising CFT. Note that the $\Lambda=19$ allowed region is drastically smaller than the $\Lambda=11$ away from the Ising point, so we can conjecture that at infinite $\Lambda$ the Ising point would appear exactly at the tip.

In Figure \ref{isingcT}, we show the results of running the OPE coefficient squared maximization algorithm for the stress tensor OPE coefficient $\lambda_{\sigma\sigma \cO_{3,2}}$, which as shown in \eqref{cTtoOPE2} is related to the stress tensor 2-point function coefficient $c_T$ as $c_T\propto \Delta_{\sigma}^2/\lambda^2_{\sigma\sigma \cO_{3,2}}$, which we then divide by the free theory value $c_T^\text{free}$ for simplicity. Again, we run these lower bounds on $c_T$ for both $\Lambda=11$ and $\Lambda=19$. The $\Lambda=11$ bounds show a kink at Ising CFT value for $\Delta_\sigma$, shown by the red line, which becomes a minimum for $\Lambda=19$. As with the scaling dimension bounds, we see that the $\Lambda=11$ and $\Lambda=19$ bounds are in general quite different, but are very similar at the value we identify with the Ising CFT. There are no accurate estimates for $c_T$ from non-bootstrap methods, so we cannot show any predictions for the y-axis Ising CFT value as we could for the scaling dimension bounds, but it is reasonable to conjecture that the Ising CFT value is at the boundary of the allowed region, which will be justified later.

\pagebreak

\subsection{Problem Set 5}
\label{hw5}
\begin{enumerate}
\item Confirm the Ising model scaling dimension upper bounds in Figure \ref{isingScal} for the $\Lambda=11$ bounds by checking a grid of points around the kink in $(\Delta_\sigma,\Delta_\epsilon)$ space. The first step is to generate the conformal bootstrap input using the mathematica wrapper \texttt{Bootstrap.m}, where each polynomial $P^{m,n}_{\ell}(x)$ is constructed from the wrapper code from the $Q^{m,n}_{\ell}(x)$ that you can efficiently compute from \texttt{scalar\_blocks} as shown in the previous problem set. You can then run the numerical bootstrap with the parameters specified in the main text again using Docker as detailed at \\ \texttt{https://github.com/davidsd/sdpb/blob/elemental/docs/Docker.md}. For $\Lambda=11$, each point should take no longer than a minute or so.
\item Change \texttt{Bootstrap.m} to add the gap $\Delta_{\epsilon'}\geq3$, and check the same grid of points for $\Lambda=11$ to see that a peninsular region is forming as shown in Figure \ref{isingScal}.
\item Change \texttt{Bootstrap.m} to run the OPE coefficient maximization algorithm for  $c_T\propto\frac{\Delta^2_\sigma}{\lambda^2_{\sigma\sigma,\cO_{3,2}}}$ with $\Lambda=11$, using the parameters specified in the main text. Since you are now looking for an optimal solution, the code will take slightly longer, but should still be no more than a couple minutes for each point. Check enough values of $\Delta_\sigma$ so that you can see the kink at the Ising CFT point as shown in Figure \ref{isingcT}.
\end{enumerate}

\pagebreak

\section{Bootstrapping global symmetries}
\label{global}

We will now generalize the discussion of the previous lectures to 4-point functions of identical scalar operators that transform in nontrivial irreps of continuous global symmetry groups, which occur in many CFTs of interest such as the critical $O(N)$ vector model discussed in the first lecture. Following \cite{Rattazzi:2010yc,Kos:2013tga}, we will find that global symmetries lead to more crossing equations, which allow us to distinguish CFTs with these symmetries in the space of all possible CFTs. All the results of previous section will hold for this case, unless specified otherwise.

\subsection{Correlation functions}
\label{4global}

For a CFT with global symmetry group $G$, all correlations functions must be invariant under both $G$ and the conformal group. Consider complex scalar operators $\phi_I(x)$ with an index $I$ in some complex irrep $\cR$ of a global symmetry group $G$. The nonzero 2-point function of such operators in an orthonormal basis is then
\es{2pointGlobe}{
\langle\phi_I(x_1)\bar\phi_{\bar J}(x_2)\rangle=\frac{\delta_{ I\bar J}}{x_{12}^{2\Delta_\phi}}\,,
}
where $\bar\phi_{\bar J}(x_2)$ transforms in the conjugate irrep $\overline \cR$. The generalization to complex operators $\cO_{\Delta,\ell,R}$ with spin $\ell$ in complex irrep $R$ of $G$ with index $a$ is
\es{2point2Globe}{
\langle\cO^{\mu_1\dots\mu_\ell}_a(x_1)\bar\cO_{\nu_1\dots\nu_\ell,\bar b}(x_2)\rangle&=C^{(\ell)}_{a\bar b}\left(\frac{I^{(\mu_1}{}_{\nu_1}(x_{12})\dotsb I^{\mu_\ell)}{}_{\nu_\ell}(x_{12})}{x_{12}^{2\Delta}}-\text{traces}\right)\,,\\
}
where $I^\mu{}_\nu$ was defined in \eqref{2point2}, and in general we unit normalize the operator by setting $C^{(\ell)}_{a\bar b}=\delta_{a\bar b}$. The 3-point function of $\phi_I(x)$, $\bar\phi_{\bar I}(x)$, and $\cO_{\Delta,\ell,R}$ is
\es{3point2Globe1}{
\langle\phi_I(x_1)\bar\phi_{\bar J}(x_2)\cO_a^{\mu_1\dots\mu_\ell}(x_3)\rangle&=\frac{\lambda_{\phi\bar\phi\cO_{\Delta,\ell,R}}{\bf T}^R{}_{I\bar Ja}(Z^{\mu_1}\dotsb Z^{\mu_\ell}-\text{traces})}{x_{12}^{2\Delta_\phi-\Delta_\ell+\ell}x_{23}^{\Delta_\ell-\ell}x_{31}^{\Delta_\ell-\ell}}\,,\\
}
where $Z^\mu$ is defined in \eqref{3point2} and ${\bf T}^R{}_{I\bar Ja}$ is an invariant tensor for $R$. All the irreps in the tensor product $\cR\times \overline\cR$ are real or come in complex conjugate pairs. We can choose ${\bf T}^R{}_{I\bar Ja}$ so that $\lambda_{\phi\bar\phi\cO_{\Delta,\ell,R}}=\bar\lambda_{\bar\phi\phi\cO_{\Delta,\ell,\bar R}}=(-1)^\ell\bar\lambda_{\phi\bar\phi\cO_{\Delta,\ell,\bar R}}$, so when $R$ is real the OPE coefficient is real for even $\ell$ and pure imaginary for odd $\ell$, while when $R$ is complex the OPE coefficients comes in complex conjugate pairs. We can also consider the conjugate 3-point functions
\es{3point2Globe2}{
\langle\phi_I(x_1)\phi_J(x_2)\cO_a^{\mu_1\dots\mu_\ell}(x_3)\rangle&=\frac{\lambda_{\phi\phi\cO_{\Delta,\ell,R}}{\bf T}^{R}{}_{IJa}(Z^{\mu_1}\dotsb Z^{\mu_\ell}-\text{traces})}{x_{12}^{2\Delta_\phi-\Delta_\ell+\ell}x_{23}^{\Delta_\ell-\ell}x_{31}^{\Delta_\ell-\ell}}\,,\\
\langle\bar\phi_{\bar I}(x_1)\bar\phi_{\bar J}(x_2)\bar\cO_{\bar a}^{\mu_1\dots\mu_\ell}(x_3)\rangle&=\frac{\lambda_{\bar\phi\bar\phi\cO_{\Delta,\ell,\bar R}}{\bf T}^{\bar R}_{\bar I\bar J\bar a}(Z^{\mu_1}\dotsb Z^{\mu_\ell}-\text{traces})}{x_{12}^{2\Delta_\phi-\Delta_\ell+\ell}x_{23}^{\Delta_\ell-\ell}x_{31}^{\Delta_\ell-\ell}}\,,\\
}
where here $a$ is an index for a complex irrep $R\in \cR\times\cR$, which implies that we can choose ${\bf T}^{R}{}_{IJa}$ and ${\bf T}^{\bar R}{}_{\bar I\bar J\bar a}$ here so that $\lambda_{\phi\phi\cO_{\Delta,\ell,R}}=\bar\lambda_{\bar\phi\bar\phi\bar\cO_{\Delta,\ell,\bar R}}$. If $R$ is in the symmetric/antisymmetric product of $\cR\times\cR$, then ${\bf T}^{R}_{IJa}$ is symmetric/antisymmetric in $IJ$, so $\ell$ must be even/odd following the argument around \eqref{realOPE}. 

For the 4-point function of two $\phi$ and two $\bar\phi$, we can take the OPE between either $\phi\times\bar\phi$ twice or $\phi\times\phi$ and its conjugate, to get the conformal block expansions
\es{4pointAgainGlobe}{
\langle\phi_I(x_1)\bar\phi_{\bar J}(x_2)\phi_K(x_3)\bar\phi_{\bar L}(x_4)\rangle=\frac{1}{x_{12}^{2\Delta_\phi}x_{34}^{2\Delta_\phi}}\sum_{\Delta,\ell}\sum_{R\in\cR\times\bar\cR}{\bf T}^R{}_{I\bar J K\bar L}|\lambda_{\phi\bar\phi\cO_{\Delta,\ell,R}}|^2g_{\Delta,\ell}(u,v)\,,
}
where $\ell$ can be even or odd for each irrep $R\in\cR\times\bar\cR$, and  
\es{4pointAgainGlobe2}{
\langle\phi_I(x_1)\phi_J(x_2)\bar\phi_{\bar K}(x_3)\bar\phi_{\bar L}(x_4)\rangle=\frac{1}{x_{12}^{2\Delta_\phi}x_{34}^{2\Delta_\phi}}\sum_{\Delta,\ell}\sum_{R\in\cR\times\cR}{\bf T}^R{}_{I J \bar K\bar L}|\lambda_{\phi\phi\cO_{\Delta,\ell,R}}|^2g_{\Delta,\ell}(u,v)\,,
}
where $\ell$ is now even/odd depending on whether $R$ is in the symmetric/antisymmetric product of $\cR\times \cR$. The tensor structures ${\bf T}^R{}_{I\bar J K\bar L}$ and ${\bf T}^R{}_{I J \bar K\bar L}$ are fixed in terms of the 3-point function tensor structures as 
\es{3to4}{
{\bf T}^R{}_{I\bar J K\bar L}={\bf T}^R_{I\bar J a}{\bf T }^{\bar R}_{K\bar L \bar a}\,,\qquad {\bf T}^R{}_{I J \bar K\bar L}={\bf T}^R_{I J a}{\bf T}^{\bar R}_{\bar K\bar L \bar a}\,,
}
where we assumed here all operators have unit normalized 2-point functions. If the 3-point function tensor structures were chosen with the OPE coefficient reality properties described above, then we find that $|\lambda_{\phi\bar\phi\cO_{\Delta,\ell,R}}|^2\geq0$ and $|\lambda_{\phi\phi\cO_{\Delta,\ell,R}}|^2\geq0$.

\subsection{Conserved Currents}
\label{moreCurs}

Recall that a local CFT has a stress tensor operator $T$ with $\Delta=d$, $\ell=2$, which we can use to define the infinitesimal generator of the conformal group, and whose 2-point function coefficient $c_T$ was related to the OPE coefficient $\lambda^2_{\phi\phi\cO_{d,2}}$ of the unit normalized $\cO_{d,2}\equiv \frac{S_d}{\sqrt{c_T}}T$ as \eqref{cTtoOPE2}. If a local CFT also has a continuous global symmetry $G$, then the spectrum of operators includes both the stress tensor in the singlet irrep of $G$, as well as a $\Delta=d-1$ and $\ell=1$ conserved current operator $J^A_\mu$ with index $a$ in the adjoint irrep $A$ of $G$. We can then define the infinitesimal generators of $G$ in terms of the canonically normalized $J$ as
\es{QfromJ}{
Q_a=-i\int_\Sigma dS_\mu J^\mu_a\,,
}
which implies that the 2-point function coefficient in \eqref{2point2Globe} is
\es{cJDef}{
C^{(1)}_{a b}=\mathfrak{t}^A\delta_{ab}\frac{c_J}{S_d^2}\,,
}
where $S_d$ was defined in \eqref{cTtoOPE}, $T_a$ are generators of $G$, $\mathfrak{t}^A$ is the index the adjoint irrep,\footnote{Defined by $\text{Tr}(T_aT_b)=\mathfrak{t}^A\delta_{ab}$} and $c_J$ is a theory dependent constant. The Ward identity that follows from \eqref{QfromJ} for a scalar operator $\phi_I$ is
\es{wardJ}{
[Q_a,\phi_I]=-(T_a)_{I\bar J}\phi_{\bar J}\,,
}
where $T_a$ are generators of $G$. This fixes $\lambda_{\phi\bar\phi\cO_{d-1,1,A}}$ and the tensor structure ${\bf T}^A_{I\bar J a}$ as defined in \eqref{3point2Globe1} for the unit normalized $\cO_{d-1,1,A}\equiv \frac{S_d}{\sqrt{\mathfrak{t}^Ac_J}}J$ to be \cite{Osborn:1993cr}
\es{cJtoOPE}{
\lambda_{\phi\bar\phi\cO_{d-1,1,A}}=\frac{1}{\sqrt{c_J}}\,,\qquad {\bf T}^A_{I\bar Ja}=-i(T_a)_{I\bar J}\,.
}
This relation can be used to enter $c_J$ into the bootstrap, just as \eqref{cTtoOPE2} was used to enter $c_T$ into the bootstrap.

 For non-local CFTs, such as the generalized free field theories discussed in section \ref{cons}, $J$ may not appear in the spectrum. The infinitesimal generators of $G$ are then just defined in terms of their algebra, and cannot be interpreted as integrals of a local quantity.

\subsection{Crossing equations}
\label{crossGlobe}

We now derive the crossing equations by demanding that \eqref{4pointAgainGlobe} and  \eqref{4pointAgainGlobe2} are invariant under permutations $\phi$ and $\bar\phi$. In each case, we add and substract the 4-point function from its crossed version, where we swap $(1,I)\leftrightarrow(3,K)$ and then relabel $I\leftrightarrow K$, to get\footnote{We could also consider swapping $(1,I)\leftrightarrow(2,J)$, but as in the non-global symmetry case this just enforces the condition of what spins can appear in which block, which we can already determine from 3-point functions.}
\es{FGlobe}{
\sum_{\Delta,\ell}\sum_{R\in \cR\times\bar\cR}|\lambda_{\phi\bar\phi\cO_{\Delta,\ell,R}}|^2\left[T^R{}_{I\bar J K\bar L}F^{\Delta_\phi}_{\pm,\Delta,\ell}(u,v)\mp T^R{}_{K\bar J I\bar L}F^{\Delta_\phi}_{\pm,\Delta,\ell}(u,v)\right]=0\,,
}
and
\es{FGlobe2}{
\sum_{\Delta,\ell}\left[\sum_{R\in \cR\times\cR}|\lambda_{\phi\phi\cO_{\Delta,\ell,R}}|^2T^R{}_{I J \bar K\bar L}F^{\Delta_\phi}_{\pm,\Delta,\ell}(u,v)\mp(-1)^\ell \sum_{R\in \cR\times\bar\cR}\lambda_{\phi\bar\phi\cO_{\Delta,\ell,R}}^2T^R{}_{\bar K J I\bar L}F^{\Delta_\phi}_{\pm,\Delta,\ell}(u,v)\right]=0\,,
}
where we define the crossing functions
\es{F22}{
F^{\Delta_\phi}_{\pm,\Delta,\ell}(u,v)\equiv v^{\Delta_\phi}g_{\Delta,\ell}(u,v)\pm u^{\Delta_\phi}g_{\Delta,\ell}(v,u)\,,
}
and the $(-1)^\ell$ factor in \eqref{FGlobe2} came from $\lambda_{\bar\phi\phi\cO_{\Delta,\ell,R}}=(-1)^\ell\lambda_{\phi\bar\phi\cO_{\Delta,\ell,R}}$ as seen from \eqref{3point2Globe1}. These $F^{\Delta_\phi}_{\pm,\Delta,\ell}(u,v)$ generalize the $F^{\Delta_\phi}_{\Delta,\ell}(u,v)=F^{\Delta_\phi}_{-,\Delta,\ell}(u,v)$ defined in \eqref{F} for the non-global symmetry case, which was the combination of blocks that appeared in the single crossing equation in that case. We can then solve \eqref{FGlobe2} in terms of $F^{\Delta_\phi}_{\pm,\Delta,\ell}(u,v)$ and the OPE coefficients to get a vector of crossing equations\footnote{Here we are not considering the case where the same irrep appear twice, in which case there will be multiple tensor structures and OPE coefficients for the same irrep, and the crossing equations must include matrix constraints, as in the mixed correlator case considered in the next lecture.}
\es{V}{
\sum_{\Delta,\ell,R}|\lambda_{\Delta,\ell,R}|^2 {\vec V}_{\Delta,\ell,R}=\vec 0\,,
}
where $\vec V$ is a linear combination of $F^{\Delta_\phi}_{\pm,\Delta,\ell}(u,v)$, and the length of $\vec V$ is the number of tensor structures that appear in the 4-point function. 

For instance, consider $\phi_I(x)$ transforming in the fundamental irrep $\cR$ of $G=O(N)$, so $\phi_I(x)$ is real and we need only consider \eqref{FGlobe}. The tensor product $\cR\times\cR$ then includes a singlet $S$, antisymmetric (adjoint) $A$, and symmetric traceless $T$, where $S$ and $T$ are in the symmetric product and so their operators have even spins, while $A$ is in the antisymmetric product so their operators have odd spins. We can write their tensor structures in terms of the basis of 4-index $O(N)$ tensors
\es{basisON}{
\delta_{IJ}\delta_{KL}\,,\qquad \delta_{IK}\delta_{JL}\,,\qquad \delta_{IL}\delta_{JK}\,,
}
by appropriately symmetrizing $IJ$ for each case, which yields
\es{tensorON}{
T^S_{IJKL}&=\delta_{IJ}\delta_{KL}\,,\qquad T^A_{IJKL}=\delta_{IK}\delta_{JL}-\delta_{IL}\delta_{JK}\,,\\
  T^T_{IJKL}&=\delta_{IK}\delta_{JL}+\delta_{IL}\delta_{JK}-\frac{2}{N}\delta_{IJ}\delta_{KL}\,.
}
The overall normalization of these tensor structures is conventional except for the sign, which must be chosen so that all OPE coefficients squared in the four-point function are positive. In principle, this can be checked by defining the 4-point tensor structures in terms of properly normalized 3-point tensor structures, but it is easier to just directly check the 4-point function in the free theory (in 3d for simplicity), where all operators are composites of $\phi_I$ with 2-point function \eqref{2pointGlobe} for real $\phi_I$ and $\Delta_\phi=\frac12$. We can then compute the free theory four-point function using Wick contractions and then expand in block as in \eqref{4pointAgainGlobe} to get
\es{4freeON}{
\langle \phi_I(x_1)\phi_J(x_2)&\phi_K(x_3)\phi_L(x_4)\rangle_\text{free}=\frac{1}{x_{12}x_{34}}\left[\delta_{IJ}\delta_{KL}+\delta_{IK}\delta_{JL}\sqrt{u}+\delta_{IL}\delta_{JK}\sqrt{\frac uv}\right]\\
&=\frac{1}{x_{12}x_{34}}\left[T^S_{IJKL}\left(g_{0,0}(r,\eta)+\frac 2Ng_{1,0}(r,\eta)+\frac{3}{8N}g_{3,2}(r,\eta)\right)\right.\\
&\left.\quad+\frac12g_{2,1}(r,\eta)T^A_{IJKL}+T^T_{IJKL}\left(g_{1,0}(r,\eta)+\frac{3}{16}g_{3,2}(r,\eta)\right)\right]+O(r^4)\,,
}
where we expanded in the $r,\eta$ variables that are related to $u,v$ by \eqref{ztouv} and \eqref{rhoToz}, and the blocks $g_{\Delta,\ell}(r,\eta)$ are computed in the conventions of Lecture \ref{blockSec}. Note that all the $\lambda^2_{\phi\phi\cO_{\Delta,\ell,R}}\geq0$, and the $S$ and $T$ channels contain only even spin operators, such as the stress tensor $\cO_{3,2,S}$, while the $A$ channel only contains odd operators, such as the current $\cO_{2,1,A}$.\footnote{The operator $\cO_{3,2,T}$ exists in the free theory, since the free theory generically contains all operators that saturate the unitarity bound in each irrep. In the interacting theory, the only conserved operators that survive are the stress tensor and flavor current.} We can then use the tensor structures \eqref{tensorON} to write \eqref{FGlobe} as a vector of crossing equations \eqref{V} with
\es{ONCross}{
\vec V_{\Delta,\ell^+,S}=\begin{pmatrix}0\\F^{\Delta_\phi}_{-,\Delta,\ell}\\F^{\Delta_\phi}_{+,\Delta,\ell}\end{pmatrix}\,,\quad \vec V_{\Delta,\ell^-,A}=\begin{pmatrix}F^{\Delta_\phi}_{-,\Delta,\ell}\\-F^{\Delta_\phi}_{-,\Delta,\ell}\\F^{\Delta_\phi}_{+,\Delta,\ell}\end{pmatrix}\,,\quad \vec V_{\Delta,\ell^+,T}=\begin{pmatrix}F^{\Delta_\phi}_{-,\Delta,\ell}\\\left(1-\frac 2N\right)F^{\Delta_\phi}_{-,\Delta,\ell}\\-\left(1+\frac2N\right)F^{\Delta_\phi}_{+,\Delta,\ell}\end{pmatrix}\,,
}
where $\ell^\pm$ denotes that only even/odd spins appear.

For $N=2$, we can consider $\phi(x)$ as a complex operator in the complex charged irrep $\cR$ under $U(1)\cong O(2)$. The tensor product $\cR\times\bar\cR$ then includes the parity even charge $\bf0_+$ irrep, which must have even spin, and the parity odd charge $\bf0_-$ irrep, which must have odd spin, while the tensor product $\cR\times\cR$ includes the charge $\bf2$ irrep, which is in the symmetric product and so must have even spin. In this presentation, there are no indices and so the tensor structures in \eqref{4pointAgainGlobe} and \eqref{4pointAgainGlobe2} become constant coefficients $T^R$, but we must still check that $T^R$ have the correct sign so that that $\lambda^2_{\phi\phi\cO_{\Delta,\ell,R}}\geq0$. We do this again using the free theory constructed from complex $\phi$ with 2-point function \eqref{2pointGlobe} and $\Delta_\phi=\frac12$. We must now consider both 4-point functions \eqref{4pointAgainGlobe} and \eqref{4pointAgainGlobe2}, which we compute using Wick contractions in 3d to get
\es{4freeO2}{
&\langle \phi(x_1)\bar\phi(x_2)\phi(x_3)\bar\phi(x_4)\rangle_\text{free}=\frac{1}{x_{12}x_{34}}\left[1+\sqrt{\frac uv}\right]\\
&=\frac{1}{x_{12}x_{34}}\left[T^{\bf 0_+}\left(g_{0,0}(r,\eta)+g_{1,0}(r,\eta)+\frac{3}{16}g_{3,2}(r,\eta)\right)-\frac12g_{2,1}(r,\eta)T^{\bf 0_-}\right]+O(r^4)\,,\\
}
which includes the irreps ${\bf0}_\pm\in{\cR}\times\bar{\cR}$, and then
\es{4freeO22}{
\langle \phi(x_1)\phi(x_2)\bar\phi(x_3)\bar\phi(x_4)\rangle_\text{free}&=\frac{1}{x_{12}x_{34}}\left[\sqrt{u}+\sqrt{\frac uv}\right]\\
&=\frac{T^{\bf 2}}{x_{12}x_{34}}\left[2g_{1,0}(r,\eta)+\frac{3}{8}g_{3,2}(r,\eta)\right]+O(r^4)\,,\\
}
which includes the irrep ${\bf2}\in{\cR}\times{\cR}$. Note that we must choose $T^{\bf0_-}<0$ to get $\lambda^2_{\phi\bar\phi\cO_{\Delta,\ell,{\bf0_-}}}>0$, so $T^R$ are nontrivial even though they contain no indices. We can choose the OPE coefficients here to have the same normalization as the $O(N)$ presentation \eqref{4freeON} with $N=2$ by setting
\es{Tset}{
T^{\bf0_+}=1\,,\qquad T^{\bf 0_-}=-1\,,\qquad T^{\bf2}=\frac12\,,
}
in which case both crossing equations \eqref{FGlobe} and \eqref{FGlobe2} are identical to \eqref{ONCross}, which was derived from the single crossing equation \eqref{FGlobe} since $\mathcal{R}$ was real, as long as we set $N=2$ and identify $S={\bf0_+}$, $A={\bf0_-}$, and $T={\bf2}$. This demonstrates that the crossing equations do not depend on how we choose to present the global symmetry, even though the intermediate calculations seem very different.

\subsection{Implementing with \texttt{SDPB}}
\label{SDPBGlobe}

As in the previous lecture \ref{bootBas}, we can now truncate each $F^{ \Delta_\phi}_{\pm,\Delta,\ell}(u,v)$ in $V^\beta_{\Delta,\ell,R}$ in terms of $m,n$ derivatives in $z,\bar z$ at the crossing symmetric point, and write them as polynomials $P_{\ell,R,\beta}^{m,n}(\Delta,\Delta_\phi)$ in $\Delta$ with positive factors $\chi_\kappa^\ell(\Delta)$, where the polynomials are now labeled by the spin $\ell$, the irrep $R$,\footnote{By $R$ we more properly mean each tensor structure, which can be different for odd and even spin.} and the crossing equation index $\beta$. The only change to the bootstrap algorithm in section \ref{bootAlg}, is that the functional $\alpha$ now acts on $V^\beta_{\Delta,\ell,R}$ instead of just a single $F^{\Delta_\phi}_{-,\Delta,\ell}(u,v)$. As a result, the list of operators $\cO_{\Delta,\ell,R}$, scaling dimension lower bounds $\Delta_{\ell,R}^B$, and OPE coefficients squared $|\lambda|_{\Delta,\ell,R}^2$ are now labeled by $R$ in addition to $\ell$. For each $\ell$ and $R$, the functional $\alpha$ acts on the $m,n$ derivatives of $ V^\beta_{\Delta,\ell,R}$ at the crossing symmetric point as 
\es{alphaV}{
\sum_{\beta}\sum_{m,n}\alpha_{m,n}^\beta \partial_z^m\partial_{\bar z}^n V^\beta_{\Delta,\ell,R}(r_c,1)\,,
}
so $\alpha_{m,n}^\beta$ now has an extra index $\beta$ that counts the extra crossing equations.

We can implement these changes in the wrapper \texttt{Bootstrap.m} used to create the \texttt{SDPB} input, by entering a different \texttt{PositiveMatrixWithPrefactor} \eqref{posMwithP} for each $\ell$ and $R$. As in the case without global symmetry, the matrix of polynomials in \texttt{PositiveMatrixWithPrefactor} is still a $1\times1$ matrix, but now our list of polynomial constraints will run over both derivatives $m,n$ as well as the crossing equation label $\beta$. Similarly, the ${\it obj}$ and ${\it norm}$ should also be entered as a list running over $m,n,\beta$. For instance, for the scaling dimension algorithm, the unit operator contribution that we normalize corresponds to $\partial_z^m\partial_{\bar z}^n V^\beta_{0,0,S}(r_c,1)$, where $S$ is the singlet irrep of whatever symmetry group we are considering.

\subsection{Single correlator bootstrap bounds with $O(N)$ symmetry}

\begin{figure}[t!]
\begin{center}
   \includegraphics[width=0.85\textwidth]{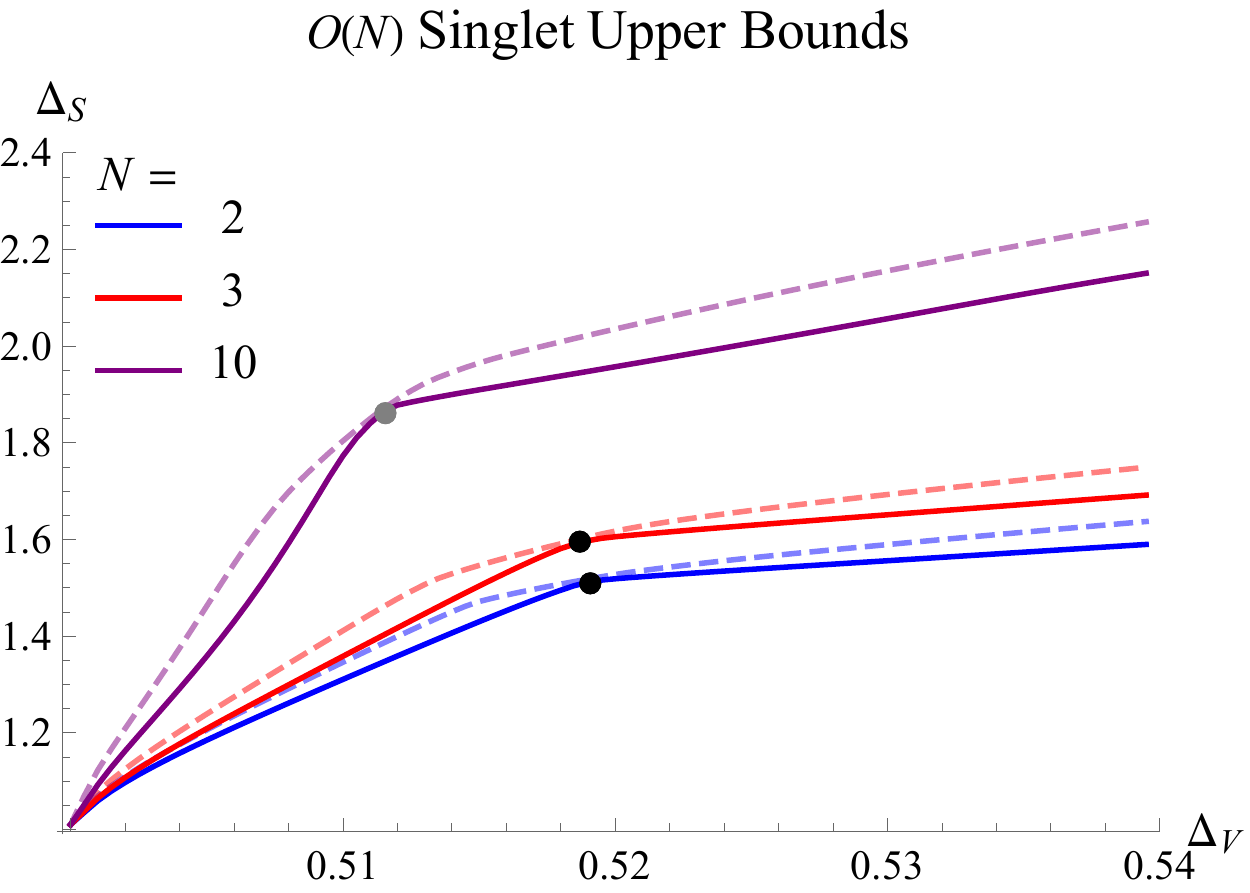}
 \caption{The solid blue, red, and purple curves shows numerical bootstrap upper bounds with $\Lambda=19$ on the scaling dimension $\Delta_{S}$ of the lowest dimension operator in the singlet irrep $S$ of $O(N)$ as a function of the scaling dimension $\Delta_{V}$ of the lowest dimension operator in the vector irrep $V$ for $N=2,3,10$, respectively. The dashed opaque curves show the analogous results with only $\Lambda=11$. For $N=2,3$, the black dots denotes the Monte Carlo values \eqref{MCON}, while for $N=10$ the black dot denotes the large $N$ value \eqref{largeNON}.}
\label{ONs}
\end{center}
\end{figure}  

\begin{figure}[t!]
\begin{center}
   \includegraphics[width=0.85\textwidth]{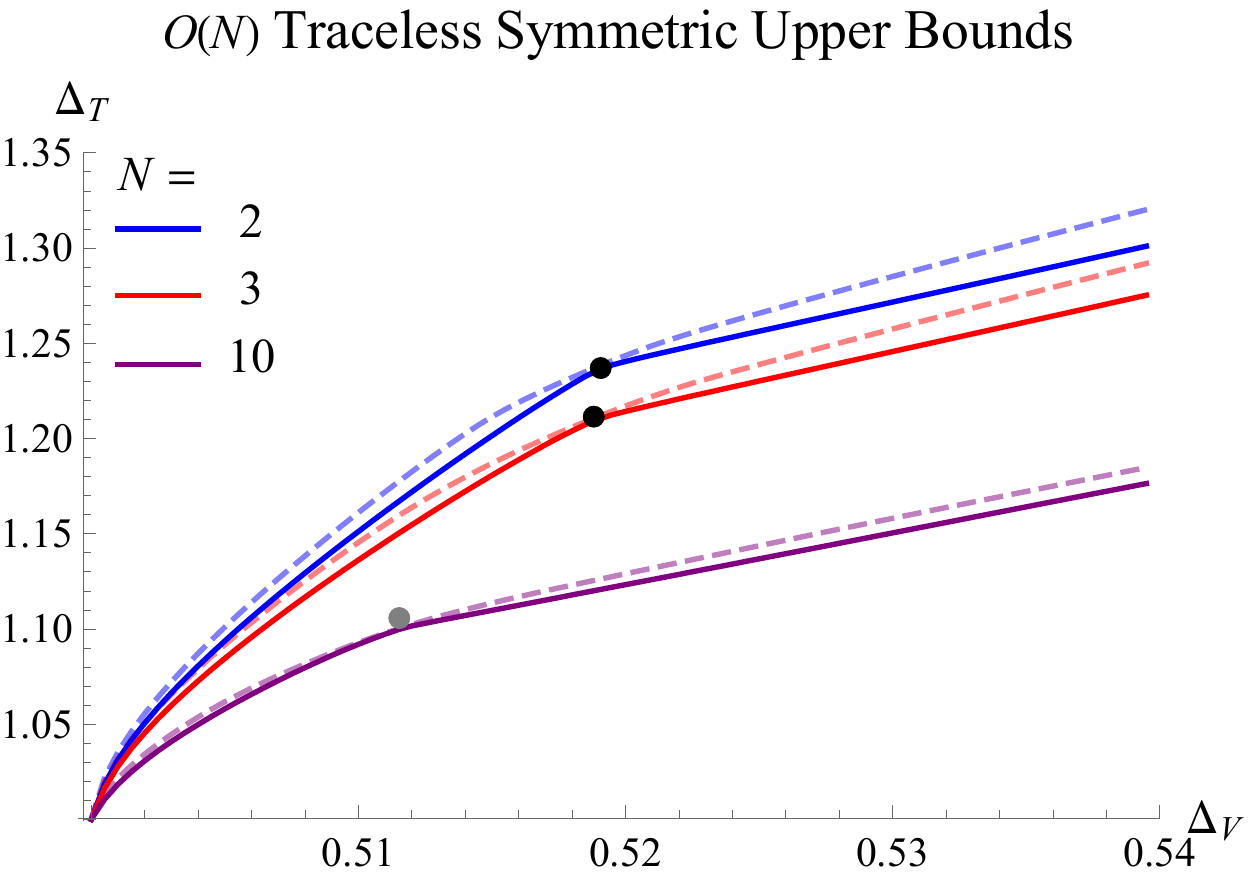}
 \caption{The solid blue, red, and purple curves shows numerical bootstrap upper bounds with $\Lambda=19$ on the scaling dimension $\Delta_{T}$ of the lowest dimension operator in the traceless symmetric irrep $T$ of $O(N)$ as a function of the scaling dimension $\Delta_{V}$ of the lowest dimension operator in the vector irrep $V$ for $N=2,3,10$, respectively. The dashed opaque curves show the analogous results with only $\lambda=11$. For $N=2,3$. the black dots denote the Monte Carlo value \eqref{MCON} for $\Delta_V$ and the pseudo-$\epsilon$ expansion values \eqref{MCON2} for $\Delta_T$, while for $N=10$ the black dot denotes the large $N$ \eqref{N10} value for all these quantities.}
\label{ONt}
\end{center}
\end{figure}  

We will now discuss the results of running the scaling dimension algorithm using the $O(N)$ crossing equations \eqref{ONCross} in 3d, as originally reported in \cite{Kos:2013tga}.\footnote{See \cite{Chester:2014gqa} for the analogous bounds in 5d.} In Figures \ref{ONs} and \ref{ONt} we show upper bounds for the lowest dimension $O(N)$ singlet scaling dimension $\Delta_S$ and $O(N)$ traceless symmetric scaling dimension $\Delta_T$, respectively, computed with $\Lambda=11$ or $\Lambda=19$ for a 4-point function of scalars in the $O(N)$ vector $V$ irrep, for $N=2,3,10$. The values of $\Delta_V$ and $\Delta_S$ for the $O(N)$ fixed point have been computed from Monte Carlo for $N=2$ \cite{Campostrini:2006ms} and $N=3$ \cite{PhysRevB.65.144520}:\footnote{The critical $O(3)$ model describes thermal phase transitions of isotropic magnets, as well as the quantum phase transition of coupled dimer antiferromagnets \cite{2004LNP...645..381S}.}
\es{MCON}{
N=2:&\qquad \Delta_V=0.51905(10)\,,\qquad \Delta_S=1.51124(22)\,,\\
N=3:&\qquad \Delta_V=0.51875(25)\,,\qquad \Delta_S=1.5939(10)\,,\\
}
while $\Delta_T$ has been computed using the pseudo-$\epsilon$ expansion \cite{Calabrese:2004ca}:
\es{MCON2}{
N=2:\qquad \Delta_T=1.237(4)\,,\qquad N=3:\qquad \Delta_T=1.211(3)\,.
}
These values have also been computed in a large $N$ expansion \cite{Moshe:2003xn,Gracey:2002qa}:
\es{largeNON}{
\Delta_V&=\frac12 + \frac{4}{3 \pi^2 N} - \frac{256}{27 \pi^4 N^2} + \frac{32 (-3188 + 3 \pi^2 (-61 + 108 \log[2]) - 3402 \zeta[3]))}{243 \pi^6 \
N^3}+O(N^{-4})\,,\\
\Delta_S&=2 - \frac{32}{3 \pi^2 N} + \frac{32}{27 \pi^4 N^2} (16 - 27 \pi^2))+O(N^{-3})\,,\\
\Delta_T&=1 + \frac{32}{3 \pi^2 N} - \frac{512}{27 \pi^4 N^2}+O(N^{-3})\,,\\
}
which for $N=10$ gives
\es{N10}{
N=10:&\qquad \Delta_V\approx0.51159\,,\qquad \Delta_S\approx1.86144\,,\qquad \Delta_T\approx1.10612\,.\\
}
We show all these previous predictions for the $O(N)$ vector model fixed point as block dots in the bootstrap plots. As in the $\mathbb{Z}_2$ bootstrap before, the upper bounds start at the free theory point $(\Delta_V,\Delta_S)=(\Delta_V,\Delta_T)=(\frac12,1)$, which must be allowed by our bounds, and then show a distinct kink at point near the $O(N)$ vector model CFT, where the large $N$ values are slightly farther from the kink than the Monte Carlo values, which is expected since the large $N$ expansion is less accurate. Also as in the Ising case, $\Lambda=19$ bounds are in general more constraining than the $\Lambda=11$ bounds, but they in fact are very close at the point that we putatively identify with the $O(N)$ CFT. 

Since $O(N)\subset O(N+1)$, we expect that the bounds should get stronger as $N$ increase, which we observe for the $\Delta_T$ bounds. For the $\Delta_S$ bounds, keep in mind that a traceless symmetric operator in $O(N+1)$ contains a singlet operator under its $O(N)$ subgroup, which generically has smaller scaling dimension than the $O(N+1)$ singlet as we see from comparing the two plots. Thus we should only expect that the $O(N+1)$ $\Delta_T$ is smaller than the $O(N)$ $\Delta_S$, which is indeed the case.

\pagebreak

\subsection{Problem Set 6}
\label{hw6}
\begin{enumerate}
\item For a general continuous global symmetry group, the tensor structures $T^R_{I \bar JK\bar L}$ in \eqref{4pointAgainGlobe} can be computed as eigenfunctions of the quadratic Casimir
\es{casGen}{
C_2\equiv\sum_{a=1}^{|G|}T^a{}_{I'\bar I}\bar T^a{}_{\bar J' J}\,,
}
where $|G|$ is the group's rank, $T^a$ are the standard generators of the group, $\bar T^a=-(T^a)^*$ are the conjugate generators, and the Casimir acts just on the first two indices of the tensor structure. Similarly, $T^R_{I J\bar K\bar L}$ in \eqref{4pointAgainGlobe2} are eigenfunctions of $\sum_{a=1}^{|G|}T^a{}_{I'\bar I} T^a{}_{ J' \bar J}$. 
\begin{enumerate}
\item Consider a 4-point function of scalar operators $\phi_I(x)$ and $\bar\phi_{\bar I}(x)$ in the fundamental and anti-fundamental, respectively, of $SU(N)$. The $|SU(N)|=N-1$ generators $T^a$ can be written with fundamental indices as
\es{SUNgen}{
T^a{}_{I\bar J}=\frac{1}{\sqrt{2a(a+1)}}\left[\left(\sum_{K=1}^a\delta_{IK}\delta_{K\bar J}\right)-a\delta_{I,a+1}\delta_{\bar J,a+1}  \right]\,,
}
so that they are normalized as $T^a_{I\bar J}T^b_{\bar I J}=\frac12\delta^{ab}$. Plug a suitable basis of tensor structures into the eigenvalue equation to derive $T^R_{I \bar JK\bar L}$ and $T^R_{I J\bar K\bar L}$ for the irreps $R$ that appear in $N\otimes \bar N$ and $N\otimes N$, respectively, up to overall normalization factors $n_R$.
\item Compute the 4-point function $\langle \phi\phi\bar\phi\bar\phi\rangle$ for a free theory in 3d with $\Delta_\phi=\frac12$ using Wick contractions. Expand in conformal blocks in the $\cR\times \bar \cR$ channel \eqref{4pointAgainGlobe} and the $\cR\times \cR$ channel \eqref{4pointAgainGlobe2}, and check that your choice of $n_R$ leads to OPE coefficients squared that are positive. Check that operators with irreps in the symmetric/antisymmetric product of $R\times R$ come only with even/odd spins in the block expansion.
\item Compute the coefficient $c_J$ of the 2-point function of canonically normalized gobal symmetry currents $J^a_\mu$, and check that it matches $\lambda^2_{\phi\bar\phi,\cO_{2,1,A}}$ for the unit normalized current $\cO_{2,1,A}\equiv \frac{S_d}{\sqrt{c_J}}J$ using the relation \eqref{cJtoOPE}.
\item Compute the crossing equations, and check that they are approximately solved in the free theory using the OPE coefficients you computed in the previous step and the block expansion up to a sufficiently higher order in $\Delta$ and $\ell$.

\end{enumerate}

\item Change \texttt{Bootstrap.m} to apply to a 4-point function of scalars in the vector irrep of $O(N)$. For $N=3$ and $\Lambda=11$, apply the scaling dimension upper bound algorithm to check a grid of points in $(\Delta_V,\Delta_S)$ and $(\Delta_V,\Delta_T)$ to confirm the kinks shown in \ref{ONs} and \ref{ONt}.
\end{enumerate}

\pagebreak

\section{Bootstrapping mixed correlators}
\label{mix}

In the previous lectures we considered 4-point functions of identical scalar operators. We will now discuss following \cite{Kos:2014bka} how to bootstrap systems of 4-point functions with different scalar operators. The main advantage of mixed system is that it gives access to operators that would not appear in a 4-point function of identical scalar operators. For simplicity, we will restrict to real scalars.

\subsection{Four-point function of mixed scalars}
\label{mixScalar}

Consider real scalar operators $\phi^I_i(x)$, were $i$ are labels for different scalars that transform in different irreps $\cR_i$ of the global symmetry group $G$, whose indices we suppress for simplicity. As shown in a previous problem set, conformal symmetry fixes the 4-point function to be
 \es{4pointMixed}{
\langle \phi_i(x_1)\phi_j(x_2)\phi_k(x_3)\phi_l(x_4)\rangle=g_{ijkl}(u,v)\frac{\left[\frac{x_{24}}{x_{14}}\right]^{{\Delta_{ij}}}  \left[\frac{x_{14}}{x_{13}}\right]^{{\Delta_{kl}}} }{x_{12}^{\Delta_i+\Delta_j}x_{34}^{\Delta_k+\Delta_l}}\,,
}
where $\Delta_{ij}\equiv \Delta_i-\Delta_j$. We can expand $g_{ijkl}(u,v)$ in conformal blocks just as we did in the single correlator case, by taking the OPE in the $12$ and $34$ channels, which yields:
\es{4pointOPEMixed}{
&g_{ijkl}(u,v)= \sum_{\Delta,\ell,R}\lambda_{\phi_i\phi_j\cO_{\Delta,\ell,R}}\lambda_{\phi_k\phi_l\cO_{\Delta,\ell,R}}T^R_{ij,kl}g^{\Delta_{ij},\Delta_{kl}}_{\Delta,\ell}(u,v)\,,
}
where $g^{\Delta_{ij},\Delta_{kl}}_{\Delta,\ell}(u,v)$ are the conformal blocks, and $T^R_{ij,kl}$ are tensor structures for the irreps $R$ that appear in the intersection of the tensor products $\cR_i\times\cR_j$ and $\cR_k\times\cR_l$. The symmetry of the OPE under exchanging $1\leftrightarrow2$ and $3\leftrightarrow4$ implies that the blocks satisfy
\es{blockId2}{
g_{\Delta,\ell}^{\Delta_{ij},\Delta_{kl}}(u/v,1/v)=(-1)^\ell v^{\frac{\Delta_{kl}}{2}}g_{\Delta,\ell}^{-\Delta_{ij},\Delta_{kl}}(u,v)=(-1)^\ell v^{\frac{-\Delta_{ij}}{2}}g_{\Delta,\ell}^{\Delta_{ij},-\Delta_{kl}}(u,v)\,,
}
which generalizes the non-mixed block identity \eqref{blockId}.

The $x$-dependent prefactor in \eqref{4pointMixed} is more complicated than the single correlator case, so passing this prefactor through the conformal Casimir as in section \ref{conCas} yields a more complicated differential equation for the blocks \cite{Dolan:2003hv}:
\es{ConfCasFinalMix}{
\cD_{\Delta_{ij},\Delta_{kl}} g^{\Delta_{ij},\Delta_{kl}}_{\Delta,\ell}(u,v)&=c_{\Delta,\ell}g^{\Delta_{ij},\Delta_{kl}}_{\Delta,\ell}(u,v)\,,\\
\cD_{\Delta_{ij},\Delta_{kl}}&\equiv(1-u-v)\partial_v\left(v\partial_v+\frac{\Delta_{kl}-\Delta_{ij}}{2}\right)+u\partial_u(2u\partial_u-d)\\
&\qquad-(1+u-v)\left(u\partial_u+v\partial_v-\frac{\Delta_{ij}}{2}\right)\left(u\partial_u+v\partial_v+\frac{\Delta_{kl}}{2}\right)\,,
}
where $c_{\Delta,\ell}$ is the same eigenvalue \eqref{conf} that appeared in the single correlator differential equation \eqref{ConfCasFinal}. Note that the external operator scaling dimensions now appear explicitly in the differential equation, unlike the single correlator case, so the blocks must be computed separately for each $\Delta_{ij}$ and $\Delta_{kl}$. In the $u\to0$ and $v\to1$ limits, however, we can check from the OPE that the blocks have the same limit \eqref{blockAss} as in the case $\Delta_{ij}=\Delta_{kl}=0$.

In even dimensions, we can solve \eqref{ConfCasFinalMix} exactly using the initial condition \eqref{blockAss}. For instance, in $d=2,4$ we can generalize the single correlator blocks \eqref{2and4} by replacing \cite{Dolan:2003hv}
\es{2and4Mix}{
k_\beta(x)\to k^{\Delta_{ij},\Delta_{kl}}_\beta(x)=x^{\frac\beta2}{}_2F_1\left(\frac{\beta-\Delta_{ij}}{2},\frac{\beta+\Delta_{kl}}{2},\beta,x\right)\,.
}
In non-even dimensions, the conformal blocks are computed most efficiently in a radial expansion in $r,\eta$ using the Zamolodchikov recursion relations \cite{Kos:2014bka}. The computation is exactly the same as the single correlator case in section \ref{Zam}, except the table of poles in \ref{tab1} is generalized to \ref{tab2}, we replace $\tilde h_{\ell}(r,\eta)$ in \eqref{leadh} by
\es{leadhMixed}{
\tilde h^{\Delta_{ij},\Delta_{kl}}_{\ell}(r,\eta) =\frac{4^\Delta\ell!C_\ell^{\frac{d-2}{2}}(\eta)}{(-2)^\ell\left(\frac{d-2}{2}\right)_\ell(1-r^2)^{\frac{d-2}{2}}(1+r^2+2r\eta)^{\frac12(1+\Delta_{ij}-\Delta_{kl})}(1+r^2-2r\eta)^{\frac12(1-\Delta_{ij}+\Delta_{kl})}}\,,
}
and the coefficients $c_{I,m}$ in \eqref{mixedBlockCoeff} by
 \es{mixedBlockCoeffMixed}{
&c_{m,1}=-\frac{m(-2)^m}{(m!)^2}\left(\frac{\Delta_{ij}+1-m}{2}\right)_m\left(\frac{\Delta_{kl}+1-m}{2}\right)_m\,,\\
&c_{m,2}=-\frac{m\ell!  (d/2+\ell)_{-m}(d/2+\ell-1)_{-m} }{(-2)^m(m!)^2(\ell-m)!(d-2+\ell)_{-m}}\left(\frac{\Delta_{ij}+1-m}{2}\right)_m\left(\frac{\Delta_{kl}+1-m}{2}\right)_m\,,\\
&c_{m,3}=\frac{-m(-1)^{m/2}((d-m)/2-1)_{m}}{2((m/2)!)^2((d-m)/2+\ell-1)_{m}((d-m)/2+\ell)_{m}} \left(\frac{\Delta_{ij}-(d+m)/2-\ell+2}{2}\right)_{\frac m2}\\
&\times\left(\frac{\Delta_{ij}+(d-m)/2+\ell}{2}\right)_{\frac m2}\left(\frac{\Delta_{kl}-(d+m)/2-\ell+2}{2}\right)_{\frac m2}\left(\frac{\Delta_{kl}+(d-m)/2+\ell}{2}\right)_{\frac m2}\,.\\
}

\begin{table}
\begin{center}
\begin{tabular}{c|c|c|c}
 $I$& $m\in\mathbb{B}_I$&  $\Delta_{I,m}$ & $\ell_{I,m}$  \\
 \hline 
1&  $1,2,3,\dots$&  $1-\ell-m$ & $\ell+m$   \\
 \hline
 2& $1,2,\dots \ell$  & $\ell+d-1-m$ & $\ell-m$   \\
  \hline
 3& $2,4,6,\dots$  & $(d-m)/2$ & $\ell$   \\
\end{tabular}
\caption{Positions of the poles $\Delta_{I,m}$ in $g_{\Delta,\ell}$, shifts of the residue conformal blocks $g^{\Delta_{ij},\Delta_{kl}}_{\Delta_{I,m}+m,\ell_{I,m}}$, and range $\mathbb{B}_{I}$ of possible values of $m$ for each $I=1,2,3$.}
\label{tab2}
\end{center}
\end{table}

\subsection{Crossing equations for mixed scalars}
\label{crossMix}

We can now derive the crossing equation by swapping $(1,i)\leftrightarrow(3,k)$ in \eqref{4pointMixed}, which yields\footnote{We can derive all the crossing constraints by permuting the $ijkl$ and then swapping $(1,i)\leftrightarrow(3,k)$, so we do not need to consider swapping $(1,i)\leftrightarrow(2,j)$.}
\es{crossing2Mixed}{
&\sum_{\Delta,\ell,R}\left[\lambda_{ij\cO_{\Delta,\ell,R}} \lambda_{kl\cO_{\Delta,\ell,R}}T^R_{ij,kl}v^{\frac{\Delta_j+\Delta_k}{2}}g_{\Delta,\ell}^{\Delta_{ij},\Delta_{kl}}(u,v)\right]\\
&\qquad\qquad=\sum_{\Delta,\ell,R'}\left[\lambda_{kj\cO_{\Delta,\ell,R'}} \lambda_{il\cO_{\Delta,\ell,R'}}T^{R'}_{kj,il}u^{\frac{\Delta_i+\Delta_j}{2}}g_{\Delta,\ell}^{\Delta_{kj},\Delta_{il}}(v,u)\right]\,,
}
where $R$ includes irreps that appear in $\cR_i\times\cR_j\cap\cR_k\times\cR_l$, while $R'$ includes irreps that appear in $\cR_k\times\cR_j\cap\cR_i\times\cR_l$. We can then add and subtract \eqref{crossing2Mixed} with its $u\leftrightarrow v$ version, like what we did in the global symmetry case, to write the crossing equations as
\es{Fmixed}{
\sum_{\Delta,\ell,R}\left[\lambda_{ij\cO_{\Delta,\ell,R}} \lambda_{kl\cO_{\Delta,\ell,R}} T^R_{ij,kl} F_{\mp,\Delta,\ell}^{ij,kl}(u,v) \right]\pm\sum_{\Delta,\ell,R'}\left[\lambda_{kj\cO_{\Delta,\ell,R'}} \lambda_{il\cO_{\Delta,\ell,R'}} T^{R'}_{kj,il}  F_{\mp,\Delta,\ell}^{kj,il}(u,v)\right]=0\,,
}
where we define the crossing functions
\es{F3}{
F^{ij,kl}_{\pm,\Delta,\ell}(u,v)\equiv v^{\frac{\Delta_k+\Delta_j}{2}}g^{\Delta_{ij},\Delta_{kl}}_{\Delta,\ell}(u,v)\pm u^{\frac{\Delta_k+\Delta_j}{2}}g^{\Delta_{ij},\Delta_{kl}}_{\Delta,\ell}(v,u)\,,
}
which are respectively symmetric and antisymmetric under swapping $u\leftrightarrow v$, and generalize the $F^{\Delta_\phi}_{\pm,\Delta,\ell}(u,V)$ defined in \eqref{F22} for the single correlator case. The $F^{ij,kl}_{\pm,\Delta,\ell}(u,v)$ satisfy the useful identities 
\es{Fidentity}{
F^{ij,kl}_{\pm,\Delta,\ell}(u,v)=F^{kl,ij}_{\pm,\Delta,\ell}(u,v)=F^{ji,lk}_{\pm,\Delta,\ell}(u,v)\,,
}
which follow from the block identities \eqref{blockId2}. These identities will reduce the number of crossing equations that would naively result from \eqref{Fmixed}.

\subsection{Matrix constraints}
\label{matrixCon}

The major difference between the mixed scalar crossing equations \eqref{Fmixed} and previous cases, is that the OPE coefficients do not appear as squares, and so while we can choose a basis such that each OPE coefficient is real just as before, there is no guarantee that $\lambda_{ij\cO_{\Delta,\ell,R}} \lambda_{kl\cO_{\Delta,\ell,R}} $ or $\lambda_{kj\cO_{\Delta,\ell,R'}} \lambda_{il\cO_{\Delta,\ell,R'}} $ will be positive, which was an important part of our bootstrap algorithm. Recall from the lecture \ref{semi} on semidefinite programming that \texttt{SDPB} can consider constraints of positive semidefinite {\it symmetric matrices} of polynomial, and not just constraints on individual polynomials. If we thus consider all 4-point functions with the scalars $\phi_{i}(x)$, $\phi_{j}(x)$, $\phi_{k}(x)$, $\phi_{l}(x)$, then we can derive a list of matrix constraints, where vectors of OPE coefficients multiply symmetric matrices of polynomials.

For instance, consider a CFT with $\mathbb{Z}_2$ symmetry, where the lowest dimension $\mathbb{Z}_2$ even operator is $\epsilon$ and the lowest dimension $\mathbb{Z}_2$ odd operator is $\sigma$. Only $\mathbb{Z}_2$ even/odd operators $\cO_{\Delta,\ell,\pm}$ can appear in the OPE of two operators with the same/different $\mathbb{Z}_2$ parity. The nonzero correlators between $\sigma$ and $\epsilon$ are then
\es{correlators}{
\langle \sigma\sigma\sigma\sigma\rangle\,,\qquad \langle \epsilon\epsilon\sigma\sigma\rangle\,,\qquad \langle \epsilon\epsilon\epsilon\epsilon\rangle\,.
}
We can apply the crossing equations \eqref{Fmixed} to each inequivalent permutation of these 4-point functions to get
\es{FmixedEx}{
0=&\sum_{\Delta,\ell^+}\lambda_{\sigma\sigma\cO_{\Delta,\ell,+}}^2F^{\sigma\sigma,\sigma\sigma}_{-,\Delta,\ell}(u,v)\,,\\
0=&\sum_{\Delta,\ell^+}\lambda_{\epsilon\epsilon\cO_{\Delta,\ell,+}}^2F^{\epsilon\epsilon,\epsilon\epsilon}_{-,\Delta,\ell}(u,v)\,,\\
0=&\sum_{\Delta,\ell}\lambda_{\sigma\epsilon\cO_{\Delta,\ell,-}}^2F^{\sigma\epsilon,\sigma\epsilon}_{-,\Delta,\ell}(u,v)\,,\\
0=&\sum_{\Delta,\ell^+}\lambda_{\sigma\sigma\cO_{\Delta,\ell,+}}\lambda_{\epsilon\epsilon\cO_{\Delta,\ell,+}}F^{\sigma\sigma,\epsilon\epsilon}_{\mp,\Delta,\ell}(u,v)\pm \sum_{\Delta,\ell}(-1)^\ell\lambda_{\sigma\epsilon\cO_{\Delta,\ell,-}}^2F^{\epsilon\sigma,\sigma\epsilon}_{\mp,\Delta,\ell}(u,v)\,,
}
where in the last equations we used $\lambda_{\epsilon\sigma\cO_{\Delta,\ell,-}}=(-1)^\ell \lambda_{\sigma\epsilon\cO_{\Delta,\ell,-}}$, and recall that $\ell^\pm$ denotes that only even/odd spins occur. We can then assemble these constraints into a vector of crossing equations
\es{Vmix}{
\sum_{\Delta,\ell}\begin{pmatrix}\lambda_{\sigma\sigma\cO_{\Delta,\ell,+}}&\lambda_{\epsilon\epsilon\cO_{\Delta,\ell,+}}\end{pmatrix} \vec V_{\Delta,\ell,+} \begin{pmatrix}\lambda_{\sigma\sigma\cO_{\Delta,\ell,+}}\\\lambda_{\epsilon\epsilon\cO_{\Delta,\ell,+}}\end{pmatrix}+\sum_{\Delta,\ell}\lambda^2_{\sigma\epsilon\cO_{\Delta,\ell,-}}\vec V_{\Delta,\ell,-}=\vec 0\,,
}
where $\vec V_{\Delta,\ell,-}$ is a vector of scalar constraints and $\vec V_{\Delta,\ell,+}$ is a vector of matrix constraints:
\es{Vmix2}{
\vec V_{\Delta,\ell,-}=\begin{pmatrix} 0 \\ 0 \\ F^{\sigma\epsilon,\sigma\epsilon}_{-,\Delta,\ell} \\ (-1)^\ell  F^{\epsilon\sigma,\sigma\epsilon}_{-,\Delta,\ell} \\ -(-1)^\ell F^{\epsilon\sigma,\sigma\epsilon}_{+,\Delta,\ell}\end{pmatrix}\,,\qquad \vec V_{\Delta,\ell,+}=\begin{pmatrix} 
\begin{pmatrix} F_{-,\Delta,\ell}^{\sigma\sigma,\sigma\sigma}&0\\0&0\end{pmatrix} 
\\ 
 \begin{pmatrix} 0&0\\0&F_{-,\Delta,\ell}^{\epsilon\epsilon,\epsilon\epsilon}\end{pmatrix} 
\\
 \begin{pmatrix} 0&0\\0&0\end{pmatrix} 
\\
  \begin{pmatrix} 0&\frac12F_{-,\Delta,\ell}^{\sigma\sigma,\epsilon\epsilon}\\\frac12F_{-,\Delta,\ell}^{\sigma\sigma,\epsilon\epsilon}&0\end{pmatrix} 
\\
    \begin{pmatrix} 0&\frac12F_{+,\Delta,\ell}^{\sigma\sigma,\epsilon\epsilon}\\\frac12F_{+,\Delta,\ell}^{\sigma\sigma,\epsilon\epsilon}&0\end{pmatrix} 
\end{pmatrix}\,,
}
where note that each $2\times2$ matrix in $\vec V_{\Delta,\ell,+}$ is symmetric.

\subsection{Implementing with \texttt{SDPB}}
\label{SDPBmixed}

As in the previous lecture, we can now truncate each $F^{ijkl}_{\pm,\Delta,\ell}(u,v)$ in $ V^\beta_{\Delta,\ell,R}$ in terms of $m,n$ derivatives in $z,\bar z$ at the crossing symmetric point, and write them as polynomials $P_{\ell,R}^{m,n,\beta}(\Delta,\Delta_{ij},\Delta_{kl})$ in $\Delta$ with positive factors $\chi_\kappa^\ell(\Delta)$, where $\beta$ denotes which crossing equation, and we must now compute a different polynomial for each $\Delta_{ij}$ and $\Delta_{kl}$ since the mixed blocks differ in each case. This makes bootstrapping different points much more computationally expensive than the single correlator case, where we only needed to compute a single block, and then the dependence on the external operator scaling dimension $\Delta_\phi$ was just a simple factor \eqref{Ftop}. The only other change to the bootstrap algorithm in section \ref{SDPBGlobe} is that the functional $\alpha_{m,n}^\beta{}$ now acts on matrices of polynomials, but recall that SDPB was set up precisely to handle such more general constraints.

We can implement these changes in the wrapper \texttt{Bootstrap.m} used to create the \texttt{SDPB} input, by entering the \texttt{PositiveMatrixWithPrefactor} \eqref{posMwithP} as matrices of polynomials when $\vec V_{\Delta,\ell,R}$ is a matrix. The objective function ${\it obj}$ and normalization ${\it norm}$ cannot be matrices, so we must impose relations between OPE coefficients to define these quantities. For instance, in the scaling dimension algorithm ${\it norm}$ is the identity operator. In the $\mathbb{Z}_2$ example above, the identity appears appears as the matrix constraint $\vec V_{0,0,+}$ in \eqref{Vmix2}. Since we know that all OPE coefficients with the identity, such as $\lambda_{\sigma\sigma\cO_{0,0,+}}$ and $\lambda_{\epsilon\epsilon\cO_{0,0,+}}$ in \eqref{Vmix}, are simply 1, we can define the scalar constraint 
\es{identityMat}{
\vec V^\text{scal}_{0,0,+}\equiv \begin{pmatrix} 1&1\end{pmatrix}\vec V_{0,0,+}\begin{pmatrix}1\\1\end{pmatrix}\,,
}
and then enter $\vec V^\text{scal}_{0,0,+}$ into ${\it norm}$. In the OPE coefficient algorithm, ${\it norm}$ is an operator whose OPE coefficient squared we want to maximize, so in general we cannot apply this algorithm to operators that appear in matrix constraints, such as $\cO_{\Delta,\ell,+}$ in \eqref{Vmix}, but we could apply it to operators that happen to have scalar constraints, such as $\cO_{\Delta,\ell,-}$ in \eqref{Vmix}. We can also apply it to operators with matrix constraints whose OPE coefficients are related, such as the stress tensor $\cO_{d,2,+}$ in \eqref{Vmix} whose OPE coefficients with any external operator are all related to the single quantity $c_T^{-1}$ by \eqref{cTtoOPE}, so we can define the scalar constraint 
\es{cTMat}{
\vec V^\text{scal}_{d,2,+}\equiv \frac{d^2}{c_T(d-1)^2}\begin{pmatrix} \Delta_\sigma&\Delta_\epsilon\end{pmatrix}\vec V_{d,2,+}\begin{pmatrix}\Delta_\sigma\\\Delta_\epsilon\end{pmatrix}\,,
}
and then maximize $c_T^{-1}$ using the OPE coefficient maximization algorithm.

\subsection{Gap assumptions}
\label{gapMix}

Without any further assumptions, the constraints described above are the best we can do. If we furthermore assume gaps, then mixed correlators allow for a richer set of constraints than single correlators. For example, if we assume a gap $\Delta^B_{2,+}>d$ and then insert the stress tensor contribution as the scalar constraint in \eqref{cTMat}, then the fact that multiple OPE coefficients are related to the same $c_T$ is a constraint that we could not access with single correlators. Unfortunately, there is no physical motivation for such a gap.

A better motivated gap is to assume that only a certain number of operators are relevant. Recall that for the Ising CFT, we know that $\sigma$ and $\epsilon$ are the only relevant operators in their respective $\mathbb{Z}_2$ channels. With single correlators, we were already able to impose a gap on the $\mathbb{Z}_2$ even sector, which turned our upper bounds into a narrow peninsula of an allowed region. With mixed correlators, we now also have access to to the $\mathbb{Z}_2$ odd sector, so we can also impose that $\sigma$ is the only relevant $\mathbb{Z}_2$ odd operator. 

Once we assume these gaps, we can impose the further constraint that the OPE coefficient $\lambda_{\sigma\sigma\epsilon}$ is invariant under permutations of its labels, since the spin is even. This allows us to combine the matrix constraint $\vec V_{\Delta_\epsilon,0,+}$ and the scalar constraint $\vec V_{\Delta_\sigma,0,-}$ into a single matrix constraint as
\es{bigCon}{
\begin{pmatrix}\lambda_{\sigma\sigma\epsilon}&\lambda_{\epsilon\epsilon\epsilon}\end{pmatrix} \vec V_{\Delta_\epsilon,0,+} &\begin{pmatrix}\lambda_{\sigma\sigma\epsilon}\\\lambda_{\epsilon\epsilon\epsilon}\end{pmatrix}+\lambda^2_{\sigma\epsilon\sigma}\vec V_{\Delta_\sigma,0,-}\\
&=\begin{pmatrix}\lambda_{\sigma\sigma\epsilon}&\lambda_{\epsilon\epsilon\epsilon}\end{pmatrix} \left[\vec V_{\Delta_\epsilon,0,+}+ \vec V_{\Delta_\sigma,0,-}\otimes \begin{pmatrix}1& 0 \\ 0 & 0 \end{pmatrix} \right]\begin{pmatrix}\lambda_{\sigma\sigma\epsilon}\\\lambda_{\epsilon\epsilon\epsilon}\end{pmatrix}\,.
}
This single constraint still allows for the possibility that there are many operators with dimensions $\Delta_\sigma$ and $\Delta_\epsilon$, with different OPE coefficients. In the Ising CFT we expect that there is only a single operator $\sigma$ and $\epsilon$ with their dimensions, which we can impose by replacing the $2\times2$ matrix constraint in \eqref{bigCon} by the scalar constraint
\es{bigCon2}{
(\lambda^2_{\sigma\sigma\epsilon}+\lambda^2_{\epsilon\epsilon\epsilon})\begin{pmatrix}\cos\theta &\sin\theta \end{pmatrix} \left[\vec V_{\Delta_\epsilon,0,+}+ \vec V_{\Delta_\sigma,0,-}\otimes \begin{pmatrix}1& 0 \\ 0 & 0 \end{pmatrix} \right]\begin{pmatrix} \cos\theta &  \sin\theta \end{pmatrix}\,,
}
where $\theta=\tan^{-1}(\lambda_{\epsilon\epsilon\epsilon}/\lambda_{\sigma\sigma\epsilon})\in[0,\pi)$ must be entered by hand. We can then scan over allowed points $(\Delta_\sigma,\Delta_\epsilon,\theta)_\text{allowed}$ that the bootstrap gives for each $\theta$, and the union of such points for all $\theta\in[0,\pi)$ can be smaller than the allowed region $(\Delta_\sigma,\Delta_\epsilon)_\text{allowed}$ that we would have found without the constraint \eqref{bigCon2}.

\subsection{Mixed correlator bootstrap bounds with $\mathbb{Z}_2$ symmetry}
\label{islands}

\begin{figure}[t!]
\begin{center}
   \includegraphics[width=0.49\textwidth]{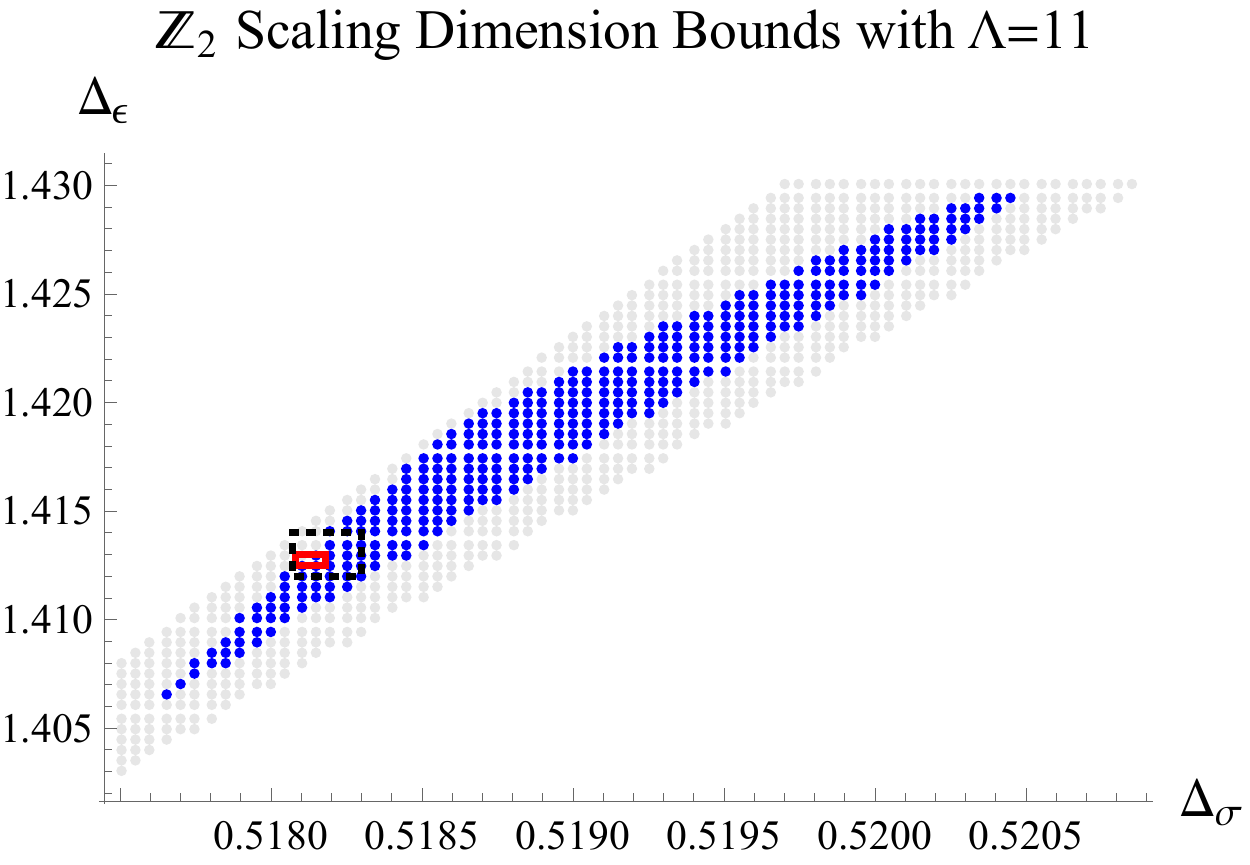} \includegraphics[width=0.49\textwidth]{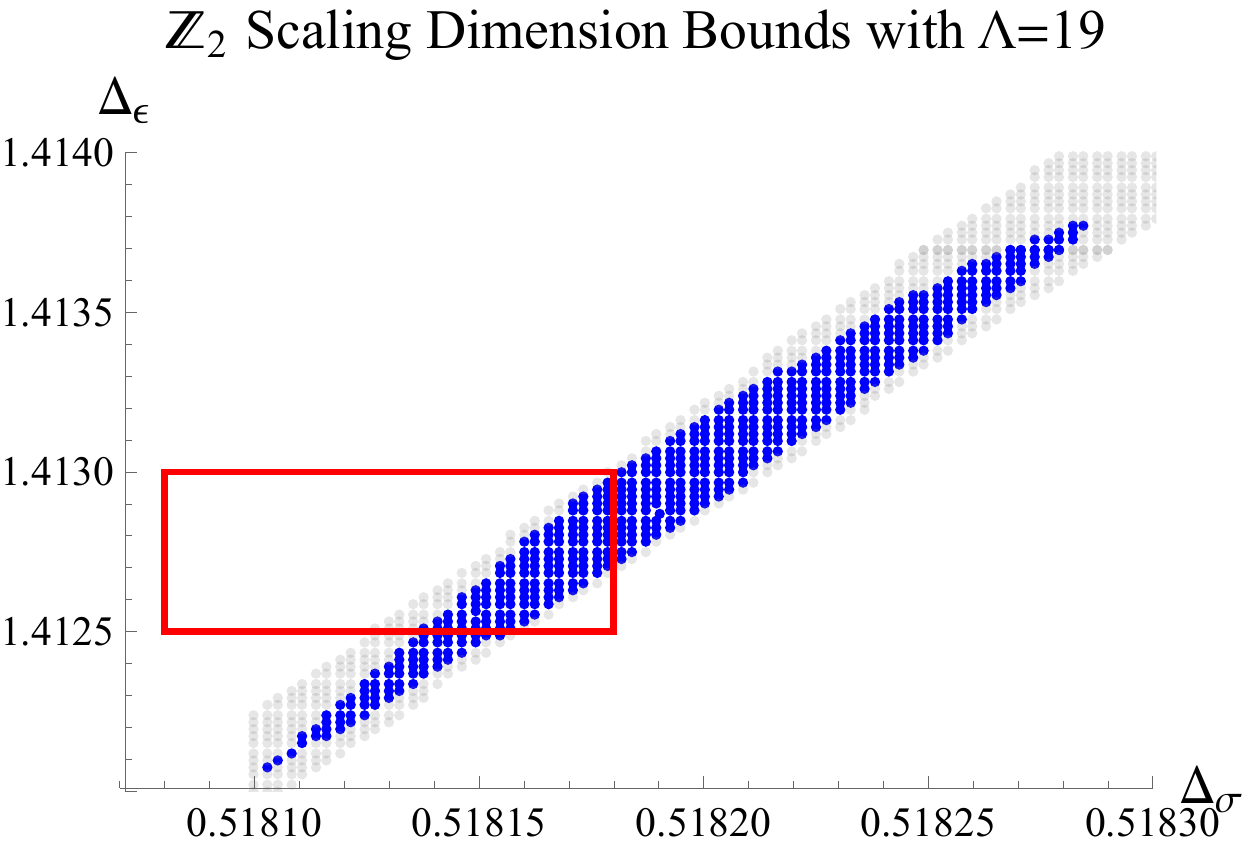}
 \caption{ 
 Blue points are values of the scaling dimensions $(\Delta_\sigma,\Delta_\epsilon)$ of the lowest dimension $\mathbb{Z}_2$ odd/even operator, respectively, allowed by the mixed correlator bootstrap with the gap assumptions that these are the only relevant operators in their sectors, while gray points are disallowed. The red rectangle is the Monte Carlo values in Table \ref{bootWin} for the Ising CFT including its error. The lefthand plot was run with $\Lambda=11$ number of derivatives, while righthand plot was run with $\Lambda=19$, and corresponds to the region in the black rectangle in the lefthand plot.
 }
\label{mixScal}
\end{center}
\end{figure}  

We now discuss the results of running the bootstrap for the $\mathbb{Z}_2$ mixed correlator system described above in 3d, which were originally reported in \cite{Kos:2014bka}.\footnote{This analysis was generalized to the $O(N)$ model in \cite{Kos:2015mba,Chester:2019ifh,Chester:2020iyt}.} If we do not assume any gaps, or only a gap above $\Delta_\epsilon$, then the results are almost identical to what we found with the single correlator bounds in Figure \ref{isingScal}. The big improvement comes from assuming a gap up to 3 above both $\Delta_\sigma$ and $\Delta_\epsilon$, in which case we find a small island around the Ising CFT as shown in Figure \ref{mixScal}. We also imposed permutation symmetry of $\lambda_{\sigma\sigma\epsilon}$ as in \eqref{bigCon}, which also shrinks the allowed region further. In the $\Lambda=11$ plot on the left, the Monte Carlo values are still much more constraining than bootstrap, but for the $\Lambda=19$ plot on the right, the bootstrap has already started to disallow some of the Monte Carlo region. The two allowed regions take roughly the same shape for $\Lambda=11$ and $\Lambda=19$, which is a general pattern once $\Lambda$ is large enough. Note that there will still exist allowed regions to the far left and above the island that we found, that are not shown in this plot.

When this same bootstrap setup is run with $\Lambda=43$ \cite{Simmons-Duffin:2015qma}, then they become much smaller than the Monte Carlo region. Even smaller bounds can be found by imposing the further constraint \eqref{bigCon2}, which drastically shrinks the allowed region. This was used in \cite{Kos:2016ysd} to find the best Ising CFT bounds to date, which as shown before in Table \ref{bootWin} are more constraining than Monte Carlo by orders of magnitude. This study involves a more computationally expensive three-dimensional scan over $\Delta_\sigma$ $\Delta_\epsilon$, and $\theta$, though, and so must be run on a computer cluster instead of just a single personal computer. 

\subsection{Conclusion}
\label{conc}

In these lectures, we demonstrated how symmetry alone, without any explicit Lagrangians or assumptions about the spectrum, can be used to derive infinite constraints on the physical content of a conformal field theory. We then truncated these equations in a controlled way that allowed us to derive rigorous bounds on the scaling dimensions and OPE coefficients of the lowest dimension operator in a given representation that appears in a given 4-point function. We saw that increasing the precision of our truncation allowed us to monotonically improve our bounds, and that kinks in the bounds corresponded to the known Ising and $O(N)$ vector model CFTs. Finally, when we imposed that only two operators are relevant in the Ising CFT, which is part of its definition, we were able to derive small islands in the space of allowed scaling dimensions of these operators, which give a more precise theoretical prediction for these values than any other method.

While these results are impressive, we have still only studied a few of the simplest conformal field theories. Many known CFTs, particularly gauge theories without supersymmetry, remain mostly unexplored, and there are probably many unknown CFTs waiting to be discovered, especially in $d>4$ where weakly coupled Lagrangians do not exist in the UV and so traditional methods fall short. There are also other constraints on 4-points functions that we have not imposed, such as modular invariance for 2d CFTs \cite{Collier:2016cls}. Even for the crossing constraints discussed in these lectures, there are many area that could be improved:
\begin{itemize}
\item Higher spin operators: We only studied operators with small spins. In fact, we were only able to truncate the infinite crossing equations to a finite numerically feasible problem since we showed that the contributions of high dimension operators, and thus also high spin operators, to the 4-point function are very small near the crossing symmetric point. To access these higher spin operators, we either need to drastically improve the precision of our numerics, or to further develop methods to bootstrap large spin data from small spin data such as \cite{Komargodski:2012ek,Alday:2015ewa,Fitzpatrick:2012yx,Simmons-Duffin:2016wlq,Albayrak:2019gnz}.
\item Higher dimension operators: Our bootstrap algorithms only applied to the lowest dimension operators with a given spin and global symmetry representation. We could have computed upper bounds on higher dimension operators by inserting all lower dimension operators, and seeing for what value of the higher dimension operator no values of the lower dimension operators are allowed. A quicker method is to use the fact that at the boundary of the allowed region of a bootstrap bound, we expect the solution to crossing given by the functional $\alpha$ to be unique. We can then read off the scaling dimensions of higher dimension operators by looking at the zeros of $\alpha$ acting on the crossing equations, and then inputting these zeros to solve for the OPE coefficients \cite{Poland:2010wg,El-Showk:2014dwa,ElShowk:2012hu,El-Showk:2016mxr}. If we think that a known theory saturates the bootstrap bounds, such as the Ising model that seemed to saturate the most general $\mathbb{Z}_2$ bounds, then this so-called extremal functional method can be used to study higher dimension operators. In the case of the 3d Ising model and ABJM theory \cite{Aharony:2008ug}, results for higher dimension operators as extracted from this method have been shown to match predictions from other methods \cite{Simmons-Duffin:2016wlq,Agmon:2017xes,Chester:2018lbz}. The downside of this method is that unlike the original bounds, this extraction is not rigorous and there is no known way yet of systematically improving the extraction.
\item Operators in other OPEs: We only studied CFT data that appears in 4-point functions of scalar operators in specific representations of the global symmetry group. Even for the mixed correlator study of the Ising model, we did not have access to CFT data such as OPE coefficients with two operators with nonzero spin. To access all the local CFT data with the numerical conformal bootstrap, we would need to study every possible 4-point function in the theory, including 4-point functions with spin \cite{Dymarsky:2017xzb,Iliesiu:2015qra,Iliesiu:2017nrv,Dymarsky:2017yzx,Karateev:2019pvw}. An alternative approach would be to look at $n$-point functions of the same operators for $n>4$, but it is not clear how to apply the bootstrap algorithm in that case, since more than two OPE coefficients would appear.
\item Gap assumptions: For the Ising CFT, the gaps we imposed were essential to deriving islands in $(\Delta_\sigma,\Delta_\epsilon)$. If we had looked at other operators in the OPE, such as spin 2 operators, we would not have any natural gap assumptions, and so we would not have been able to derive islands in their scaling dimensions. For most strongly coupled CFTs, we do not even know how many relevant operators there should be, so no gap assumption is justified. For CFTs with enough supersymmetry, however, certain operators have protected scaling dimensions, so we can compute both upper and lower bounds on their OPE coefficients \cite{Poland:2011ey}, and thereby derive small islands in OPE space \cite{Agmon:2017xes,Chester:2014fya,Chester:2014mea,Binder:2020ckj}.
\end{itemize}

The conformal bootstrap program disdains to conceal its views and aims. It openly declares that its ends can be attained only by numerical solutions to all existing CFTs. Let perturbative methods tremble at a bootstrap revolution. Physicists have nothing to lose but their Lagrangians. They have a non-perturbative world to win. 

\begin{center}
Working field theorists of all disciplines, bootstrap!
\end{center}

\pagebreak

\subsection{Problem Set 7}
\label{hw7}

The goal of this problem set is to reproduce the bootstrap islands in $(\Delta_\sigma,\Delta_\epsilon)$ space for the 3d Ising model as computed with $\Lambda=11$ in Figure \ref{mixScal}. Since the crossing equations \eqref{bigCon2} include crossing functions $F_{\pm,\Delta,\ell}^{\epsilon\sigma,\sigma,\epsilon}$ and $F_{\pm,\Delta,\ell}^{\sigma\epsilon,\sigma,\epsilon}$ that are constructed from mixed blocks $g_{\Delta,\ell}^{\Delta_{\epsilon\sigma},\Delta_{\sigma\epsilon}}$ and $g_{\Delta,\ell}^{\Delta_{\sigma\epsilon},\Delta_{\sigma\epsilon}}$, respectively, we must compute a different block for each $(\Delta_\sigma,\Delta_\epsilon)$, unlike the single correlator case where a single block sufficed for each external operator scaling dimension. 

\begin{enumerate}

\item Compute mixed conformal blocks $g_{\Delta,\ell}^{\Delta_{ij},\Delta_{kl}}(r,\eta)$ for $\ell=0$, $\Delta_{ij}=\frac12$, and $\Delta_{kl}=-\frac12$ up to order $r_\text{max}=20$ using the recursion relation in Section \ref{mixScalar}.  
\item Replace poles as in Section \ref{polyCross} applied to mixed blocks with $\kappa=10$, and plot $\bar p^{\Delta_{ij},\Delta_{kl}}_{\Delta,\ell}(r_c,1)$ versus $ p^{\Delta_{ij,\Delta_{kl}}}_{\Delta,\ell}(r_c,1)$ for $1\leq\Delta\leq4$ to see how accurate the replacement is at the crossing symmetric point.
\item Make a table of derivatives 
\es{Q2}{
Q^{m,n}_\ell(\Delta,\Delta_{ij},\Delta_{kl})\equiv\left(r^{-\Delta}\partial^m_z\partial^n_{\bar z}\left[r^\Delta\bar p^{\Delta_{ij},\Delta_{kl}}_{\Delta,\ell}(r,\eta) \right]\right)\big\vert_{r=r_c,\eta=1}\,,
}
 for $m+n\leq3$, $\ell=0$, $\Delta_{ij}=\frac12$, and $\Delta_{kl}=-\frac12$. Compare to the values computed using the \texttt{scalar\_blocks} package on \texttt{Docker}, which outputs $\frac{(-2)^\ell(d/2-1)_\ell}{(d-2)_\ell}Q^{m,n}_\ell(x=\Delta-\ell-1,\Delta_{ij},\Delta_{kl})$, as described in Problem Set \ref{hw4}.
\item Change \texttt{Bootstrap.m} to apply to a mixed system of 4-point functions with $\mathbb{Z}_2$ symmetry as described in this lecture. Impose that $\sigma$ and $\epsilon$ are the only relevant operators in their respective sectors, and that $\lambda_{\sigma\sigma\epsilon}$ is permutation invariant as in \eqref{bigCon}. Reproduce the $\Lambda=11$ island in $(\Delta_\sigma,\Delta_\epsilon)$ shown in Figure \ref{mixScal} by checking an appropriate grid of a few points that show an allowed region surrounded by a disallowed region (you do not need to reproduce each point in Figure \ref{mixScal}). 

Note that for each $(\Delta_\sigma,\Delta_\epsilon)$, you will need to compute a different table of derivatives $P_{\ell,R}^{m,n,\beta}(\Delta,\Delta_{\sigma\epsilon},\Delta_{\sigma\epsilon})$ and $P_{\ell,R}^{m,n,\beta}(\Delta,\Delta_{\epsilon\sigma},\Delta_{\sigma\epsilon})$, which can be assembled from the appropriate $Q^{m,n}_\ell(\Delta,\Delta_{ij},\Delta_{kl})$ that can be efficiently computed using \texttt{scalar\_blocks}. The crossing equations also depend on $P_{\ell,R}^{m,n,\beta}(\Delta,\Delta_{\sigma\sigma},\Delta_{\sigma\sigma})$, $P_{\ell,R}^{m,n,\beta}(\Delta,\Delta_{\epsilon\epsilon},\Delta_{\epsilon\epsilon})$, and $P_{\ell,R}^{m,n,\beta}(\Delta,\Delta_{\sigma\sigma},\Delta_{\epsilon\epsilon})$, which depend on non-mixed blocks and so need only be computed once as in the single correlator case.

\end{enumerate}

\section*{Acknowledgments} 

We thank Ofer Aharony for carefully reading through the manuscript, and him and Petr Kravchuk, Walter Landry, Silviu Pufu, David Simmons-Duffin, Ryan Thorngren and Emilio Trevisani for useful discussions.   Special thanks to Alessandro Vichi for the original wrapper code that \texttt{Bootstrap.m} is based on. We also thank all the members of the Weizmann 2019 Bootstrap class for pointing out many errors and typos, and especially Ohad Mamroud, Rohit Reghupathy, and Erez Urbach for troubleshooting the problem sets. SMC is supported by the Zuckerman STEM Leadership Fellowship.

\pagebreak

\bibliographystyle{ssg}
\bibliography{BootLecs}

\end{document}